\documentstyle[12pt,epsfig,amssymb]{article}
\voffset     = - 0.10 in
\hoffset     = - 0.00 in
\begin{document}
\thispagestyle{empty}

%{\large\bf
%\hfill
%\begin{tabular}{r}
%ATLAS Internal Note                \\
%TILE-CAL-NO-\mbox{~~~~}             \\
%\today                             \\
%\end{tabular}
%}

\vspace*{\fill}

\begin{center}
{\Large\bf
Non-Compensation of the Barrel Tile Hadron Module-0 Calorimeter 
}
\end{center}

\vspace*{\fill}

\begin{center}
{\large\bf Y.A.~Kulchitsky} 

\bigskip

{\it 
Institute of Physics, National Academy of Sciences,  Minsk,  Belarus \\
\& JINR,  Dubna,  Russia}

\bigskip
\bigskip

{\large\bf V.B.~Vinogradov}

\bigskip

{\it JINR,  Dubna,  Russia}

\end{center}

\vspace*{\fill}

%%%%%%%%%%%%%%%%%%%%%%%%%%%%%%%%%%%%%%%%%%%%%%%%%%%%%%%%%%%%%%%%%%%%%%%%%%
\begin{abstract}
The detailed experimental 
information about the electron and pion responses,  the
electron energy resolution
and the $e / h$ ratio as a function of incident energy $E$,  impact point $Z$
and incidence
angle $\Theta$ of the Module-0 
of the iron-scintillator barrel hadron calorimeter with the 
longitudinal tile configuration is presented.
The results are based on
the electron and pion beams data for 
E = 10,  20,  60,  80,  100 and  180  GeV at
$\eta = -0.25$ and $-0.55$, 
which have been obtained during the test beam period in 1996.
The results are compared with the existing experimental data
of TILECAL 1m prototype modules, various iron-scintillator calorimeters
and with some Monte Carlo calculations.
\end{abstract}

\vspace*{\fill}
\newpage

%%%%%%%%%%%%%%%%%%%%%%%%%%%%%%%%%%%%%%%%%%%%%%%%%%%%%%%%%%%%%%%%%%%%%%%%%%
\section{Introduction}

The ATLAS Collaboration is building a general-purpose pp detector
which is designed to exploit the full discovery potential of the CERN's
Large Hadron Collider (LHC),  a super-conducting ring to provide
proton -- proton collisions around 14 TeV \cite{atcol}.
LHC will open up new physics horizons,  probing interactions between proton
constituents at the 1 TeV level,  where new behavior is expected to reveal
key insights into the underlying mechanisms of Nature \cite{lhcnews}.

\begin{figure*}[tbph]
     \begin{center}
\mbox{\epsfig{figure=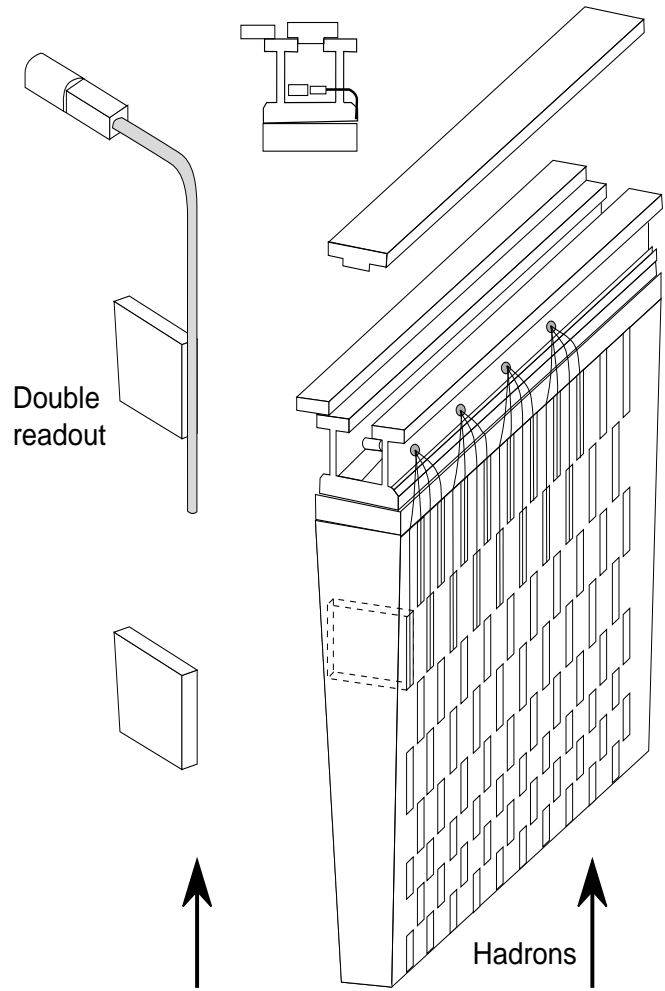,width=0.7\textwidth,height=0.8\textheight}}
     \end{center}
       \caption{
       Principle of the tile hadronic calorimeter.
       \label{f2-28tp}}
\end{figure*}

The bulk of the hadronic calorimetry in the ATLAS detector is provided
by a large (11 m in length,  8.5 m in outer diameter,  2 m in thickness, 
10000 readout channels)
scintillating tile hadronic barrel calorimeter (TILECAL).
The technology for this calorimeter is based on a sampling technique
using steel absorber material and scintillating plates readout by
wavelength shifting fibres.
An innovative feature of this design is the orientation of the scintillating
tiles which are placed in planes perpendicular to the colliding beams
staggered in depth \cite{gild-91} (Fig.~\ref{f2-28tp}).

In order to test this concept five 1m prototype modules and the
Module-0 
were built and exposed to high energy pion,  electron and muon beams at the
CERN Super Proton Synchrotron.

In the following we consider two test beam setups.
The setup 1, shown in Fig.\ 3-1 given in \cite{tdr96},
consists of five 1m prototype modules.
The obtained results about the electron and pion
responses and the $e / h$ ratio \cite{budagov95} 
for this setup  are used in this paper for comparison.
The setup in question (setup 2), shown in Fig.\ 3-2 given in \cite{tdr96}, 
has as the basis Module-0.

In this work the detailed experimental information is presented about the
electron and pion responses and the $e / \pi$ and $e / h$ ratios 
(an intrinsic non-compensation) of the Tile calorimeter Module-0.

%%%%%%%%%%%%%%%%%%%%%%%%%%%%%%%%%%%%%%%%%%%%%%%%%%%%%%%%%%%%%%%%%%%%%%%%%%
\section{The 1m Prototype Modules}

Each module spans 100 cm in the $Z$ direction, 
180 cm in the $X$ direction (about 9 interaction lengths at $\eta=0$
or about 80 effective radiation lengths)
and with  a front face of 100 $\times$ 20 cm$^2$ \cite{berger}.
The iron structure of each module consists of 57 repeated "periods".
Each period is 18~mm thick and consists of four layers. 
The first and third layers are formed by large trapezoidal steel plates
(master plates),  and spanning the full longitudinal dimension of the
module.
In the second and fourth layers,  smaller trapezoidal steel plates
(spacer plates) and scintillator tiles alternate along the $X$ direction.
These layers consist of 18 different trapezoids of steel or scintillator, 
each spanning 100 mm along X.

The master plates,  spacer plates and scintillator tiles are 5 mm,  4 mm and
3 mm thick,  respectively.
The iron to scintillator ratio is 4.67:1 by volume.

Wavelength shifting fibres collect scintillation light from the tiles at both
of their open edges and bring it to photo-multipliers (PMTs) at the periphery
of the calorimeter.
Each PMT views a specific group of tiles through the corresponding
bundle of fibres.

The modules are longitudinally segmented into four depth segments by
grouping fibers from different tiles.
As a result,  each module is divided into
$5\ (along\ Z) \times 4\ (along\ X)$
separate cells.
The readout cells have the lateral dimensions
$ 200\ mm\ (along\ Z) \times (200-380)\ mm$
(along Y,  depending on a depth number)
and the longitudinal dimensions 300,  400,  500,  600 mm for depths 1 -- 4, 
corresponding to 
1.5,  2,  2.5 and 3 $\lambda_{I}$ at $\eta =0$.
At the output we have  200 values of responses $Q_{ijkl}$
from PMT properly calibrated \cite{berger} with pedestal subtracted,  
for each event.
Here
$i = 1,  \ldots,  5$ is the column of cells (tower) number, 
$j = 1,  \ldots,  5$ is the module number, 
$k = 1,  \ldots,  4$ is the depth number and $l = 1,  2$ is the PMT number.

%%%%%%%%%%%%%%%%%%%%%%%%%%%%%%%%%%%%%%%%%%%%%%%%%%%%%%%%%%%%%%%%%%%%%%%%%%
\section{The Module-0}

The layout of the readout cell geometry for the Module-0 is shown
in Fig.\ 3-3 given in \cite{tdr96}.
The Module-0 has three depth segmentations.
The thickness of the Module-0 at $\Theta = 0^{o}$ is 1.5 $\lambda$ in
the first depth sampling,  4.2 $\lambda$ in the second and 1.9 $\lambda$
in the third with a total depth of 7.6 $\lambda$.
The 
Module-0 samples the shower with 11 tiles varying in depth from 97 to 187 mm.
The front face area is of $560 \times 22\ cm^{2}$.

In the setup 2 (see Fig.3-2 given in \cite{tdr96})
the 1m prototype modules are placed on a scanning table
on top and at the bottom of the Module-0
with a 10 cm gap between them.
This scanning table allowed movement in any direction.
Upstream of the calorimeter,  a trigger counter telescope (S1-S3)
 was installed, 
defining a beam spot of 2~cm in diameter.
Two delay-line wire chambers (BC1-BC2), 
each with $Z$,  $Y$  readout, 
allowed  the impact point of beam particles on the
calorimeter face to be reconstructed to better than $\pm 1$~mm 
\cite{ariz-94}.
A helium \v{C}erenkov threshold counter was used to tag
$\pi$-mesons and electrons for $E$ =10 and 20 $GeV$.
For the measurements of the hadronic shower longitudinal and lateral
leakages back ($80\times80 cm^2$) and side ($40\times115 cm^2$) "muon walls"
were placed behind and on the side of the calorimeter.

%%%%%%%%%%%%%%%%%%%%%%%%%%%%%%%%%%%%%%%%%%%%%%%%%%%%%%%%%%%%%%%%%%%%%%%%%%
\section{Data Taking and Event Selection}

Data were taken with electron and pion beam of
E = 10,  20,  60,  80,  100 and  180  GeV at
$\eta = -0.25$ and $-0.55$, 
The following 6 cuts were used.
The cuts 1 and 2 removed beam halo.
The cut 3 removed muons and non-single-track events.
The cuts 4, 5 and 6 carried out the electron-pion separation
The cut 4 is connected with \v{C}erenkov counter amplitude.
Cut 5 is the relative shower energy deposition in the first two
calorimeter depths: 
\begin{equation}
\label{ci}
C_{i} = \sum_{selected\  i}\  \sum_{j = 3} \sum_{k = 1}^{2}
\sum_{l = 1}^{2} Q_{ijkl} / E, 
\end{equation}
where 
\begin{equation}
\label{en}
E = \sum_{ijkl} Q_{ijkl}.
\end{equation}
and the indexes $i$ and $k$ in $Q_{ijkl}$ determine
the regions of electromagnetic shower development.
The values $C_i$ depend on a particle's entry angle $\Theta$.
The basis for the electron-hadron separation by using the cut 5
 is the very different longitudinal
energy deposition for electrons and hadrons.

The cut 6 is related with the lateral shower spread
\cite{juste95}:
\begin{equation}
\label{ev5}
E_{cut} = 
\frac{ \sqrt{ \sum_c (E_c^{\alpha} - \sum_c E_c^{\alpha}/N_{cell})^2 }}
{ \sum_c E_c^{\alpha} } , 
\end{equation}
where $1 \le c \le N_{cell}$ and $N_{cell}$
is the used cells number.
The power parameter $\alpha = 0.6$ have been tuned in \cite{juste95}
to achieve maximum separation efficiency. 

\begin{figure*}[tbph]
     \begin{center}
        \begin{tabular}{cc}
        \mbox{\epsfig{figure=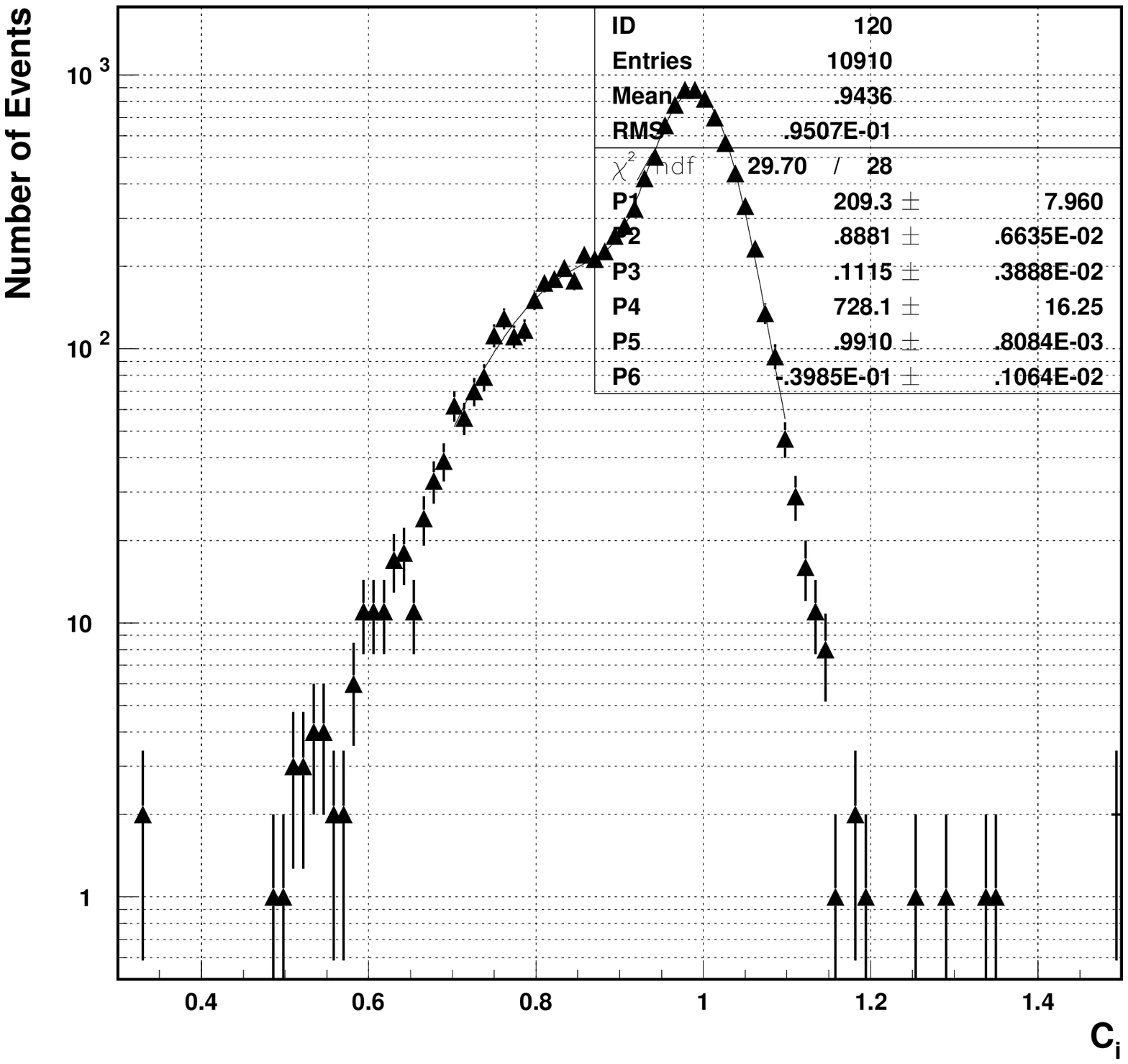,width=0.45\textwidth,height=0.25\textheight}}
        &
        \mbox{\epsfig{figure=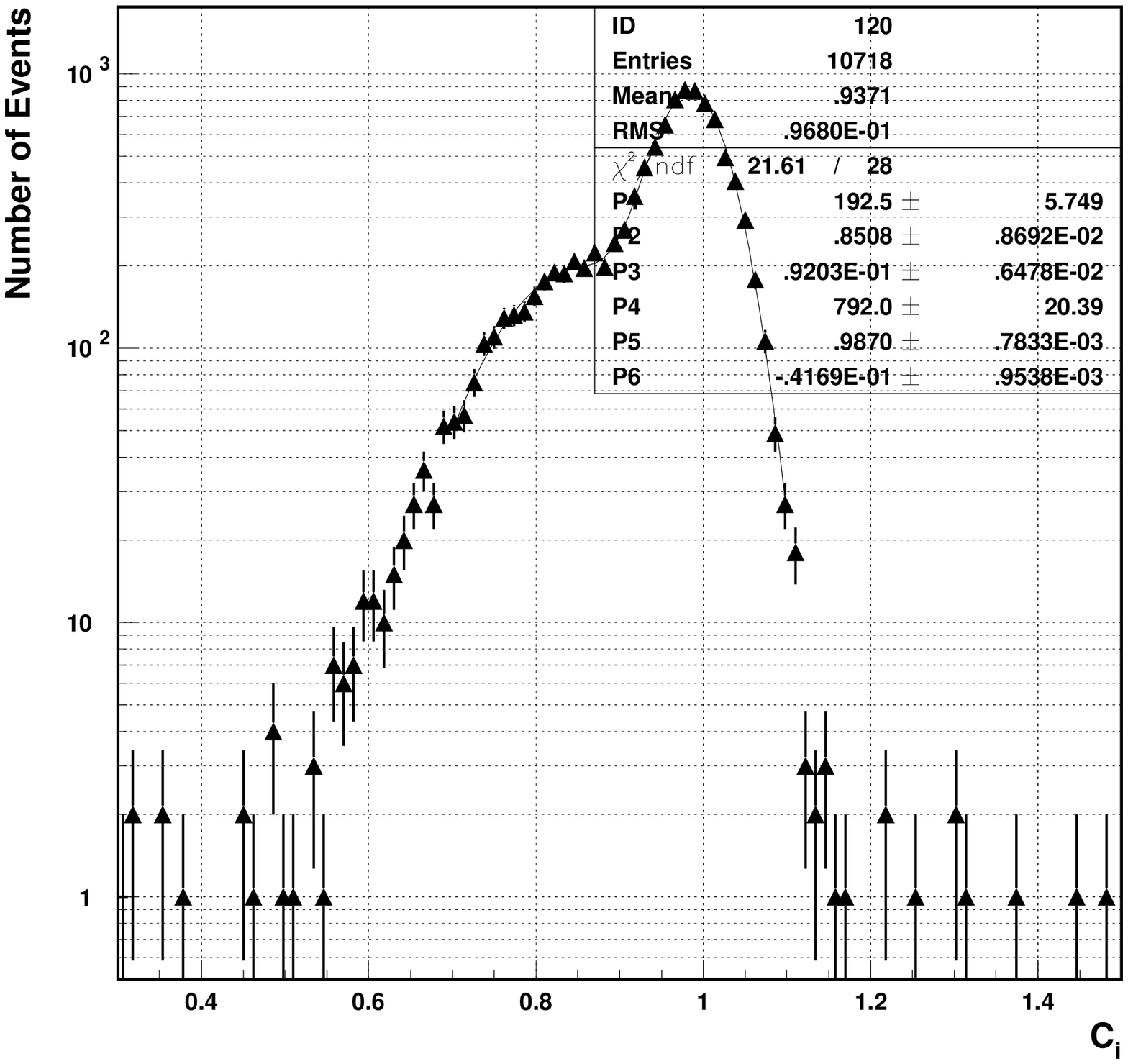,width=0.45\textwidth,height=0.25\textheight}}
        \\
        \mbox{\epsfig{figure=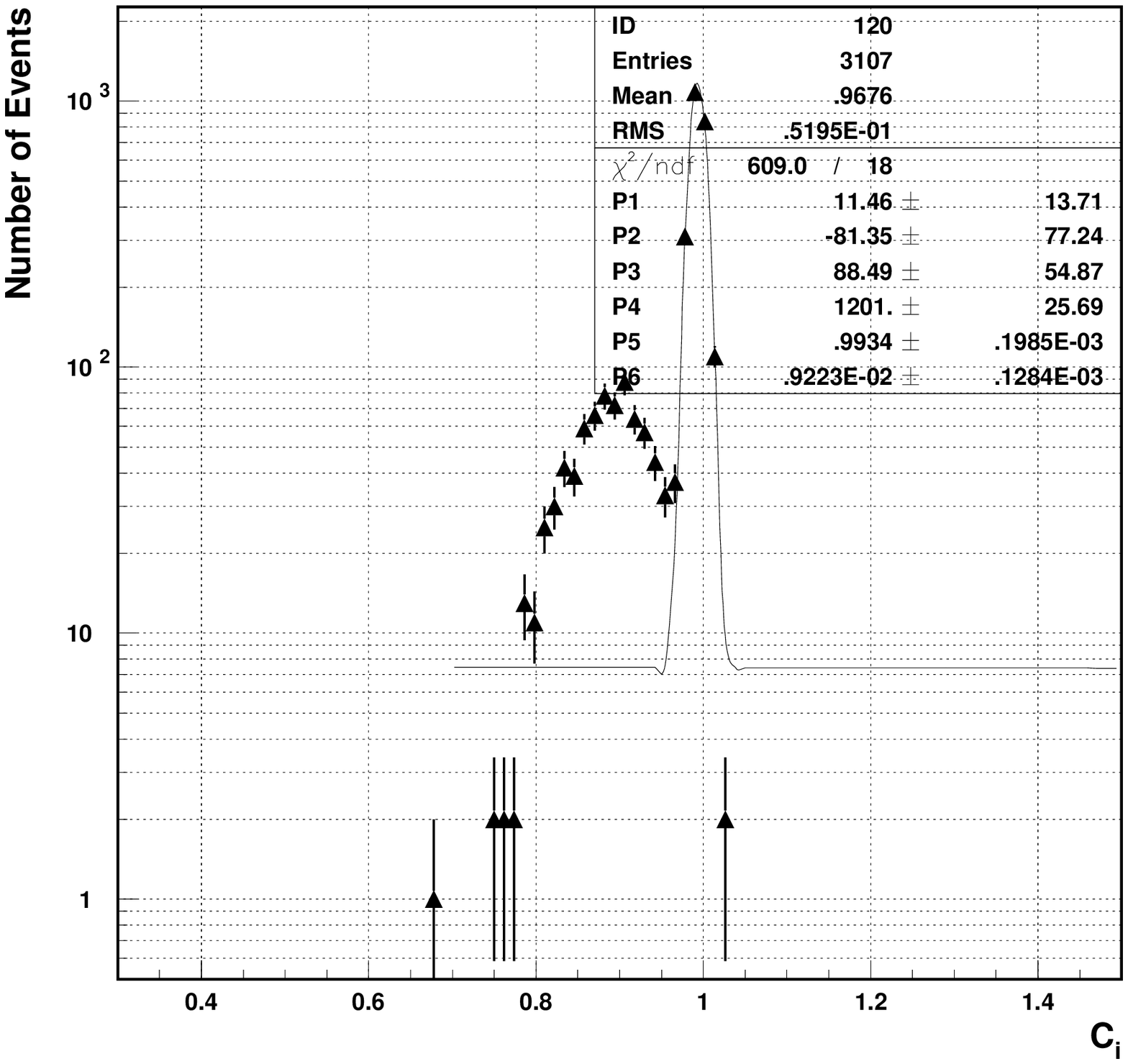,width=0.45\textwidth,height=0.25\textheight}}
        &
        \mbox{\epsfig{figure=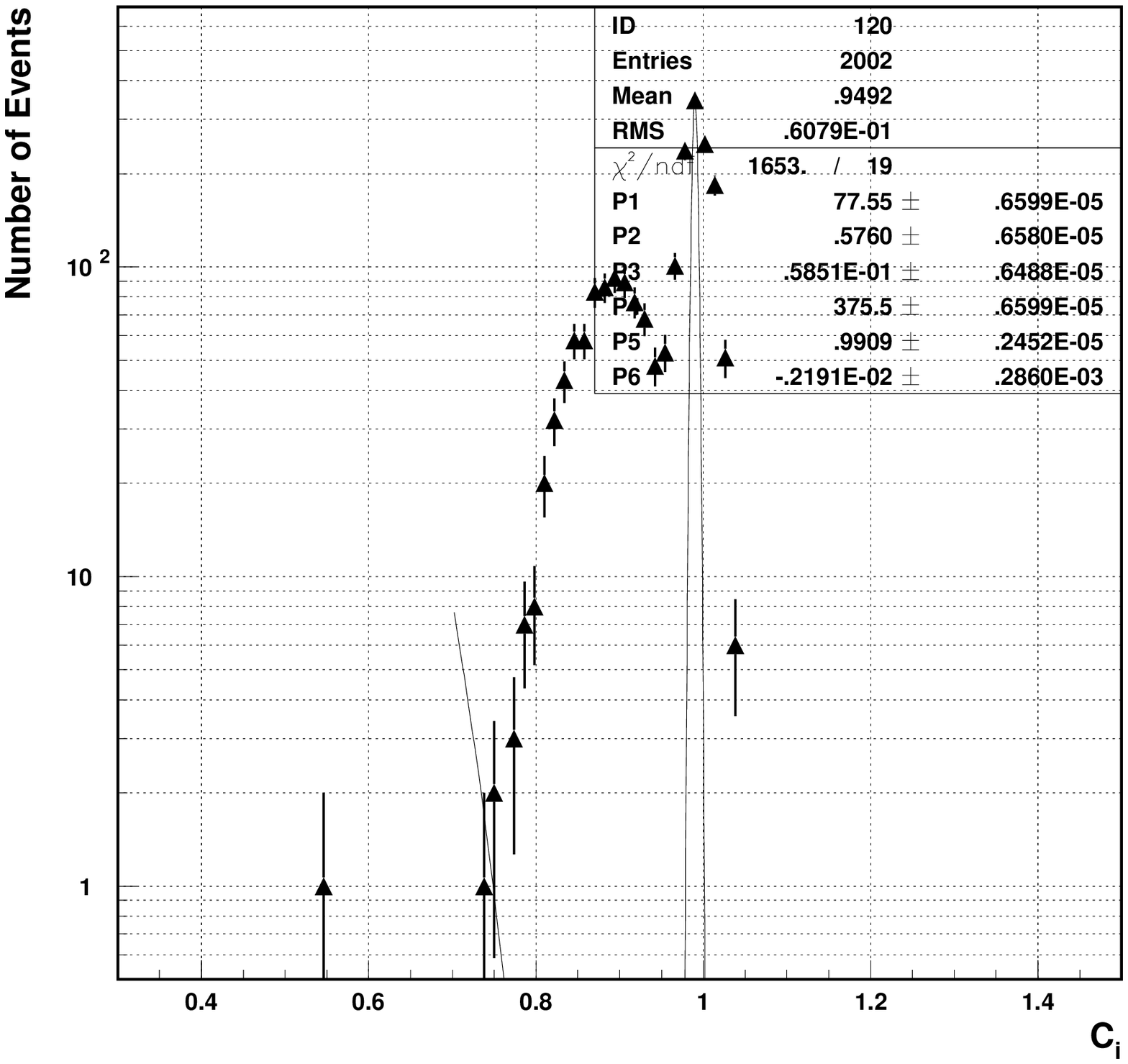,width=0.45\textwidth,height=0.25\textheight}}
        \\
        \mbox{\epsfig{figure=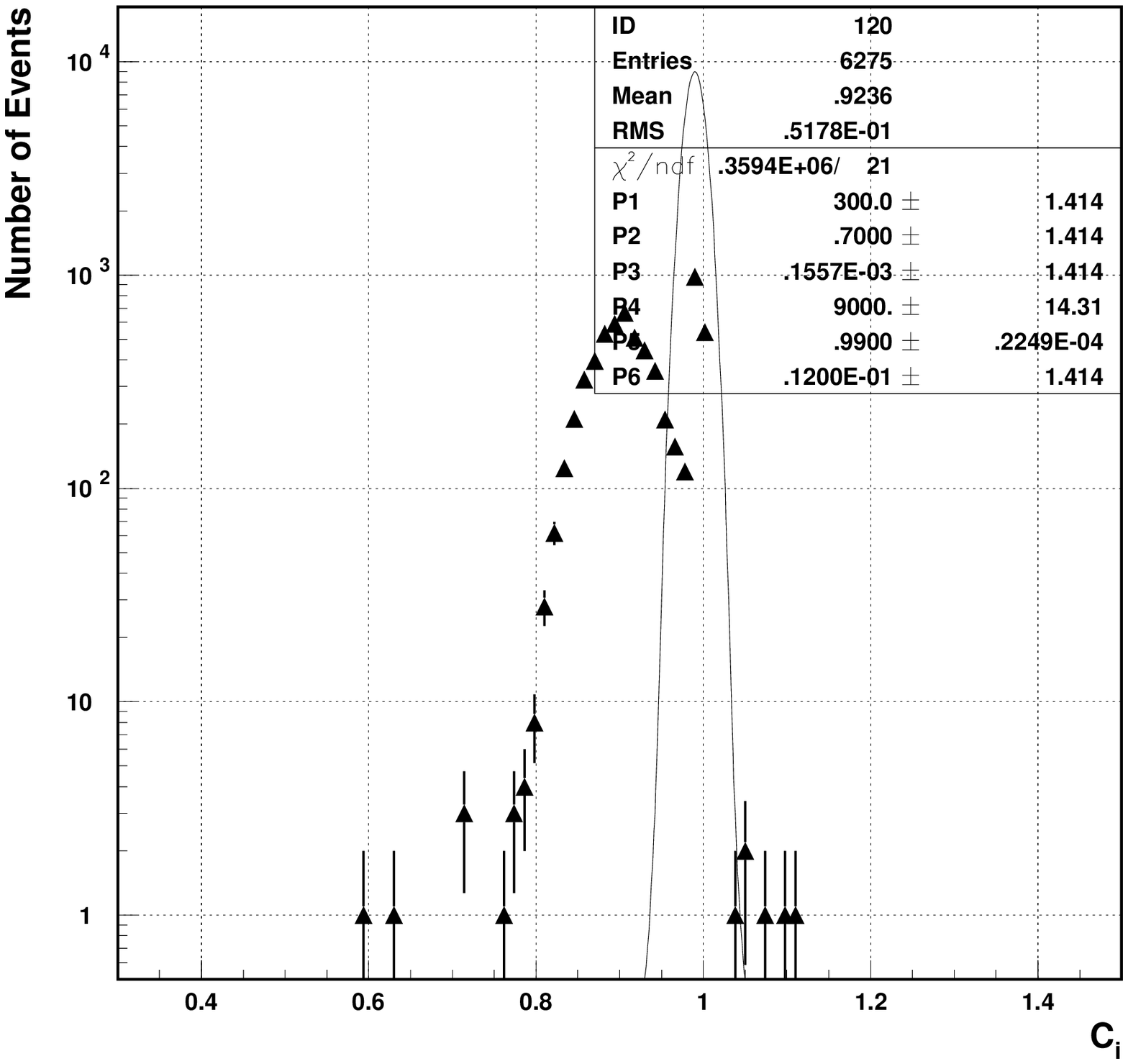,width=0.45\textwidth,height=0.25\textheight}}
        &
        \mbox{\epsfig{figure=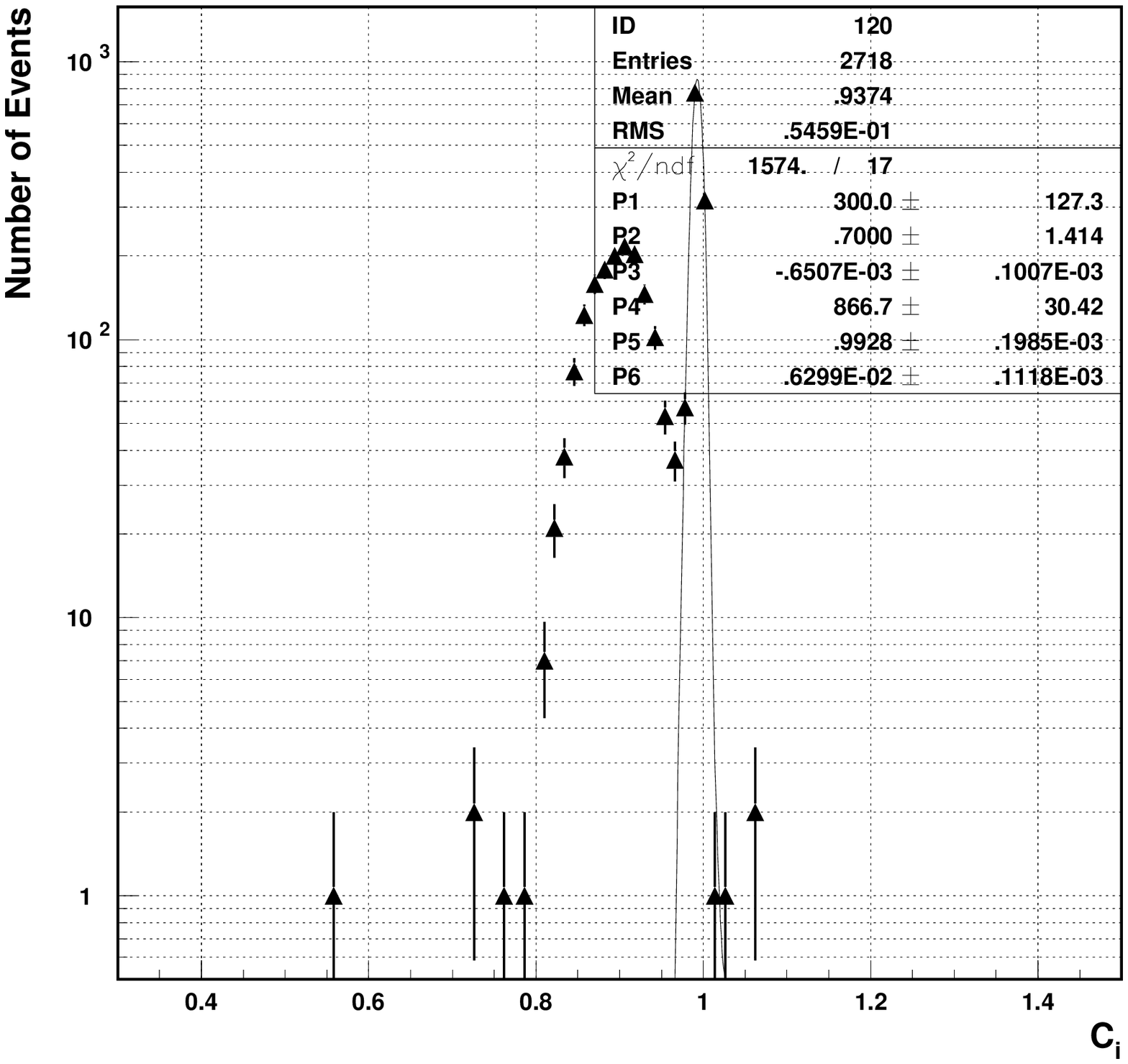,width=0.45\textwidth,height=0.25\textheight}}
        \\
        \end{tabular}
     \end{center}
       \caption{
        The distributions of the events as a function of $C_i$ for
	for E = 20, 100, 180 GeV
	at $\eta = -0.25$ (left column, up to down)   and
	at $\eta = -0.55$    (right column, up to down).
                      \label{fv29}}
\end{figure*}

\begin{figure*}[tbph]
     \begin{center}
      \begin{tabular}{cc}
      \mbox{\epsfig{figure=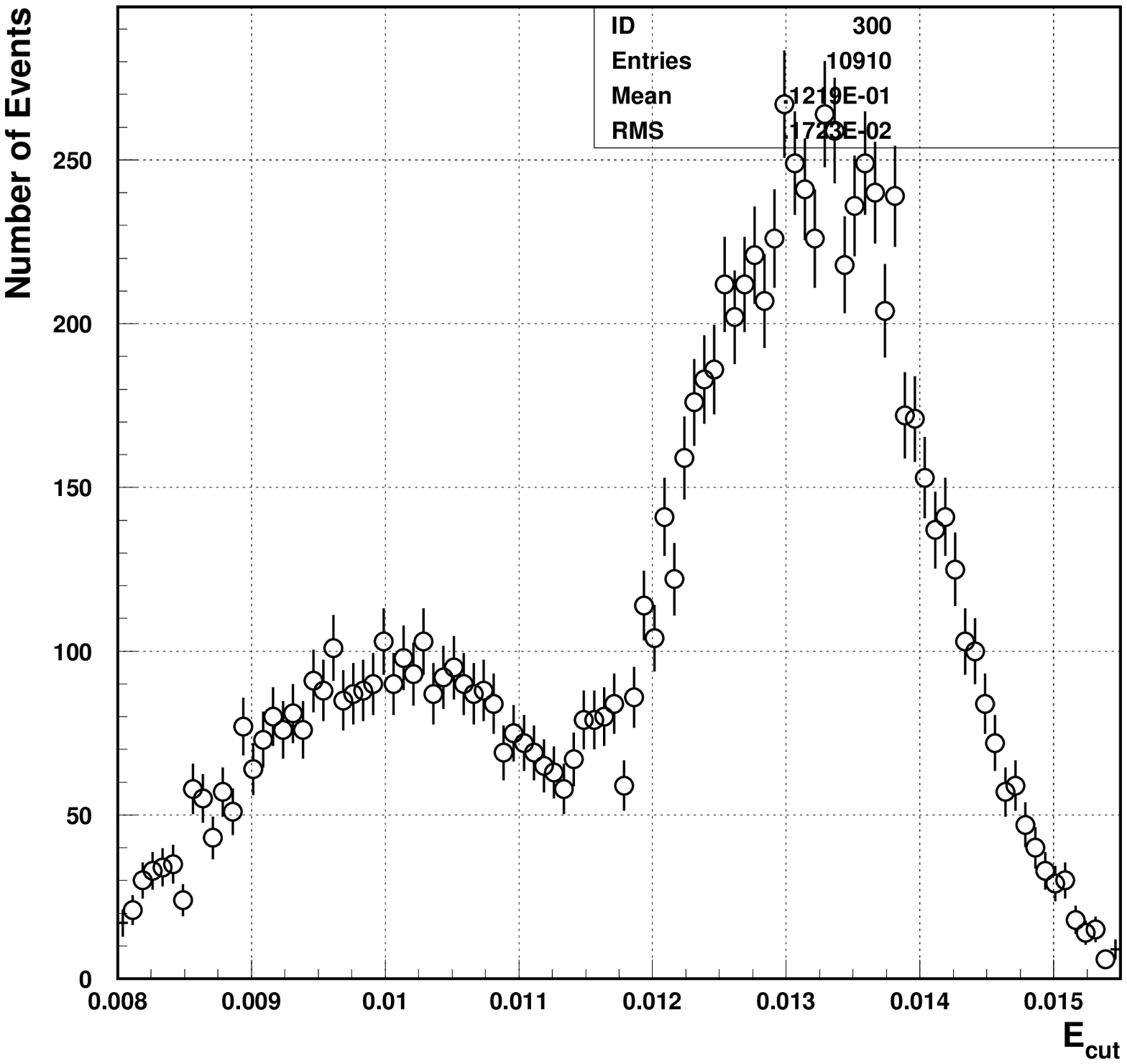,width=0.45\textwidth,height=0.25\textheight}}
      &
      \mbox{\epsfig{figure=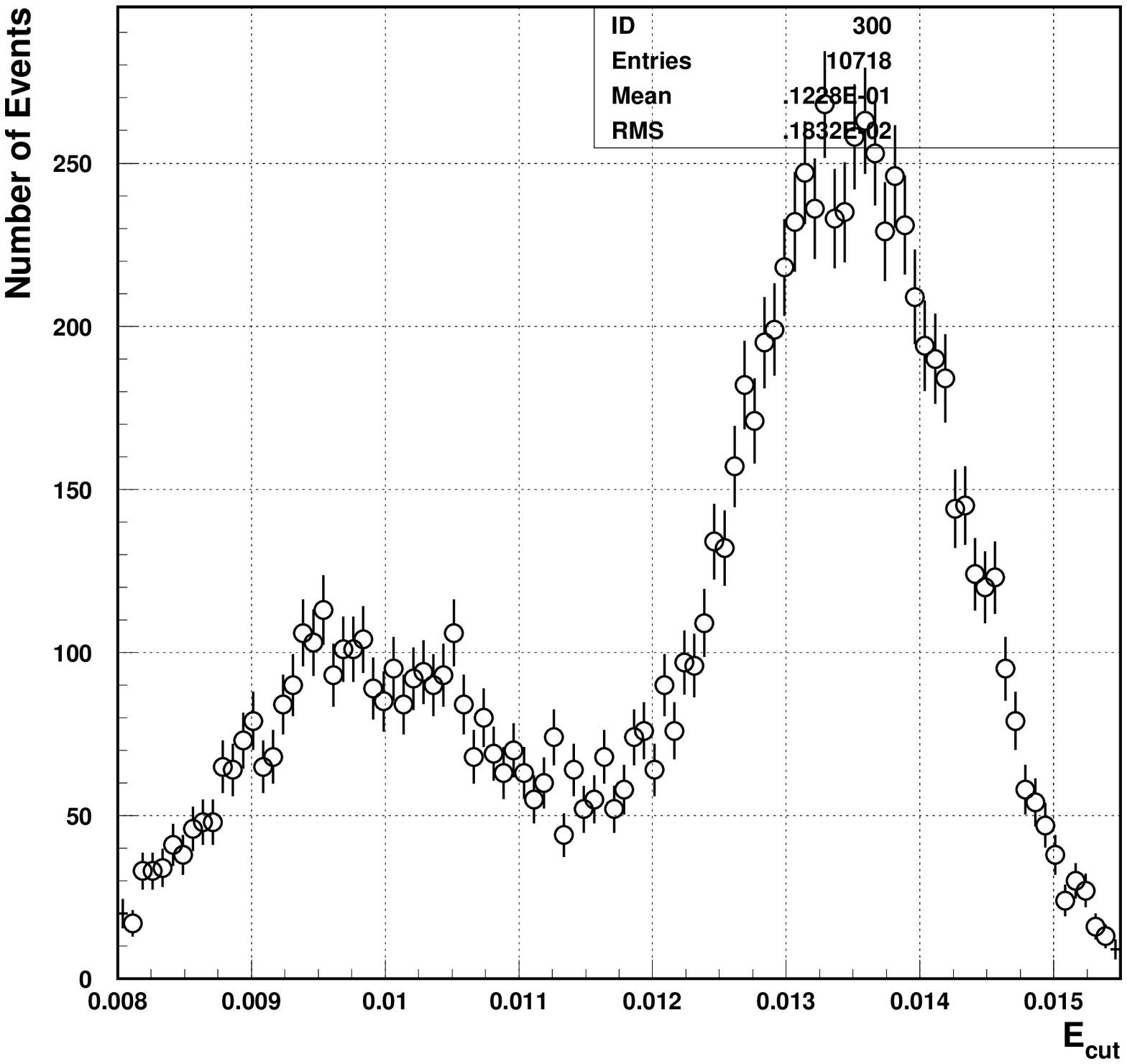,width=0.45\textwidth,height=0.25\textheight}}
      \\
      \mbox{\epsfig{figure=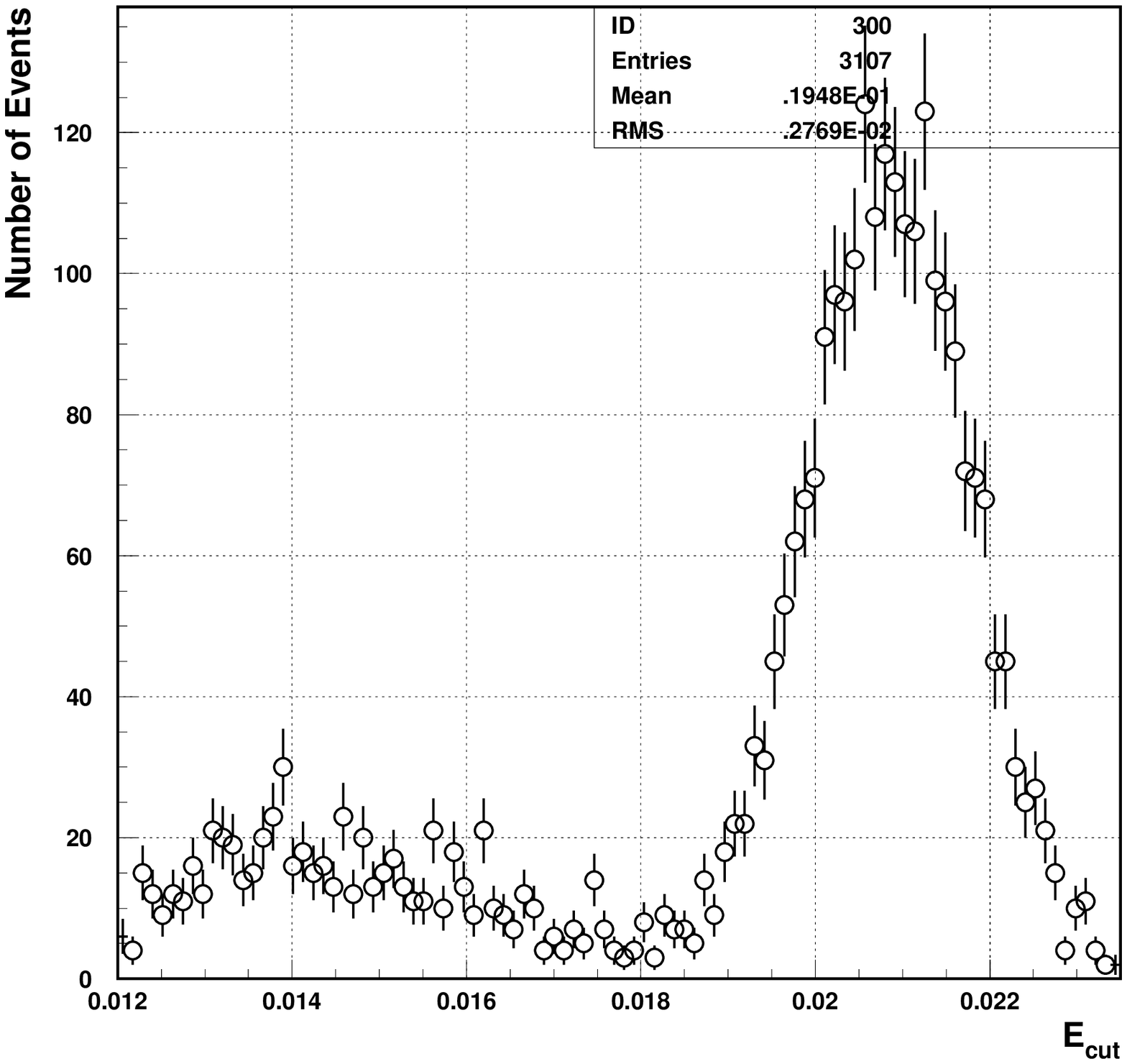,width=0.45\textwidth,height=0.25\textheight}}
      &
      \mbox{\epsfig{figure=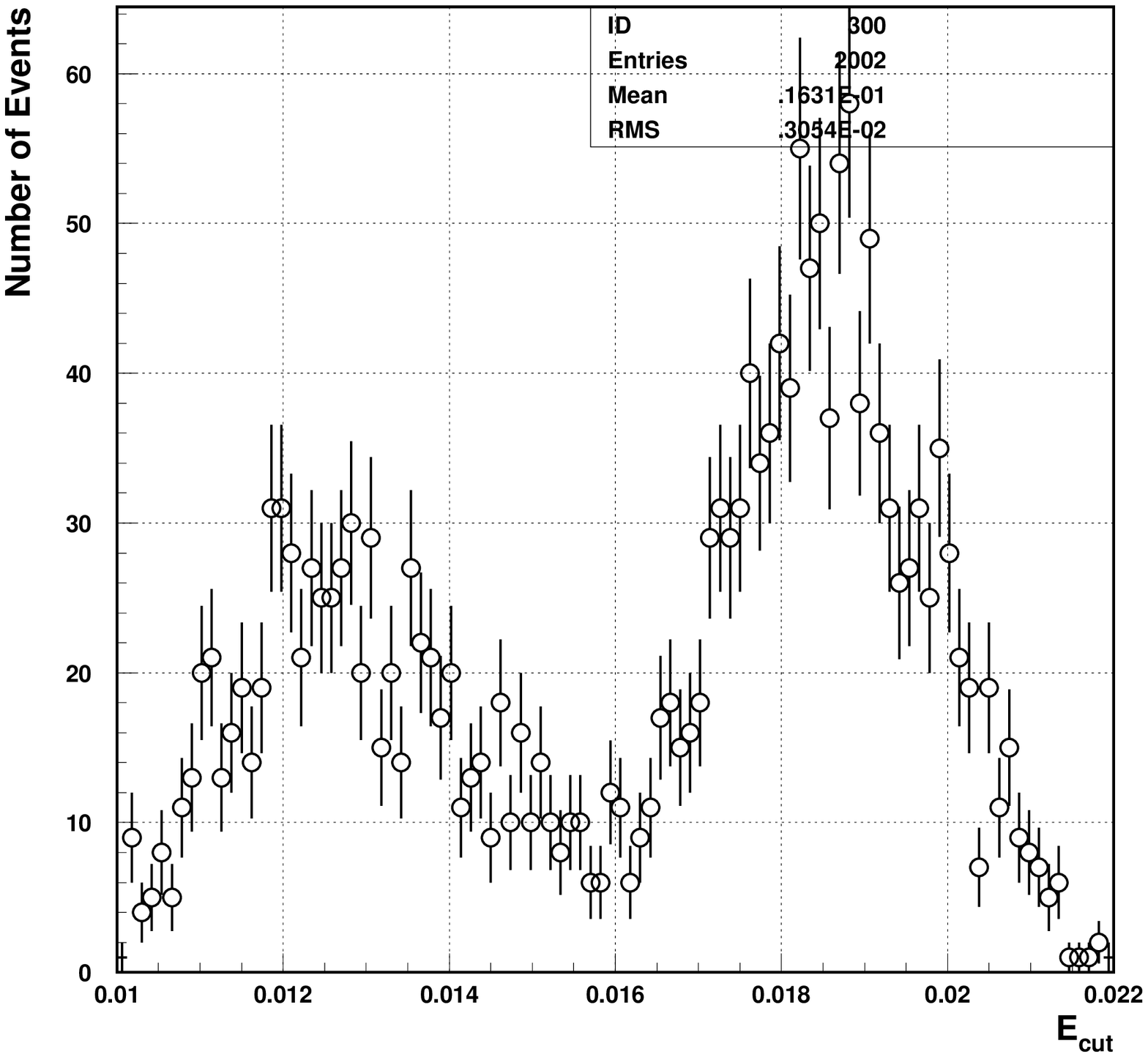,width=0.45\textwidth,height=0.25\textheight}}
      \\
      \mbox{\epsfig{figure=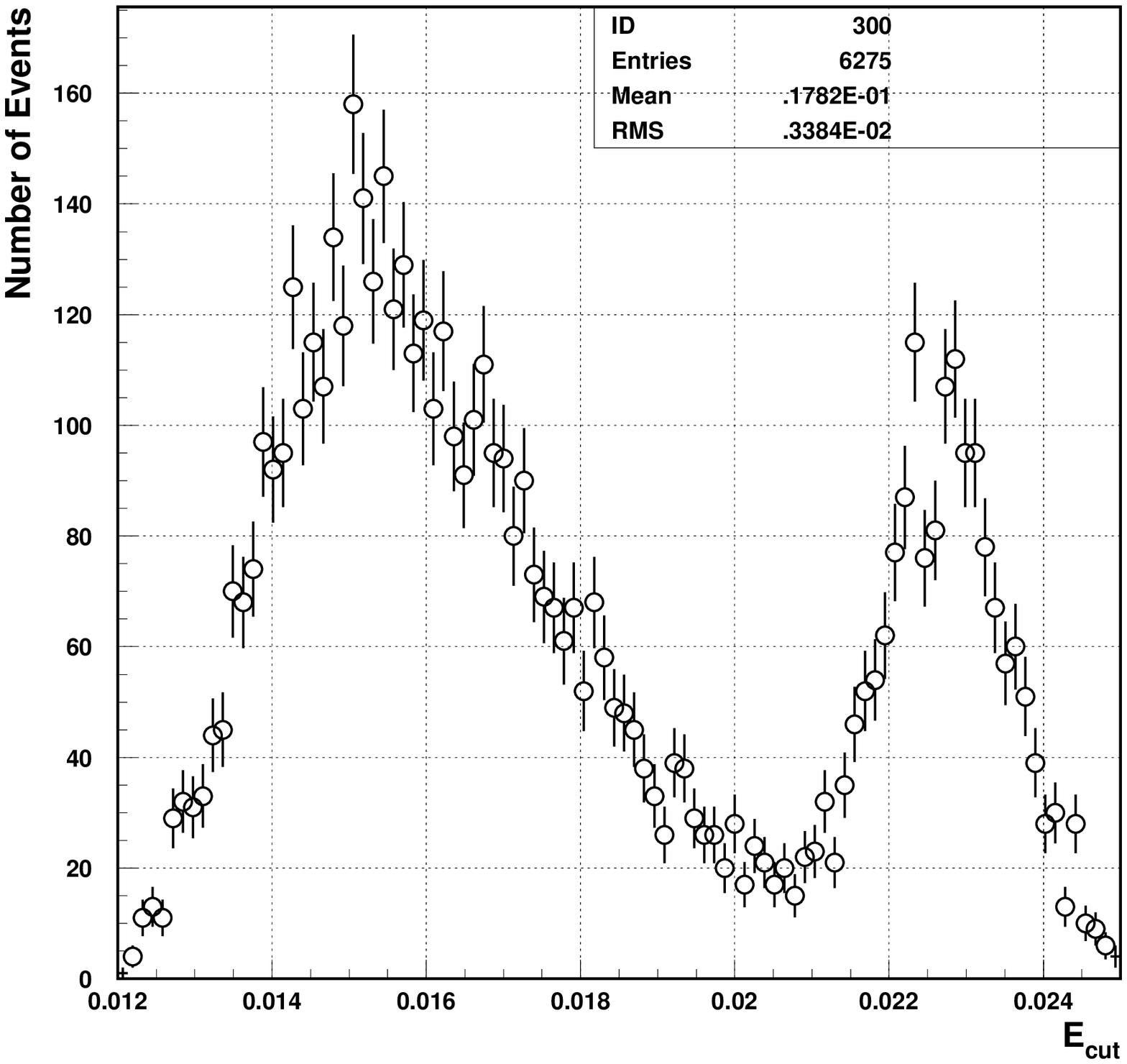,width=0.45\textwidth,height=0.25\textheight}}
      &
      \mbox{\epsfig{figure=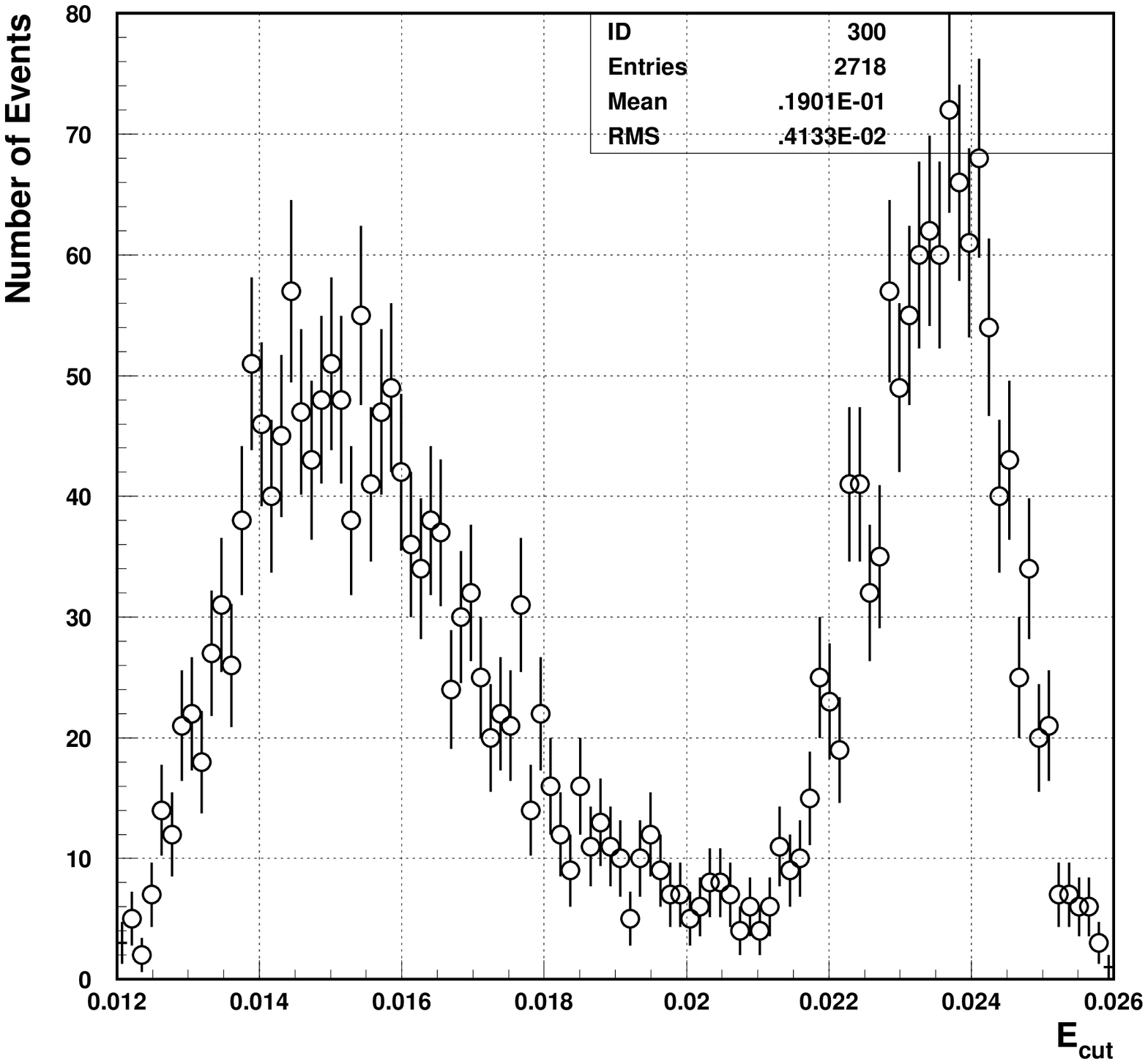,width=0.45\textwidth,height=0.25\textheight}}
      \\
         \end{tabular}
     \end{center}
       \caption{ 
	The distributions of the events as a function of $E_{cut}$
	for E = 20, 100, 180 GeV
	at $\eta = -0.25$ (left column, up to down)   and
	at $\eta = -0.55$    (right column, up to down).
       \label{fv28}}
\end{figure*}

\begin{figure*}[tbph]
     \begin{center}
      \begin{tabular}{cc}
      \mbox{\epsfig{figure=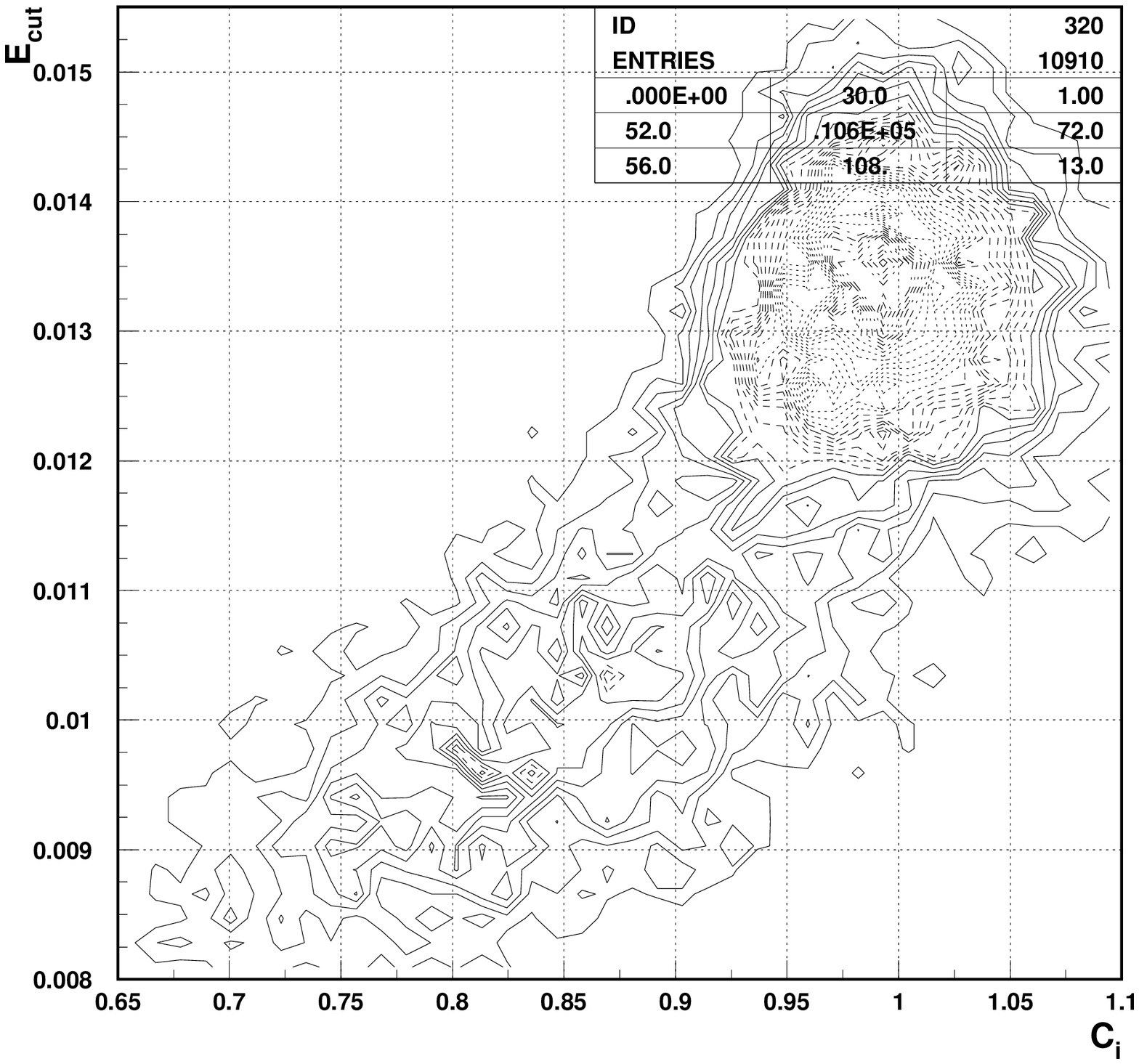,width=0.45\textwidth,height=0.25\textheight}}
      &
      \mbox{\epsfig{figure=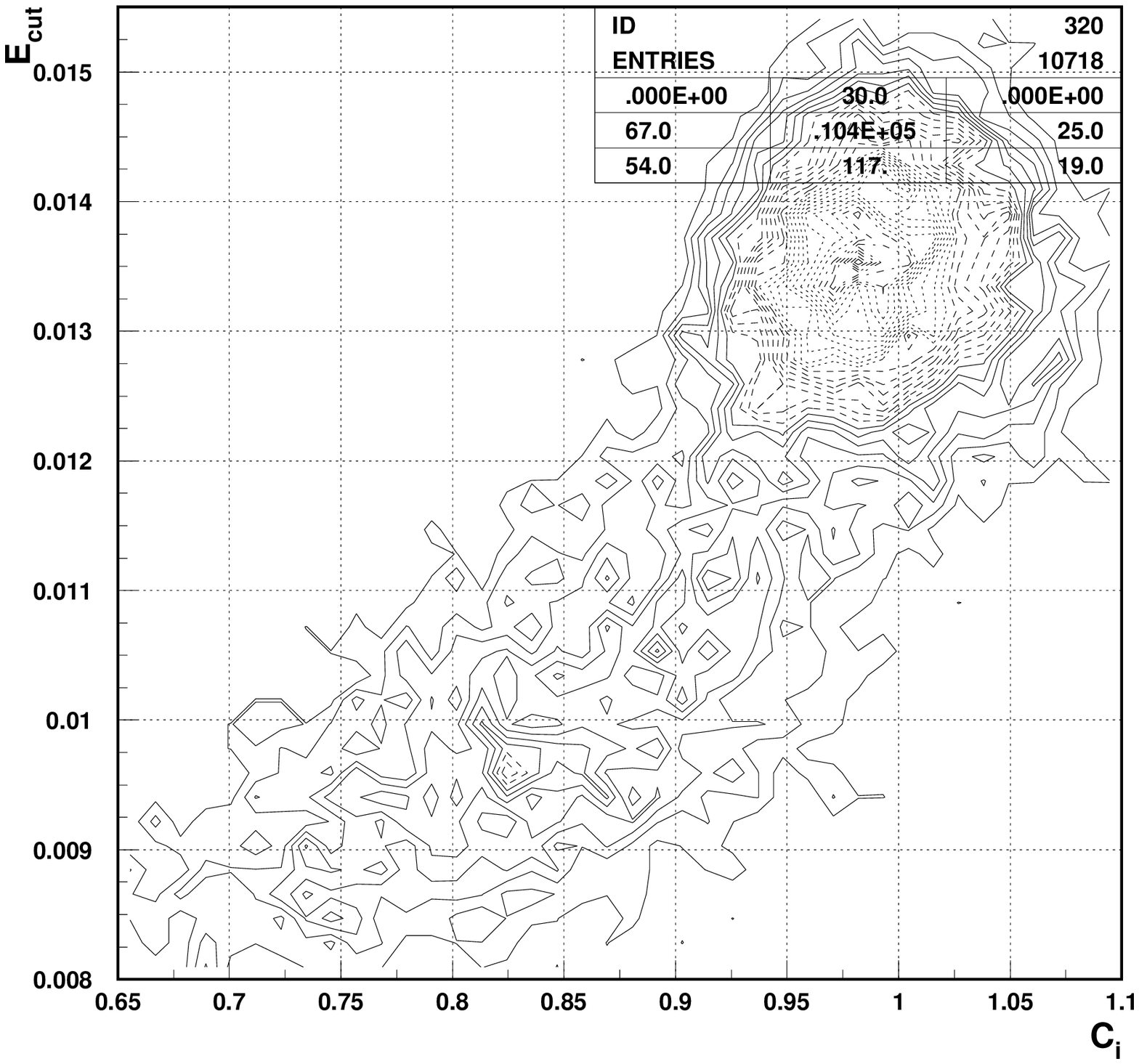,width=0.45\textwidth,height=0.25\textheight}}
      \\
      \mbox{\epsfig{figure=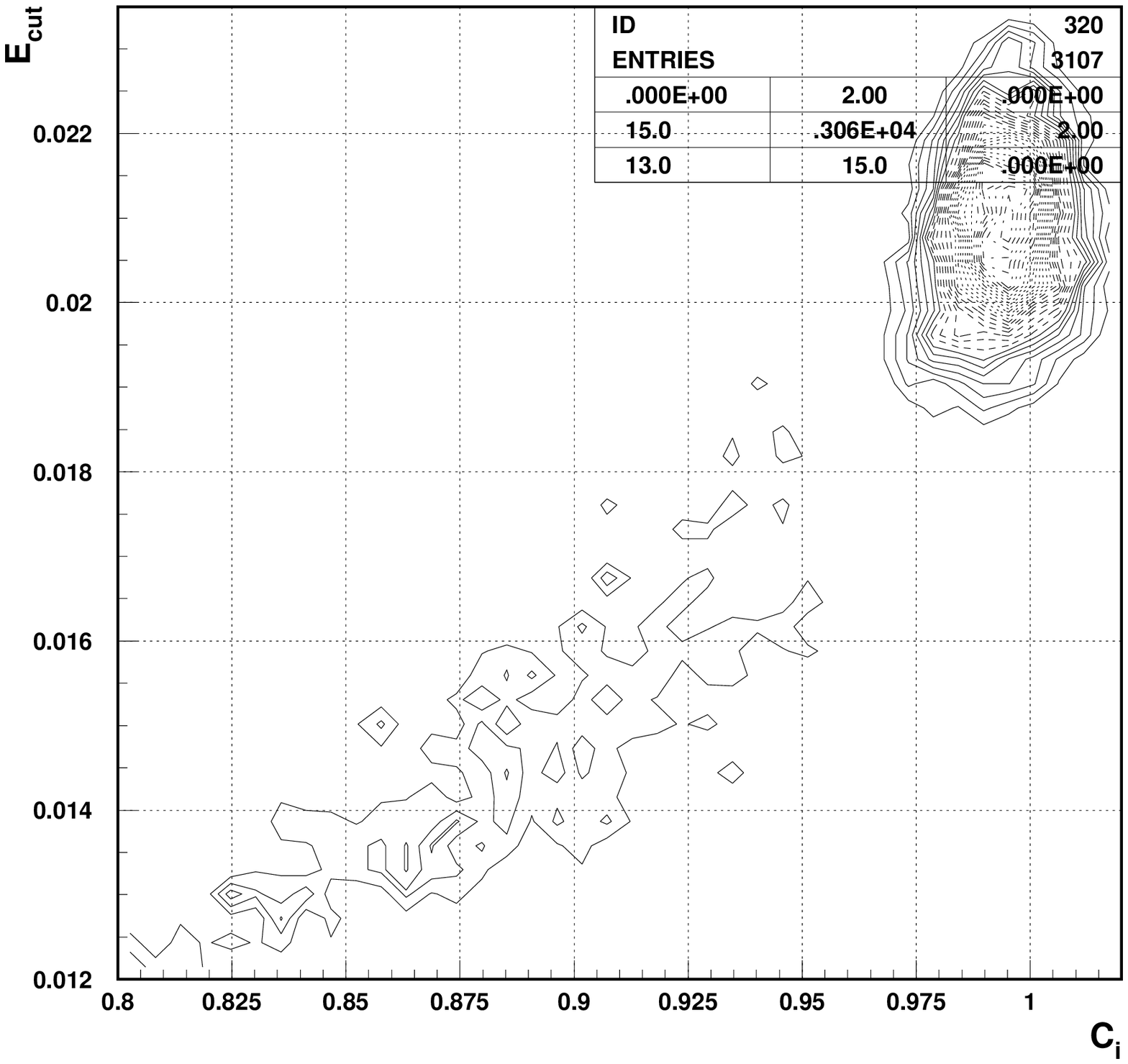,width=0.45\textwidth,height=0.25\textheight}}
      &
      \mbox{\epsfig{figure=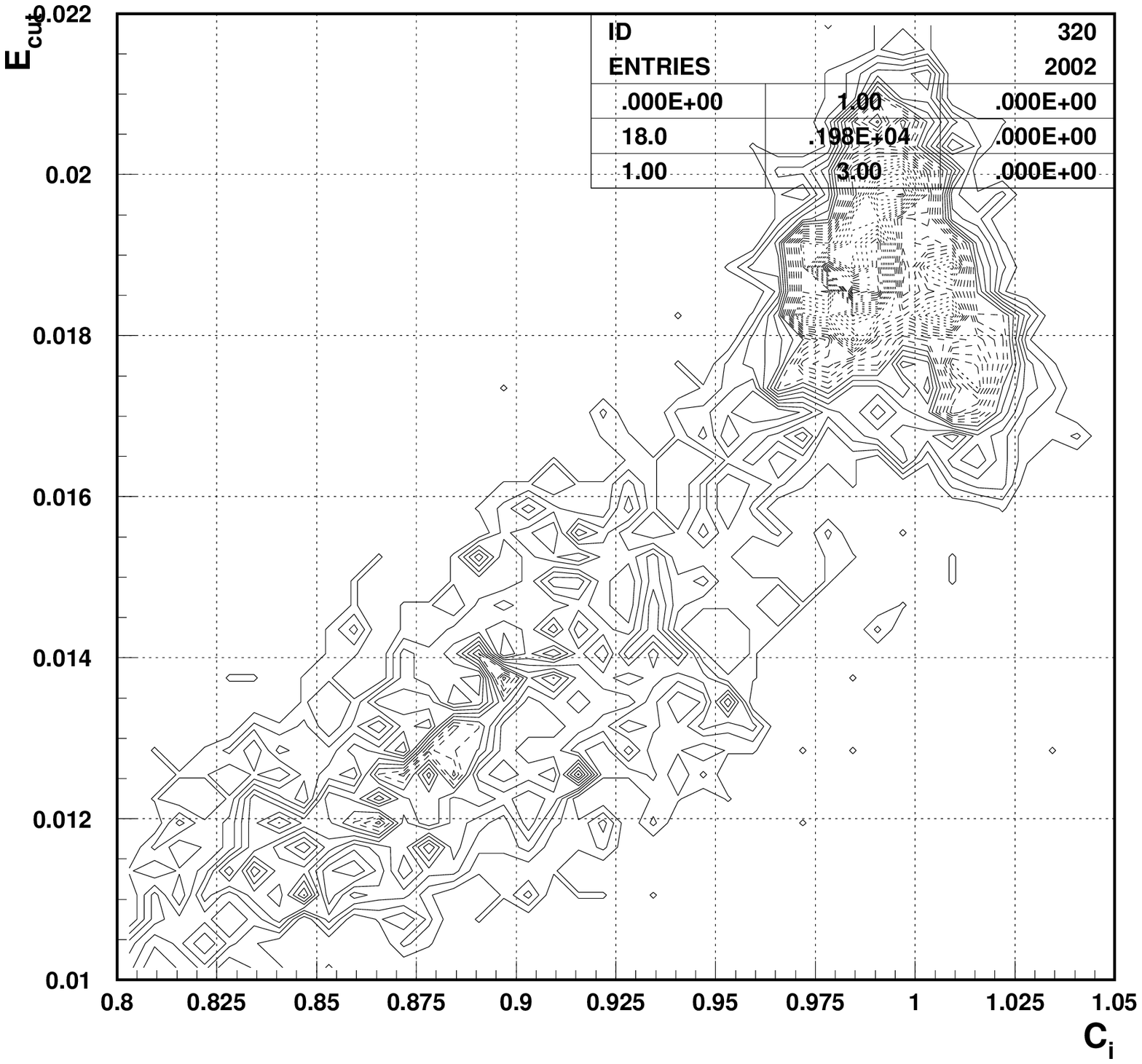,width=0.45\textwidth,height=0.25\textheight}}
      \\
      \mbox{\epsfig{figure=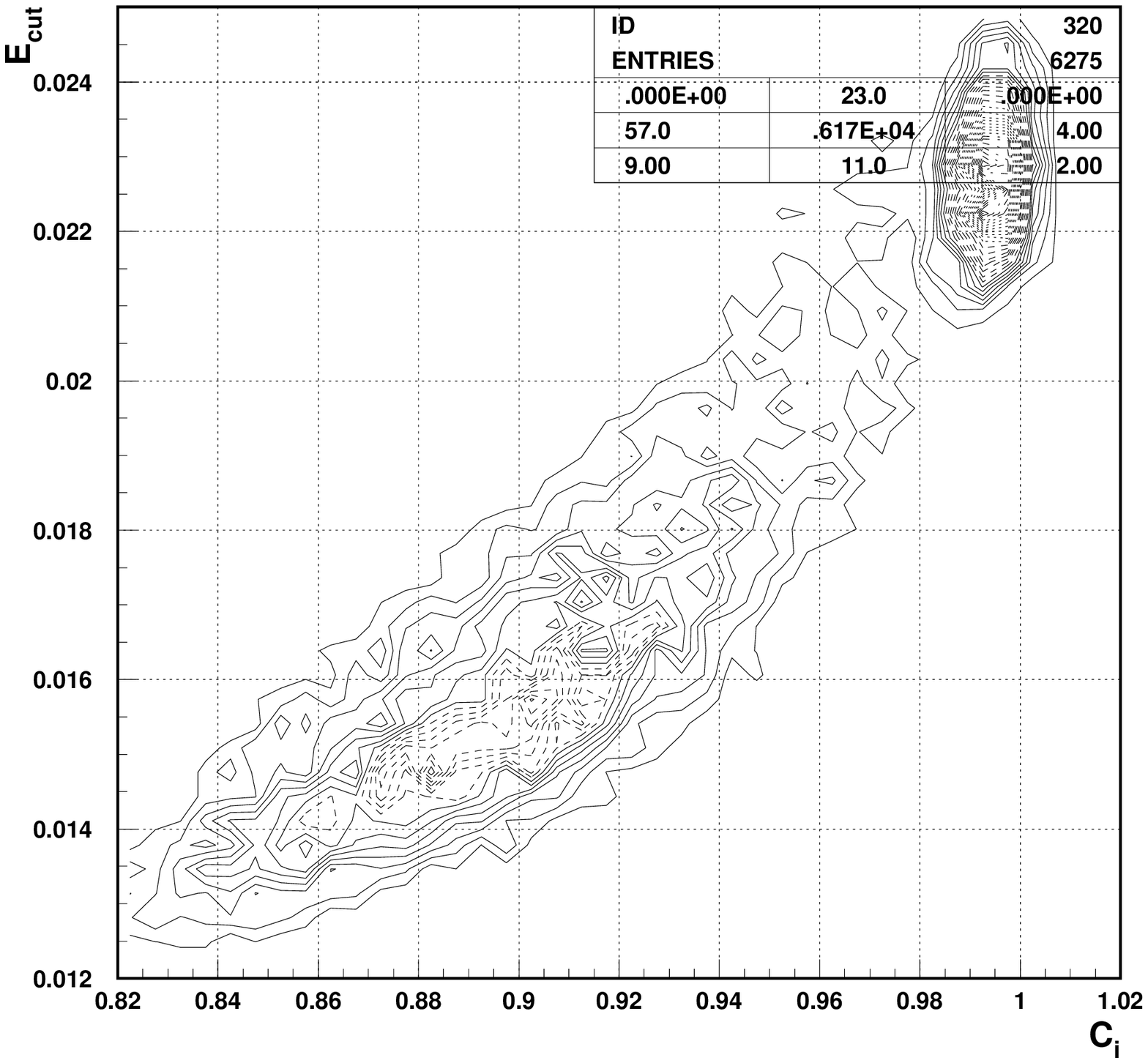,width=0.45\textwidth,height=0.25\textheight}}
      &
      \mbox{\epsfig{figure=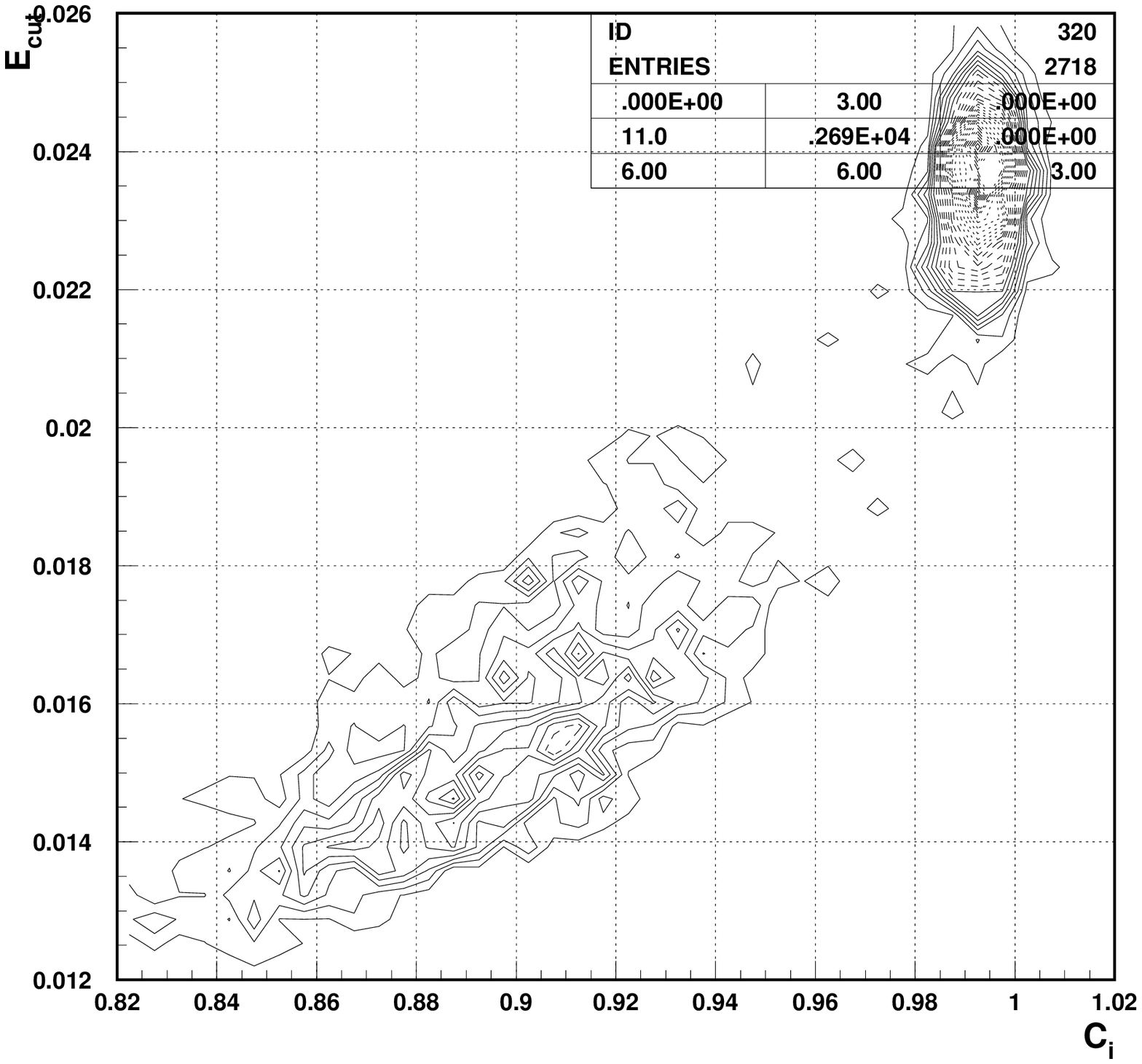,width=0.45\textwidth,height=0.25\textheight}}
      \\
         \end{tabular}
     \end{center}
       \caption{
	The scatter plots  $E_{cut}$ versus $C_i$
	for E = 20, 100, 180 GeV
	at $\eta = -0.25$ (left column, up to down)   and
	at $\eta = -0.55$    (right column, up to down).
       \label{fv23}}
\end{figure*}

The distributions of events as a function of $C_i$ and $E_{cut}$ 
for various energies at  $\eta = -0.25$ and $\eta = -0.55$ are shown in
Fig.~\ref{fv29} and Fig.~\ref{fv28}.
Fig.~\ref{fv23} shows the scatter plots  $E_{cut}$ versus $C_i$.
Two groups of events are clearly separated:
the left group corresponds to pions,  
the right group corresponds to electrons.

%%%%%%%%%%%%%%%%%%%%%%%%%%%%%%%%%%%%%%%%%%%%%%%%%%%%%%%%%%%%%%%%%%%%%%%%%%
\section{Electrons Response}

As to the electron response our calorimeter is very complicated object.
It may be imagined as a continuous set of calorimeters with the
variable absorber and scintillator thicknesses
(from $t$ = 58 to 28 mm and from $s$ = 12 to 6 mm
for $14^{o} \leq \Theta \leq 30^{o}$), 
where $t$ and $s$ are the thicknesses of absorber and scintillator
respectively.

\begin{figure*}[tbph]
     \begin{center}
        \begin{tabular}{cc}
        \mbox{\epsfig{figure=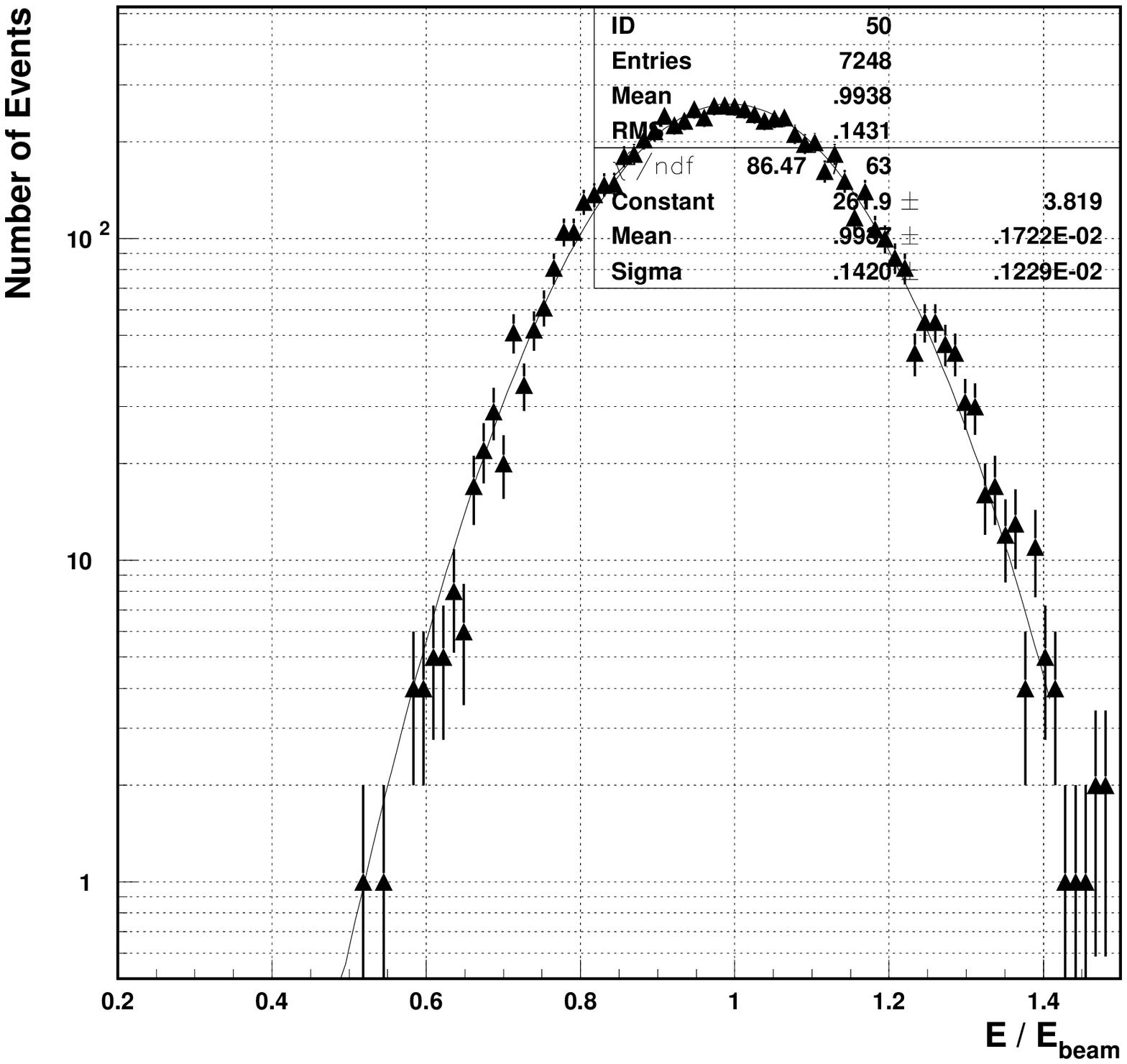,width=0.45\textwidth,height=0.25\textheight}}
        &
        \mbox{\epsfig{figure=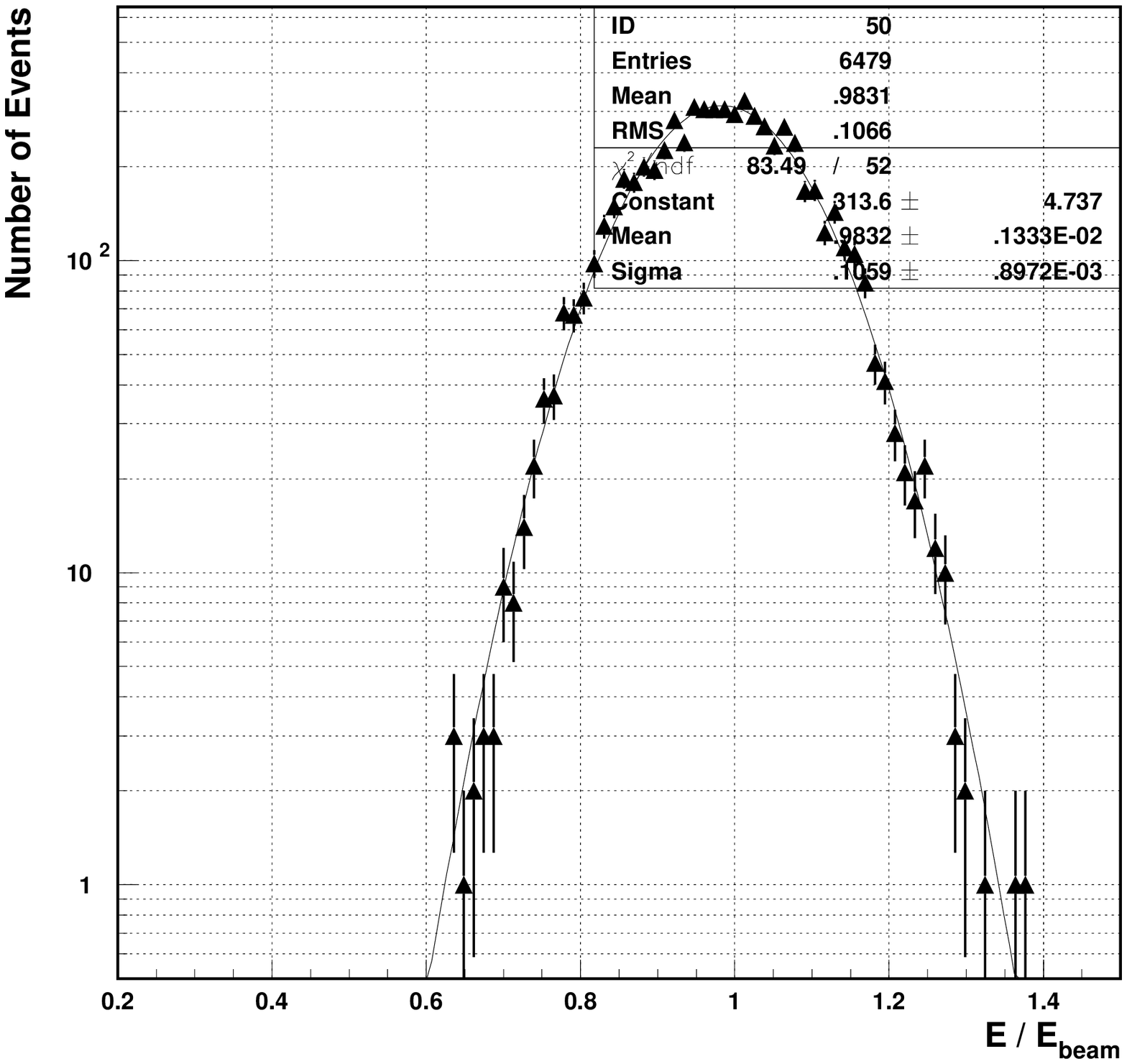,width=0.45\textwidth,height=0.25\textheight}}
        \\
        \mbox{\epsfig{figure=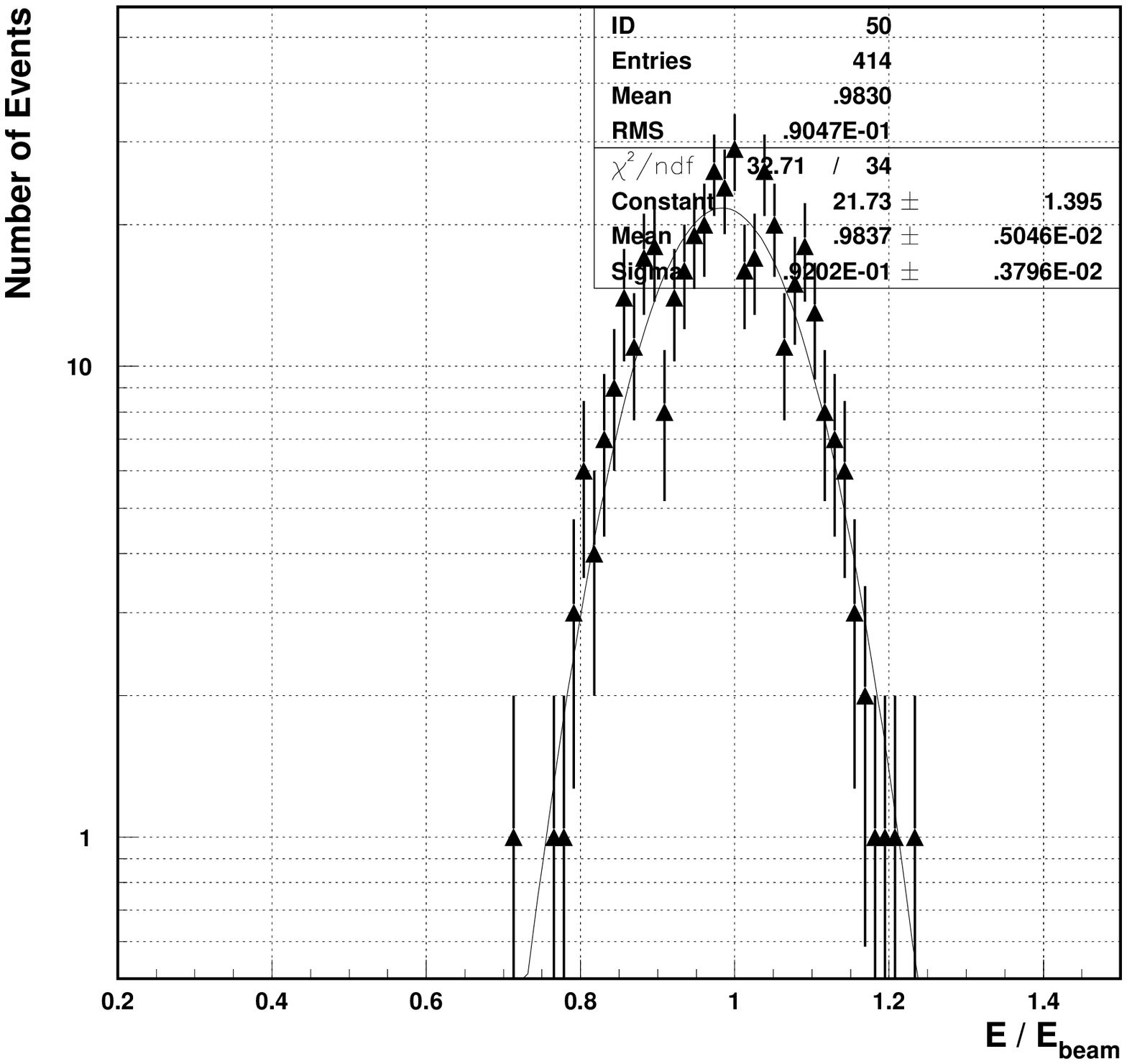,width=0.45\textwidth,height=0.25\textheight}}
        &
        \mbox{\epsfig{figure=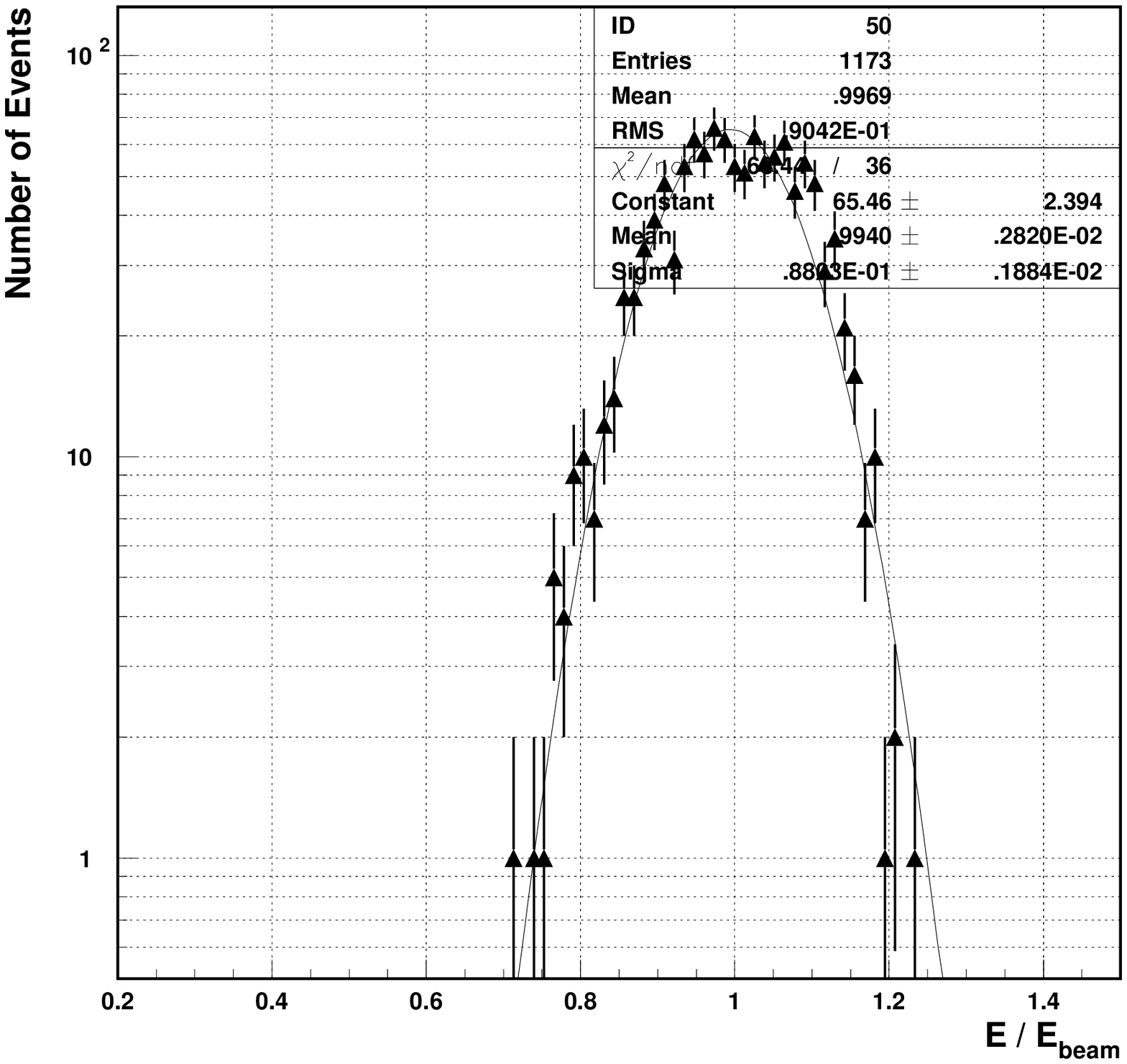,width=0.45\textwidth,height=0.25\textheight}}
        \\
        \mbox{\epsfig{figure=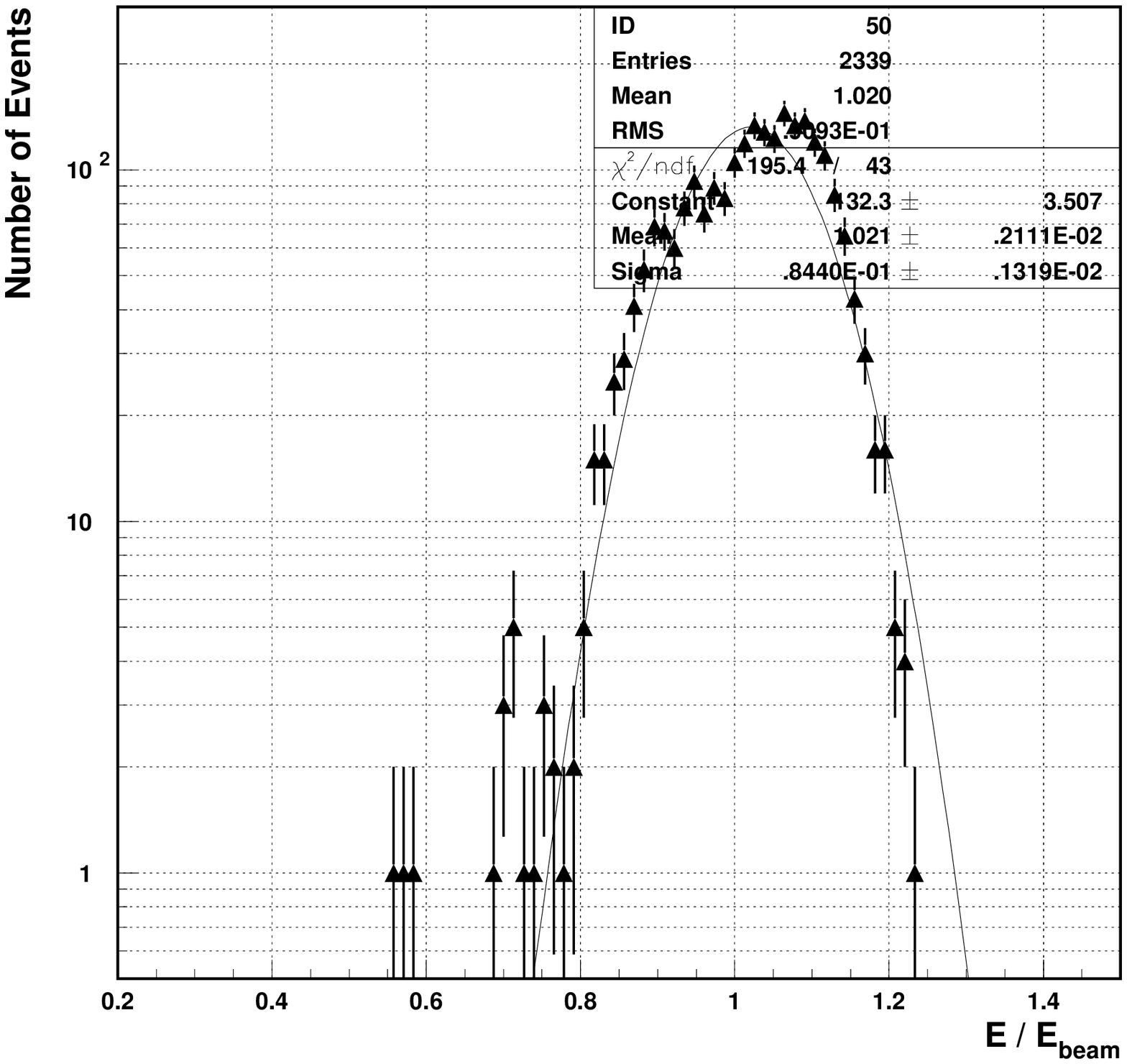,width=0.45\textwidth,height=0.25\textheight}}
        &
        \mbox{\epsfig{figure=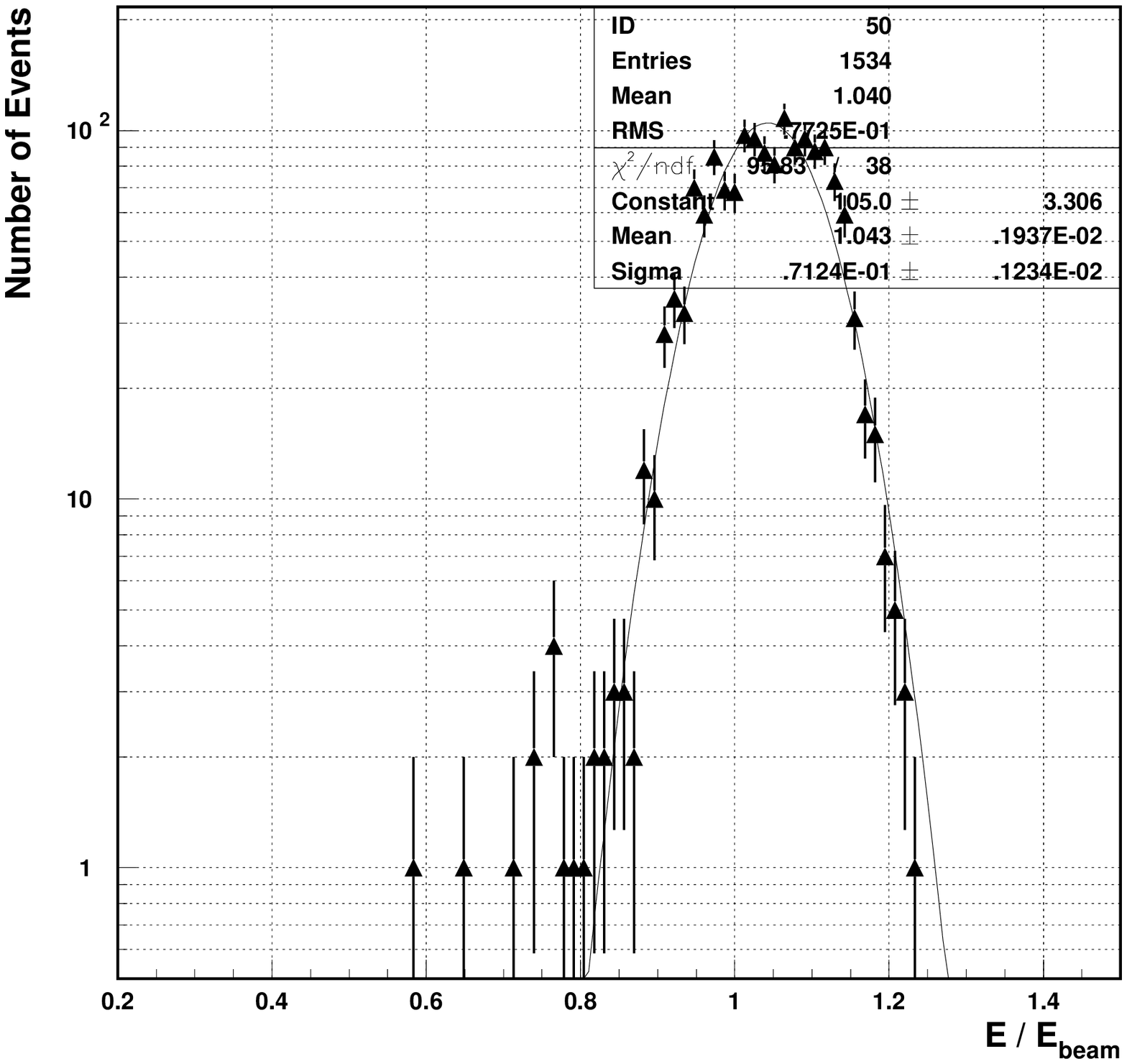,width=0.45\textwidth,height=0.25\textheight}}
        \\
        \end{tabular}
     \end{center}
       \caption{
	The normalized electron response ($E_e / E_{beam}$)
 	for E = 10, 60, 100 GeV (left column, up to down)
  	and
	E = 20, 80, 180 GeV (right column, up to down)
 	at $\eta = -0.25$.
       \label{fv34}}
\end{figure*}

\begin{figure*}[tbph]
     \begin{center}
        \begin{tabular}{cc}
        \mbox{\epsfig{figure=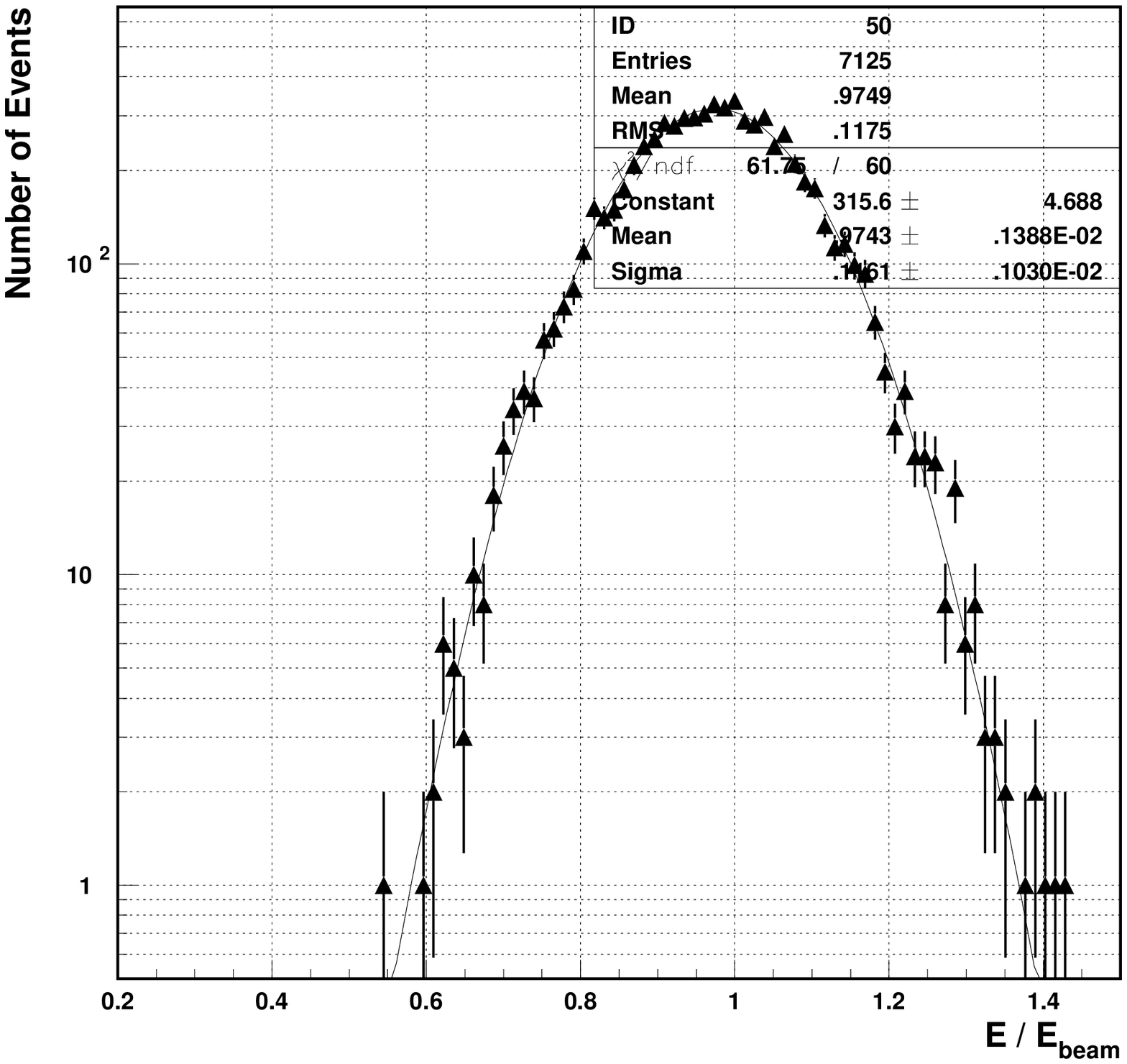,width=0.45\textwidth,height=0.25\textheight}}
        &
        \mbox{\epsfig{figure=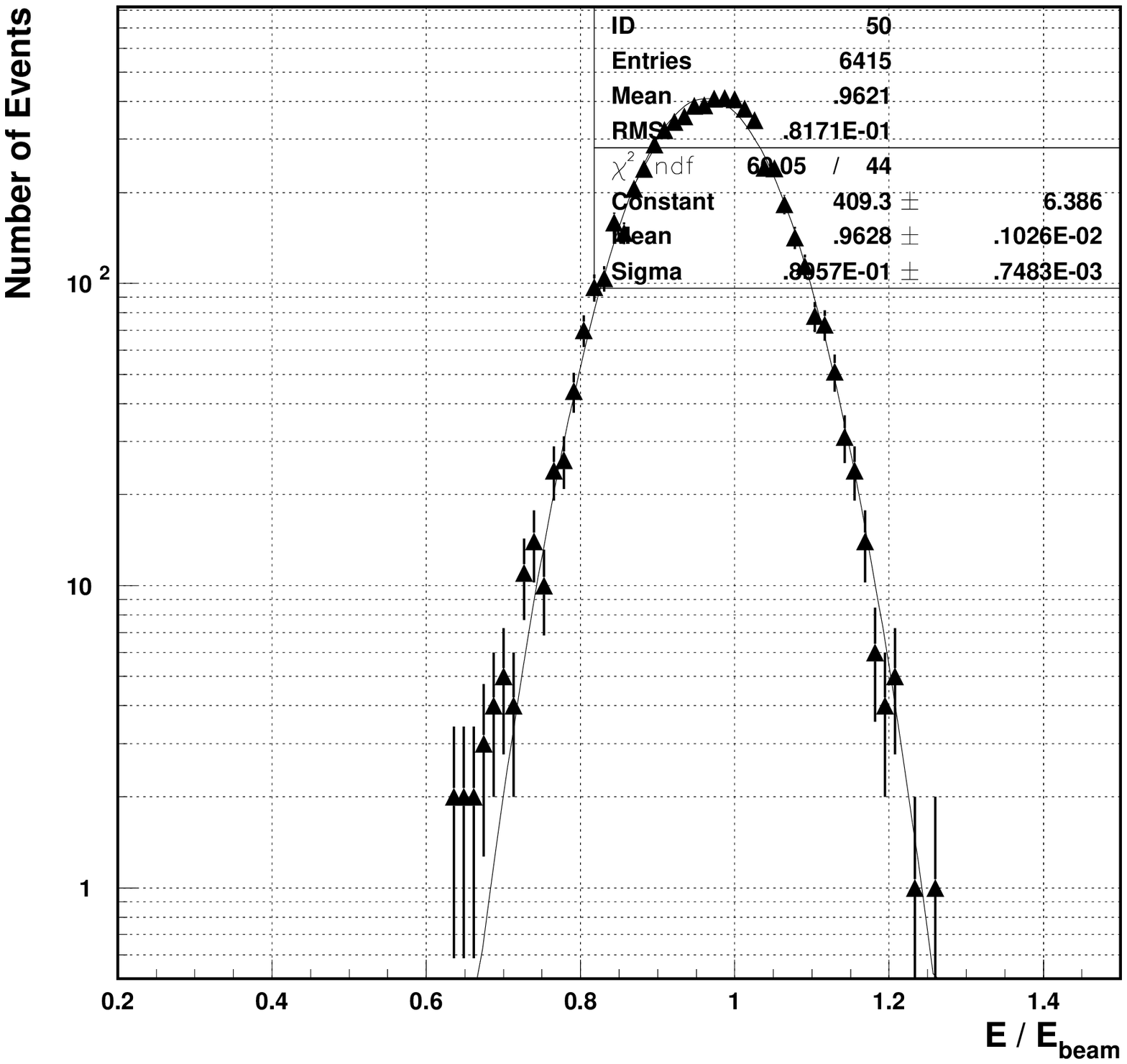,width=0.45\textwidth,height=0.25\textheight}}
        \\
        \mbox{\epsfig{figure=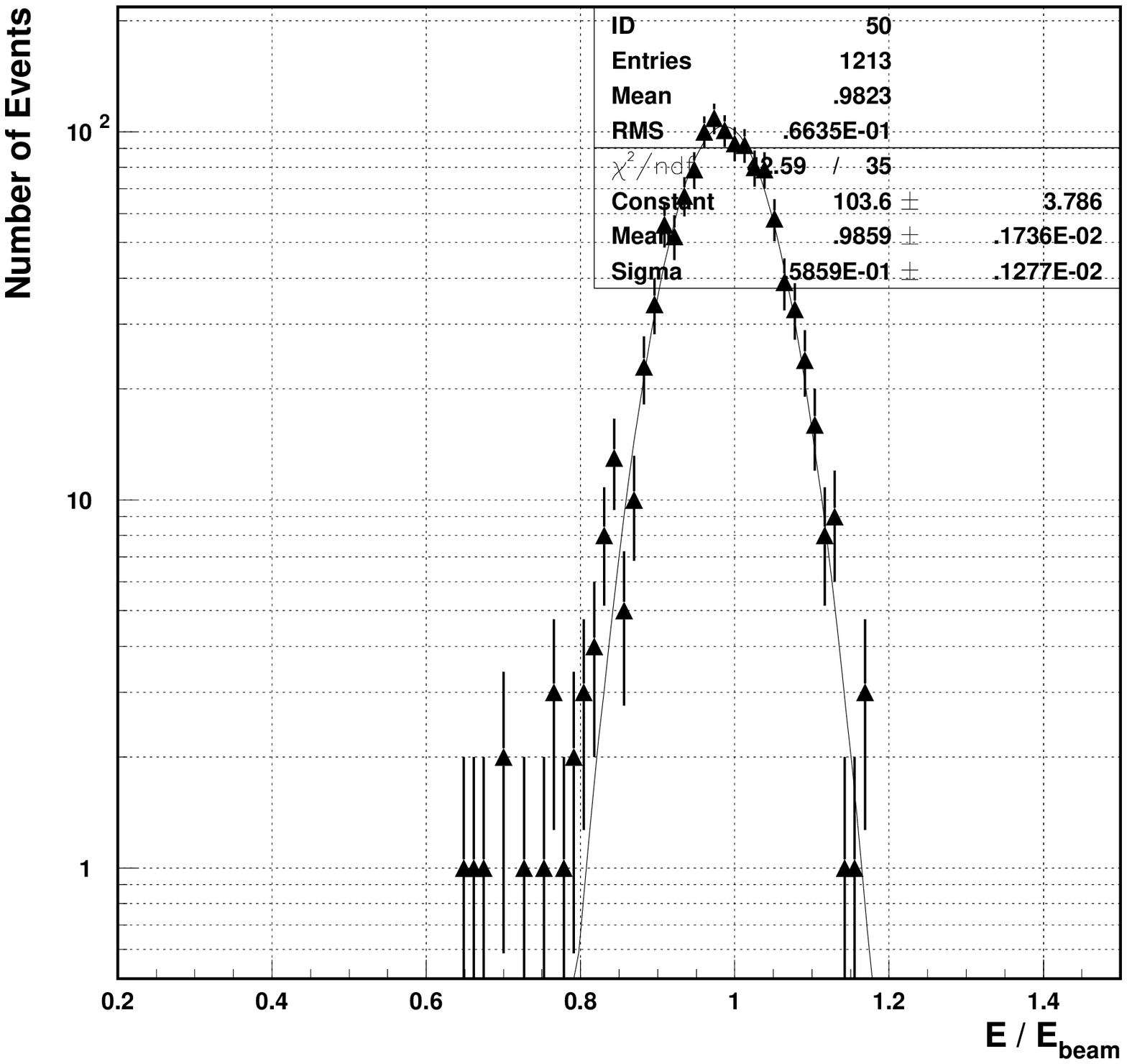,width=0.45\textwidth,height=0.25\textheight}}
        &
        \mbox{\epsfig{figure=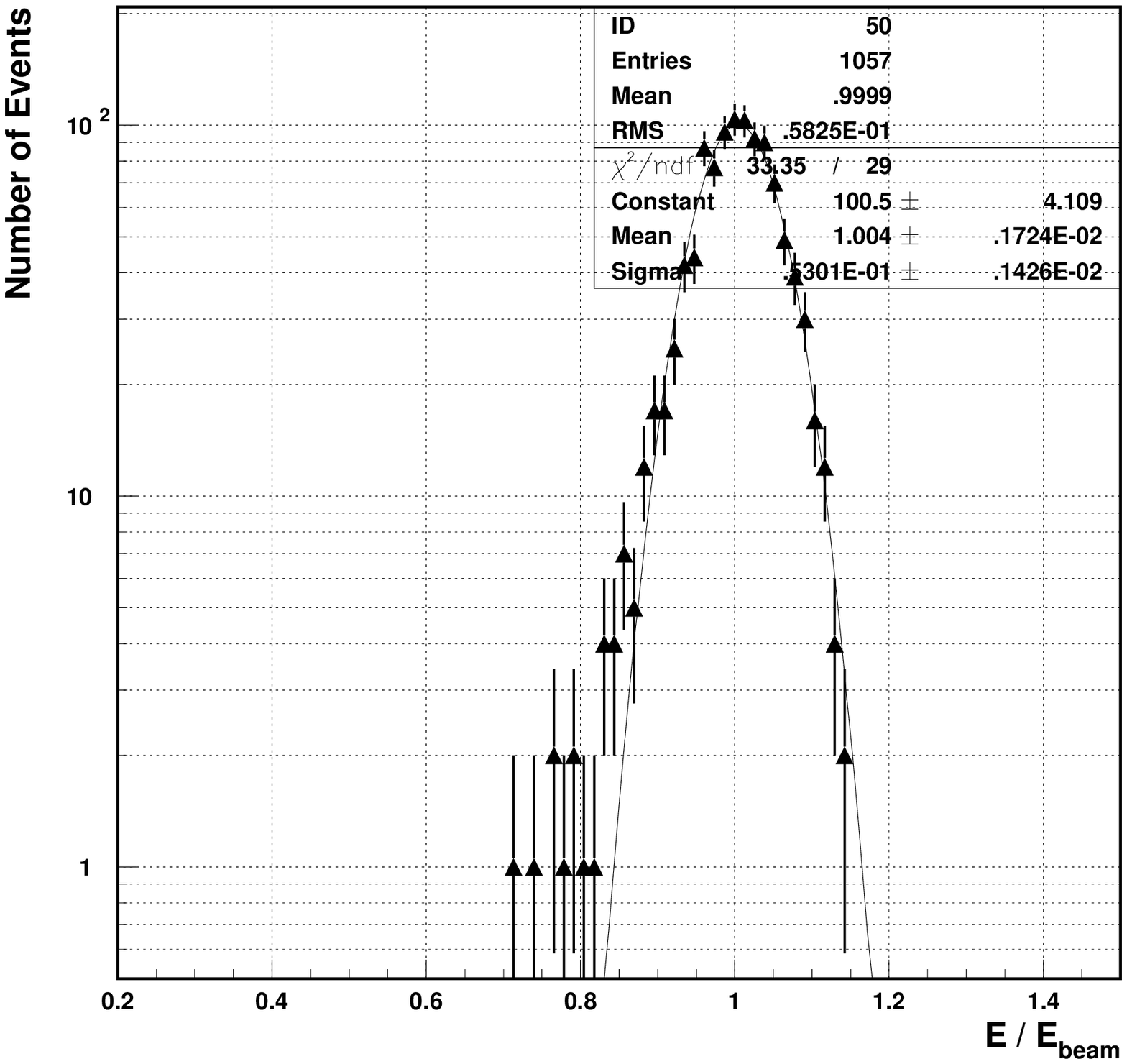,width=0.45\textwidth,height=0.25\textheight}}
        \\
        \mbox{\epsfig{figure=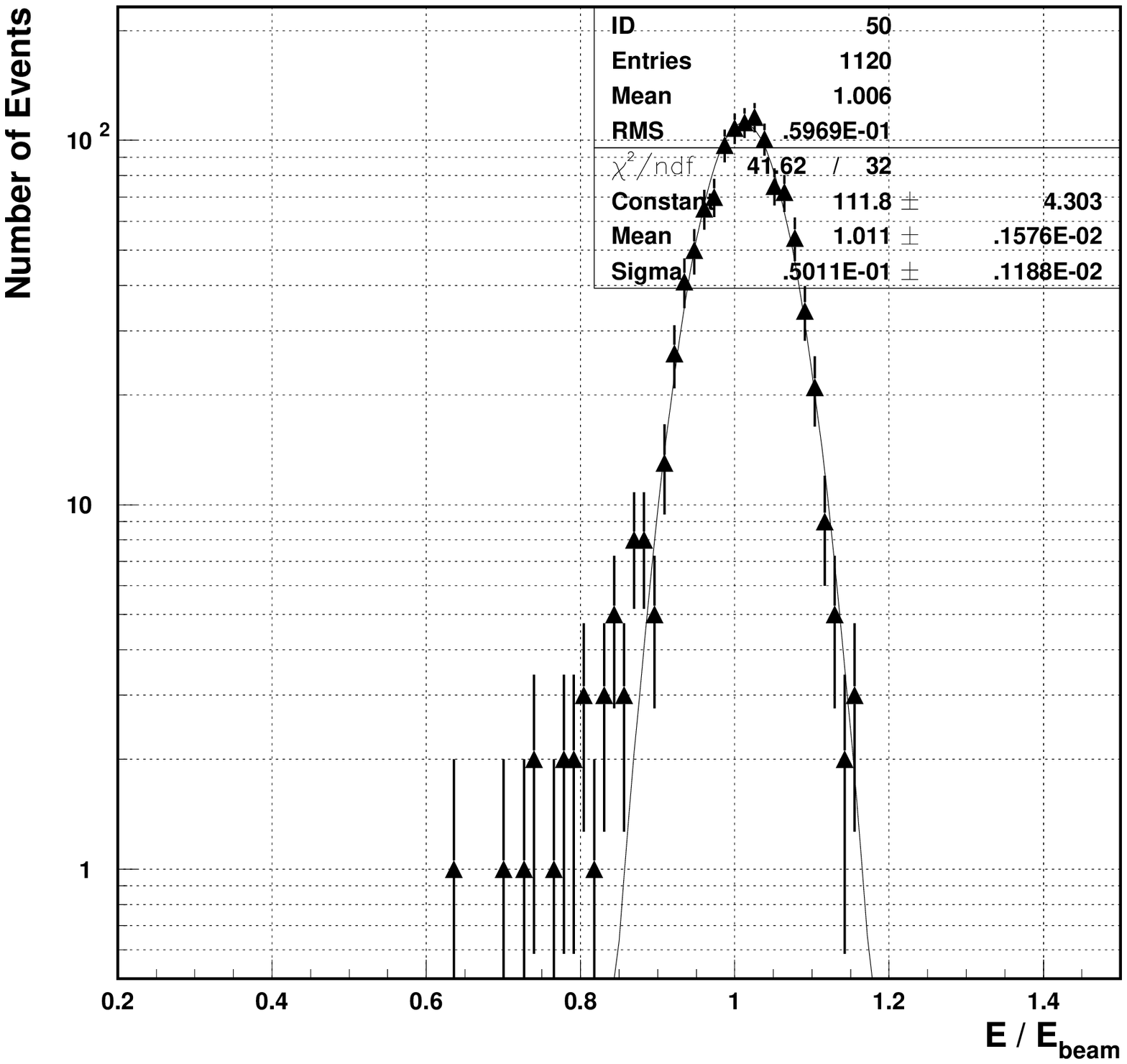,width=0.45\textwidth,height=0.25\textheight}}
        &
        \mbox{\epsfig{figure=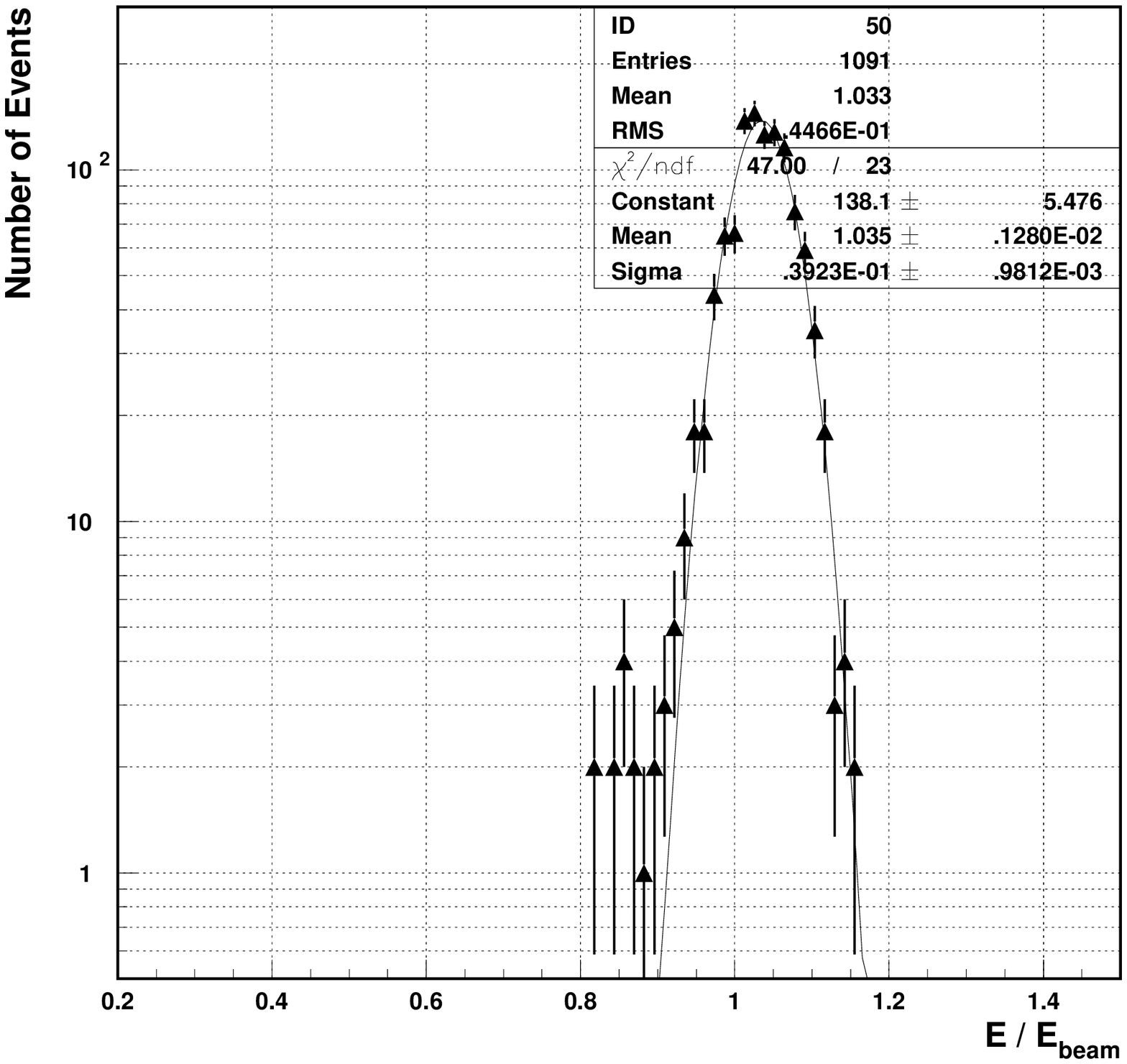,width=0.45\textwidth,height=0.25\textheight}}
        \\
        \end{tabular}
     \end{center}
       \caption{
	The normalized electron response ($E_e / E_{beam}$)
	for E = 10, 60, 100 GeV (left column, up to down)
  	and
	E = 20, 80, 180 GeV (right column, up to down)
 	at $\eta = -0.55$.
       \label{fv35}}
\end{figure*}

\begin{figure*}[tbph]
     \begin{center}
        \begin{tabular}{cc}
        \mbox{\epsfig{figure=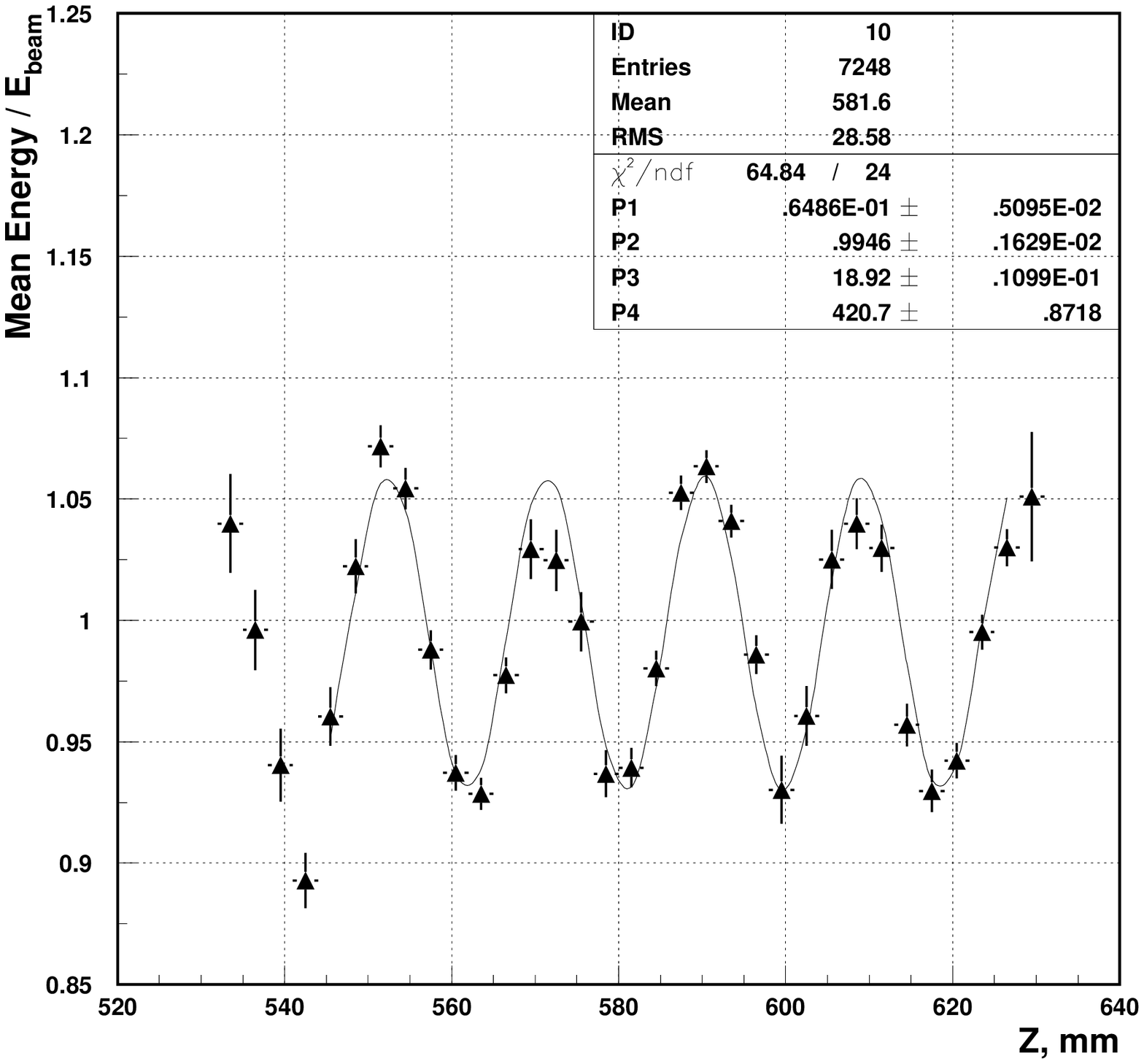,width=0.45\textwidth,height=0.25\textheight}}
        &
        \mbox{\epsfig{figure=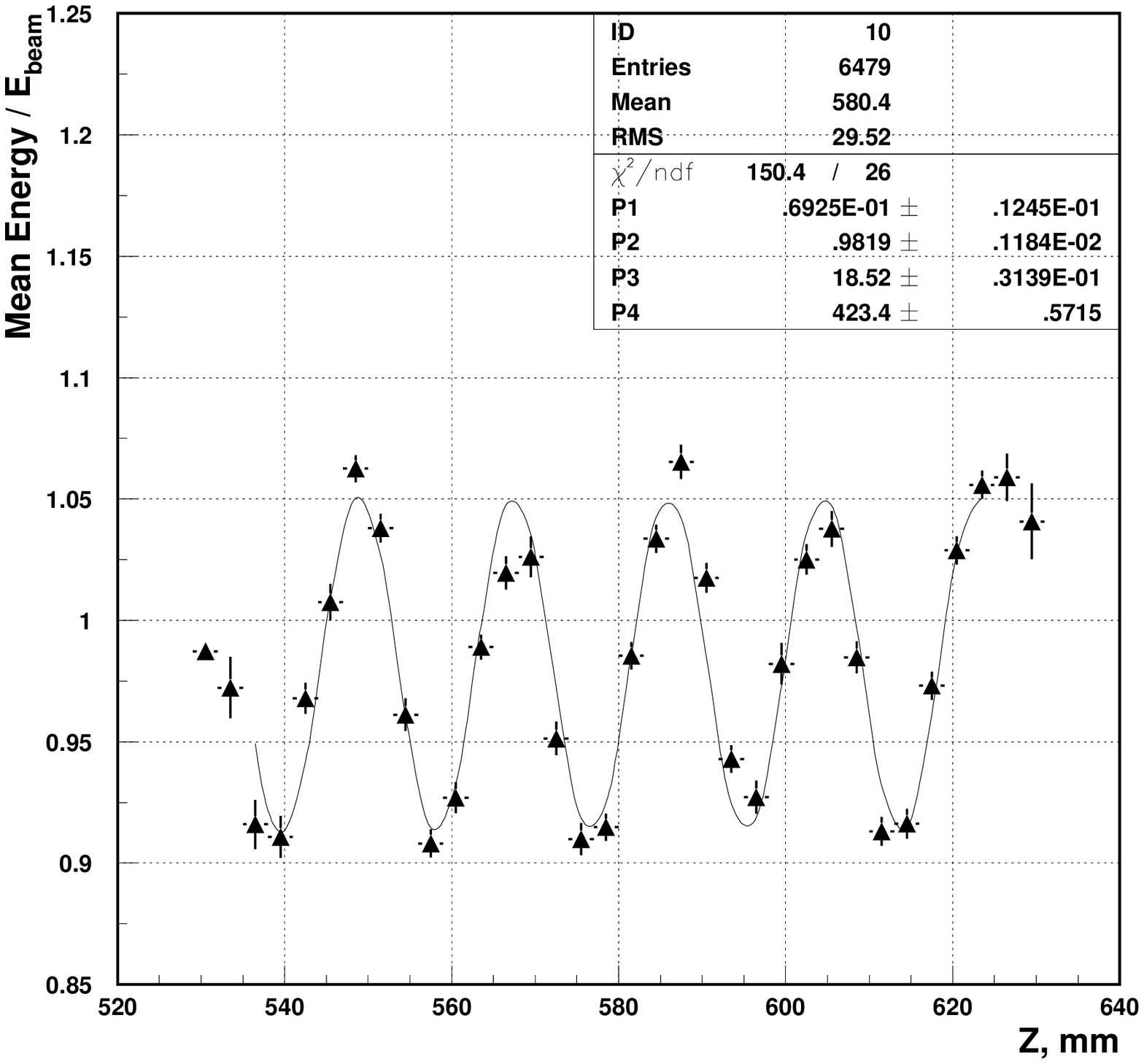,width=0.45\textwidth,height=0.25\textheight}}
        \\
        \mbox{\epsfig{figure=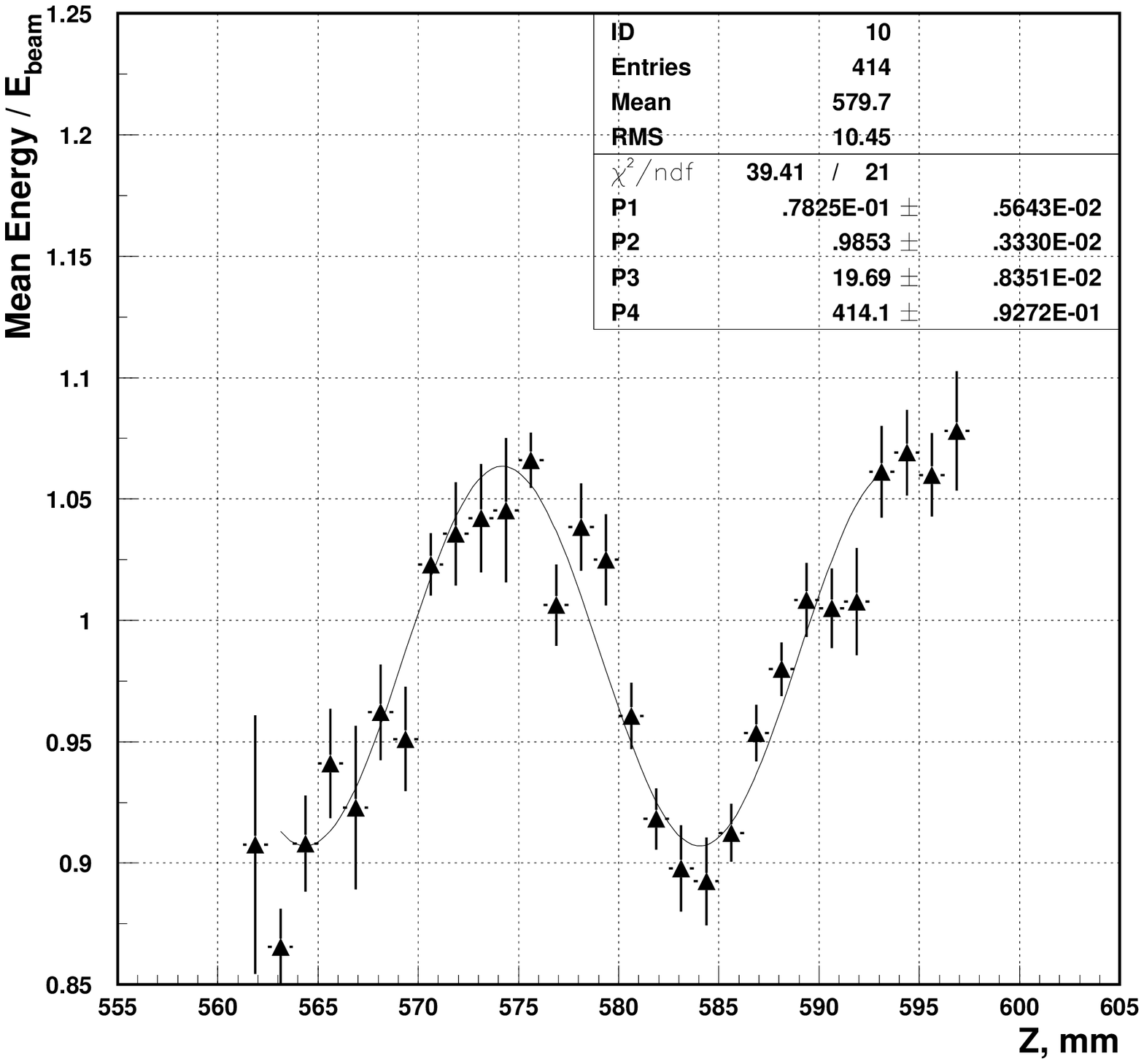,width=0.45\textwidth,height=0.25\textheight}}
        &
        \mbox{\epsfig{figure=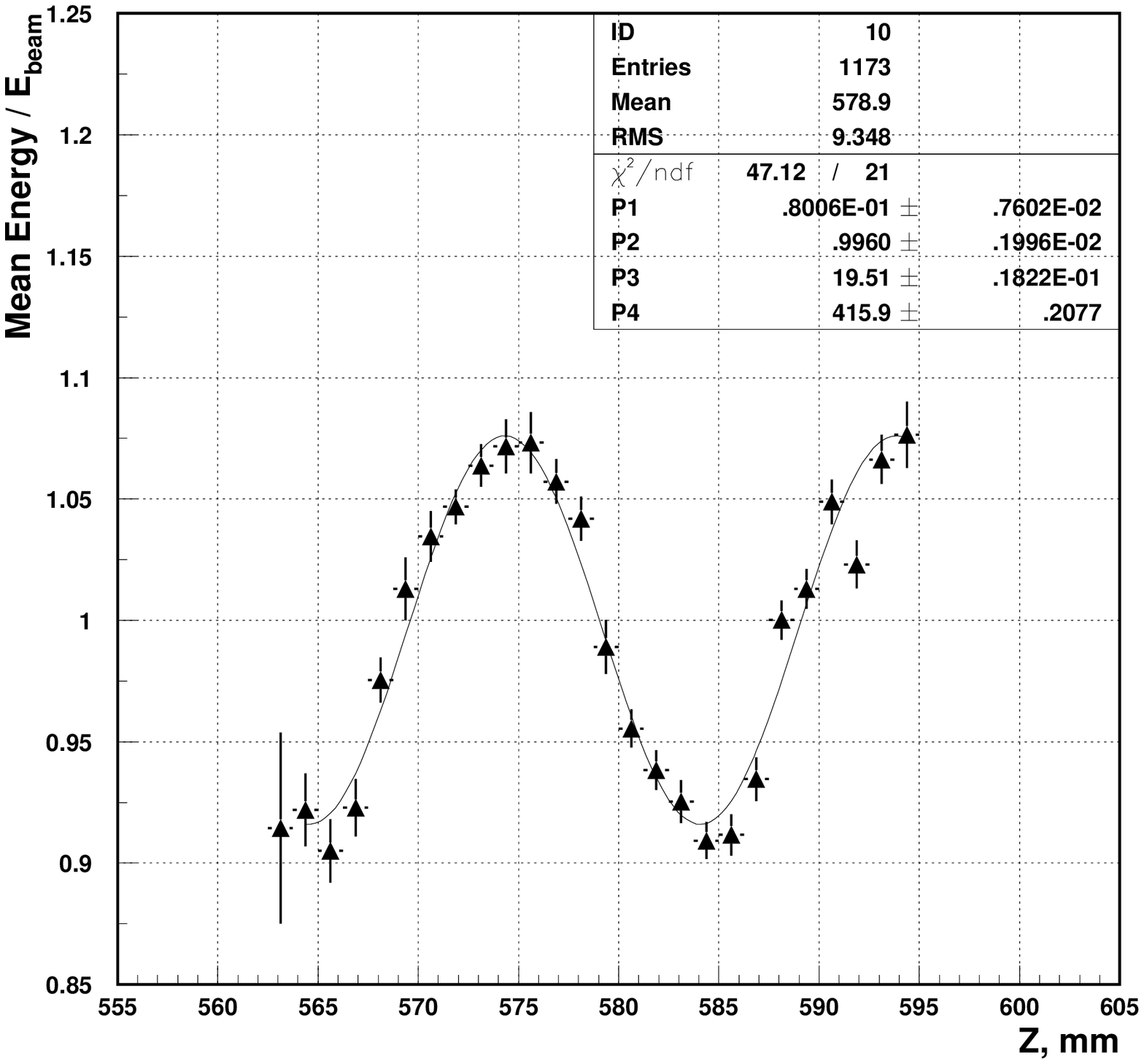,width=0.45\textwidth,height=0.25\textheight}}
        \\
        \mbox{\epsfig{figure=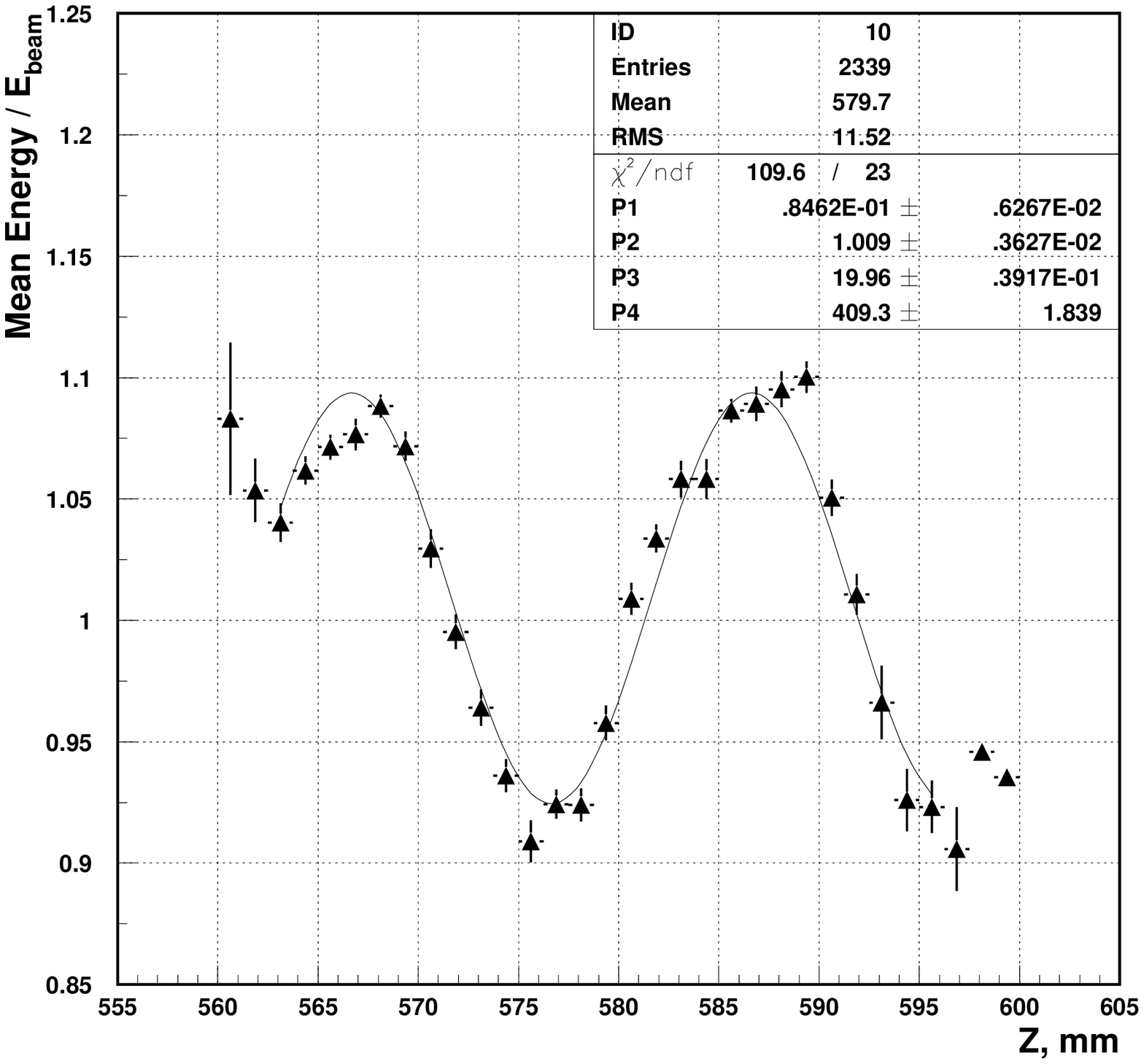,width=0.45\textwidth,height=0.25\textheight}}
        &
        \mbox{\epsfig{figure=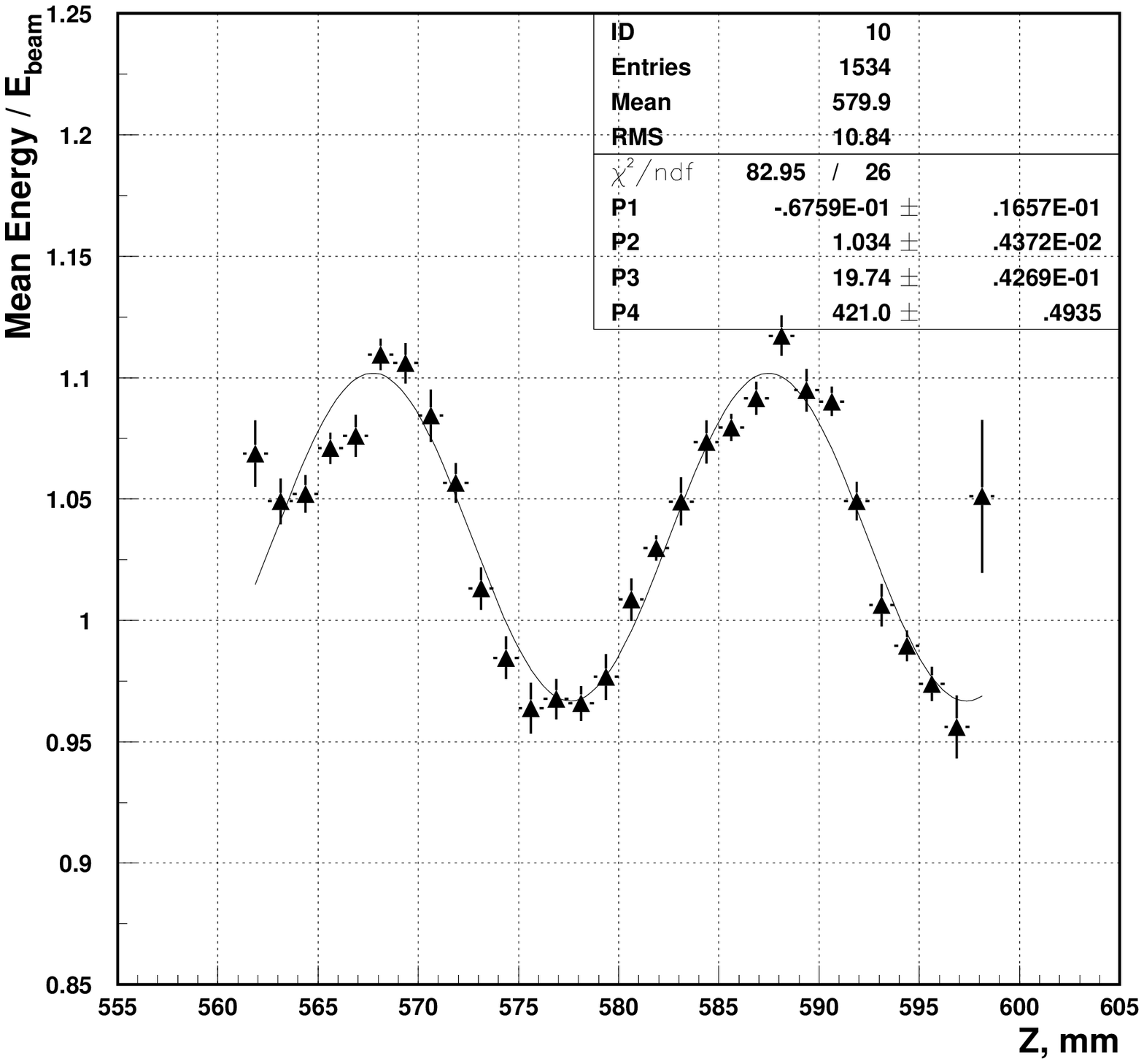,width=0.45\textwidth,height=0.25\textheight}}
        \\
        \end{tabular}
     \end{center}
       \caption{
	The normalized electron response ($E_e / E_{beam}$)
 	for E = 10, 60, 100 GeV (left column, up to down)
  	and
	E = 20, 80, 180 GeV (right column, up to down)
 	at $\eta = -0.25$
	as a function of impact point $Z$ coordinate.
       \label{fv36}}
\end{figure*}

\begin{figure*}[tbph]
     \begin{center}
        \begin{tabular}{cc}
        \mbox{\epsfig{figure=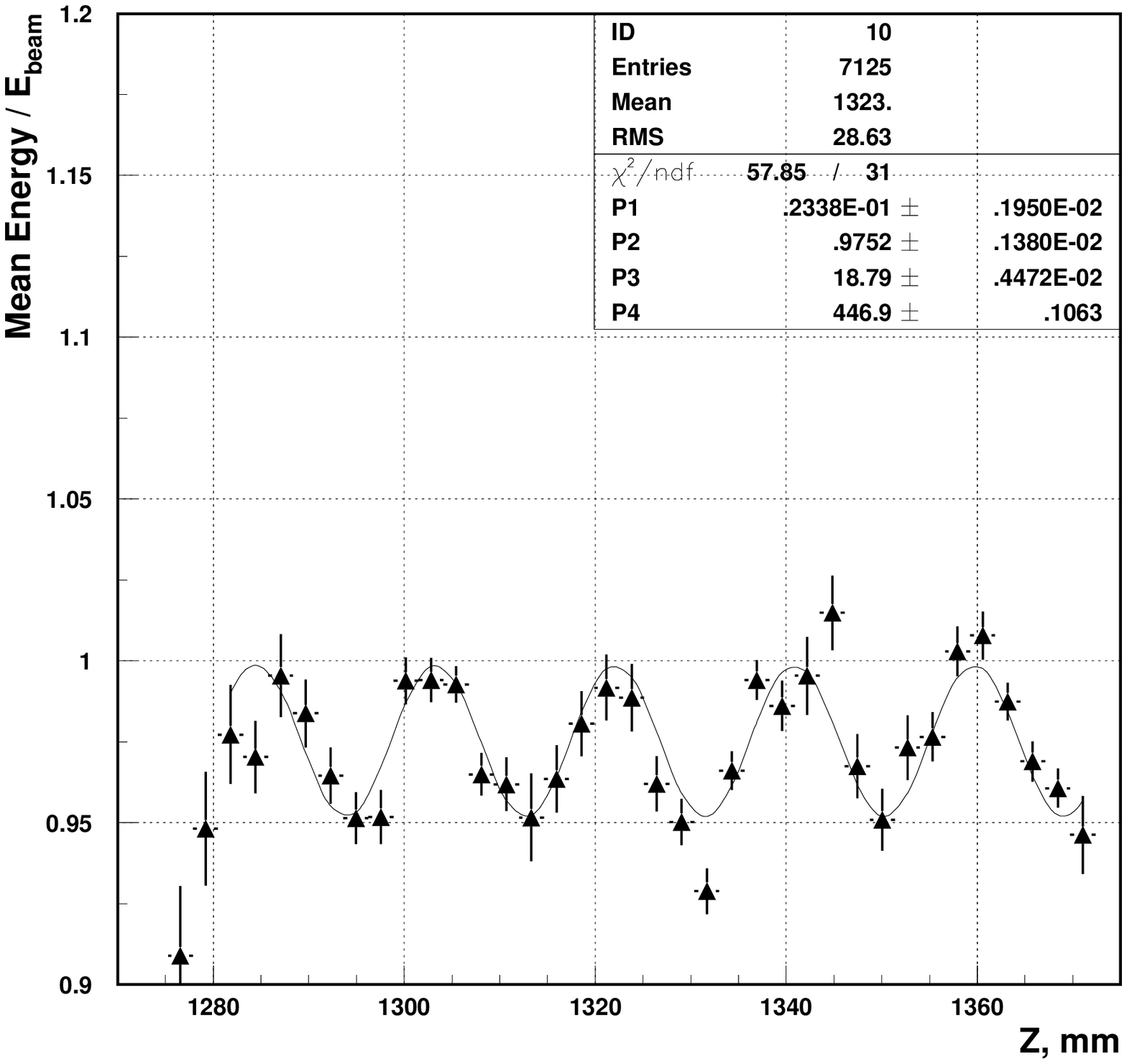,width=0.45\textwidth,height=0.25\textheight}}
        &
        \mbox{\epsfig{figure=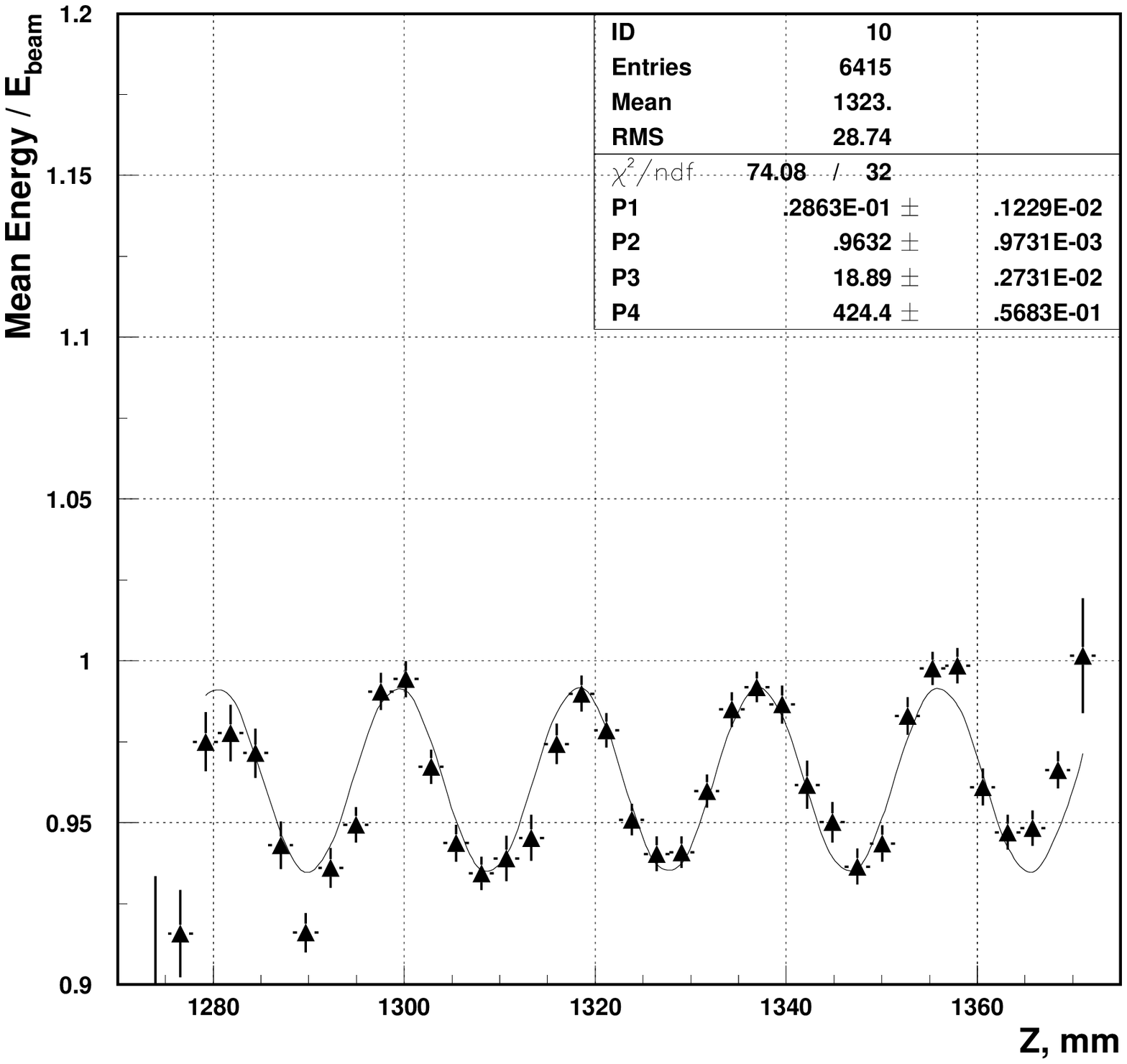,width=0.45\textwidth,height=0.25\textheight}}
        \\
        \mbox{\epsfig{figure=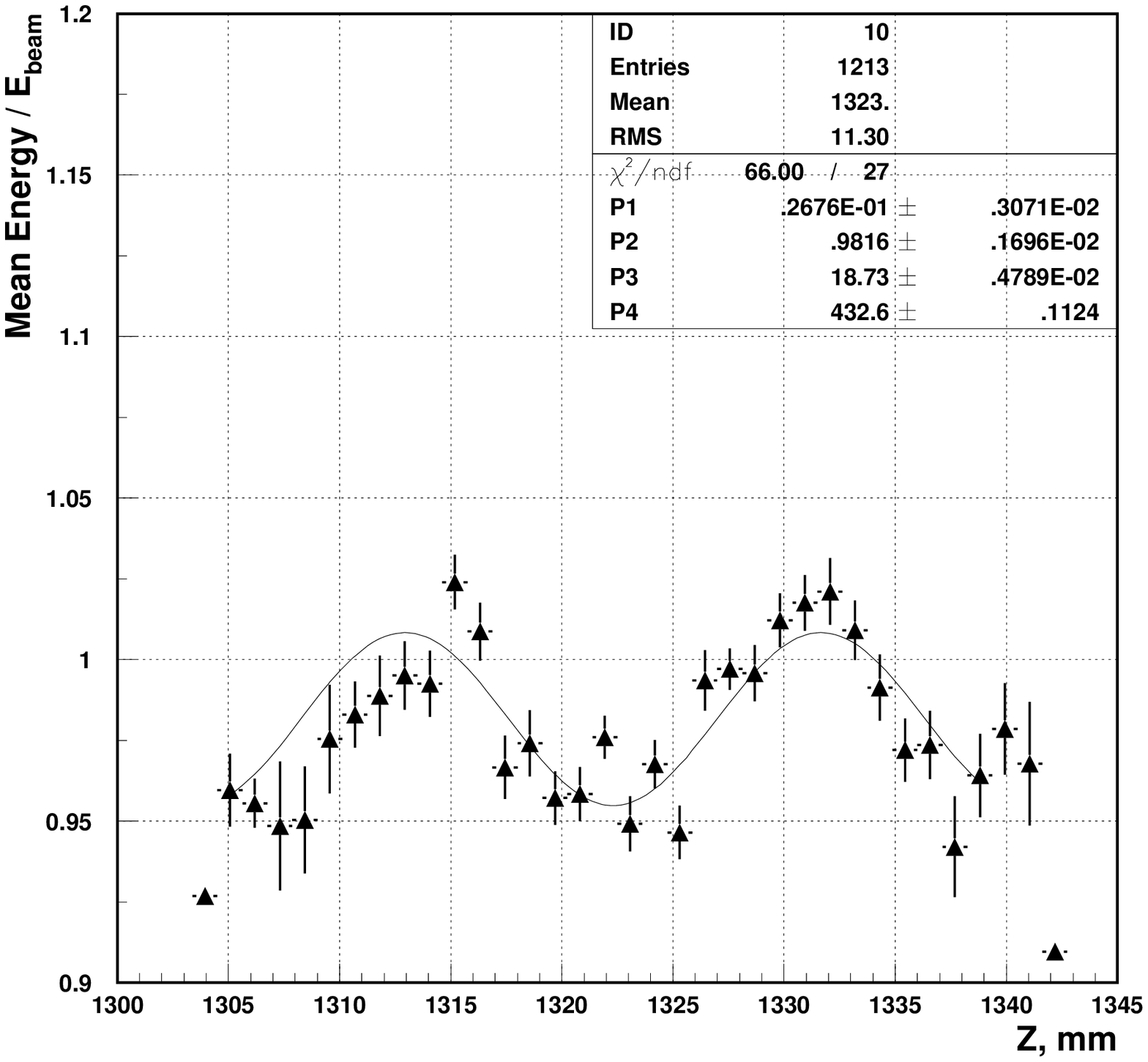,width=0.45\textwidth,height=0.25\textheight}}
        &
        \mbox{\epsfig{figure=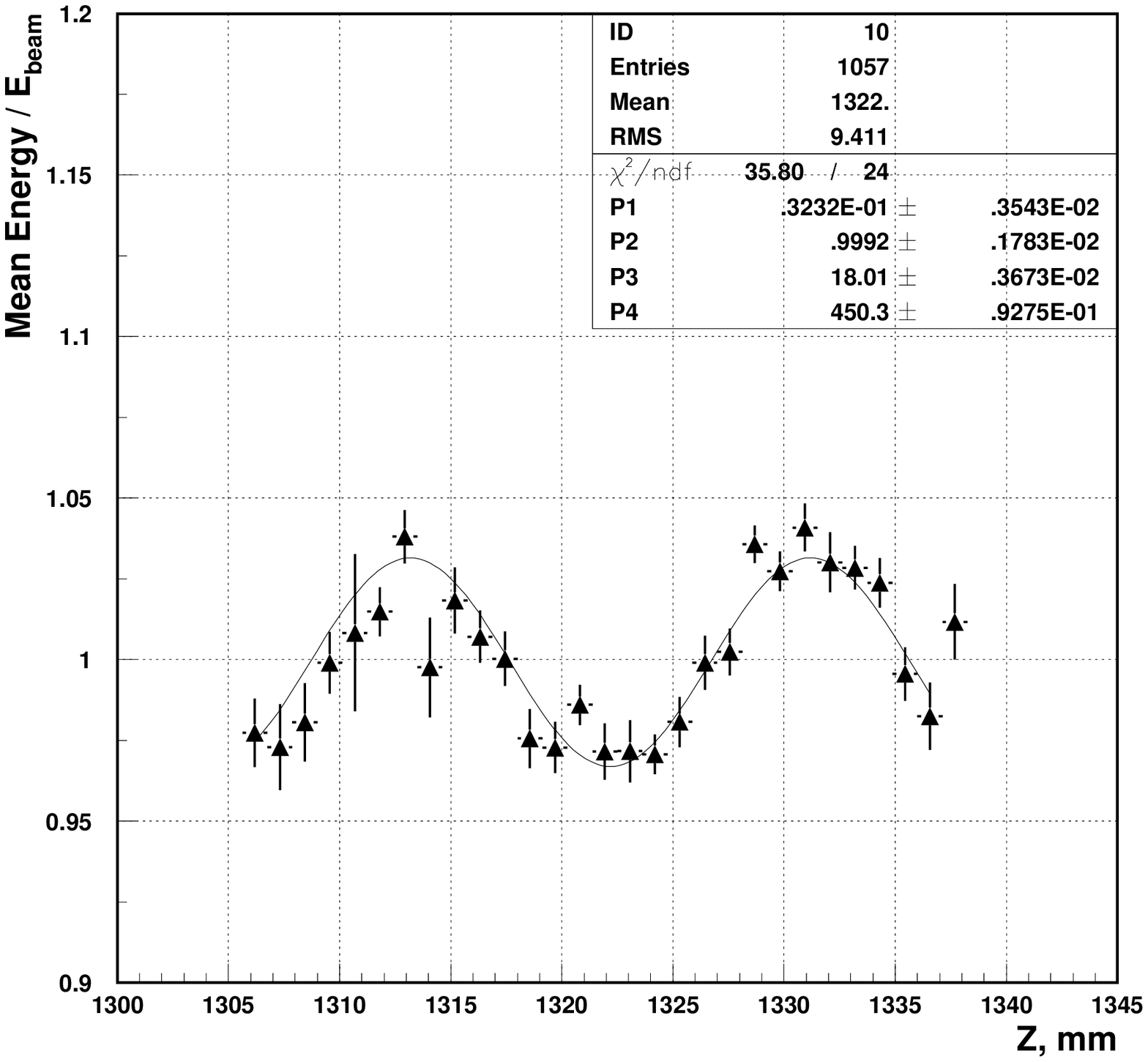,width=0.45\textwidth,height=0.25\textheight}}
        \\
        \mbox{\epsfig{figure=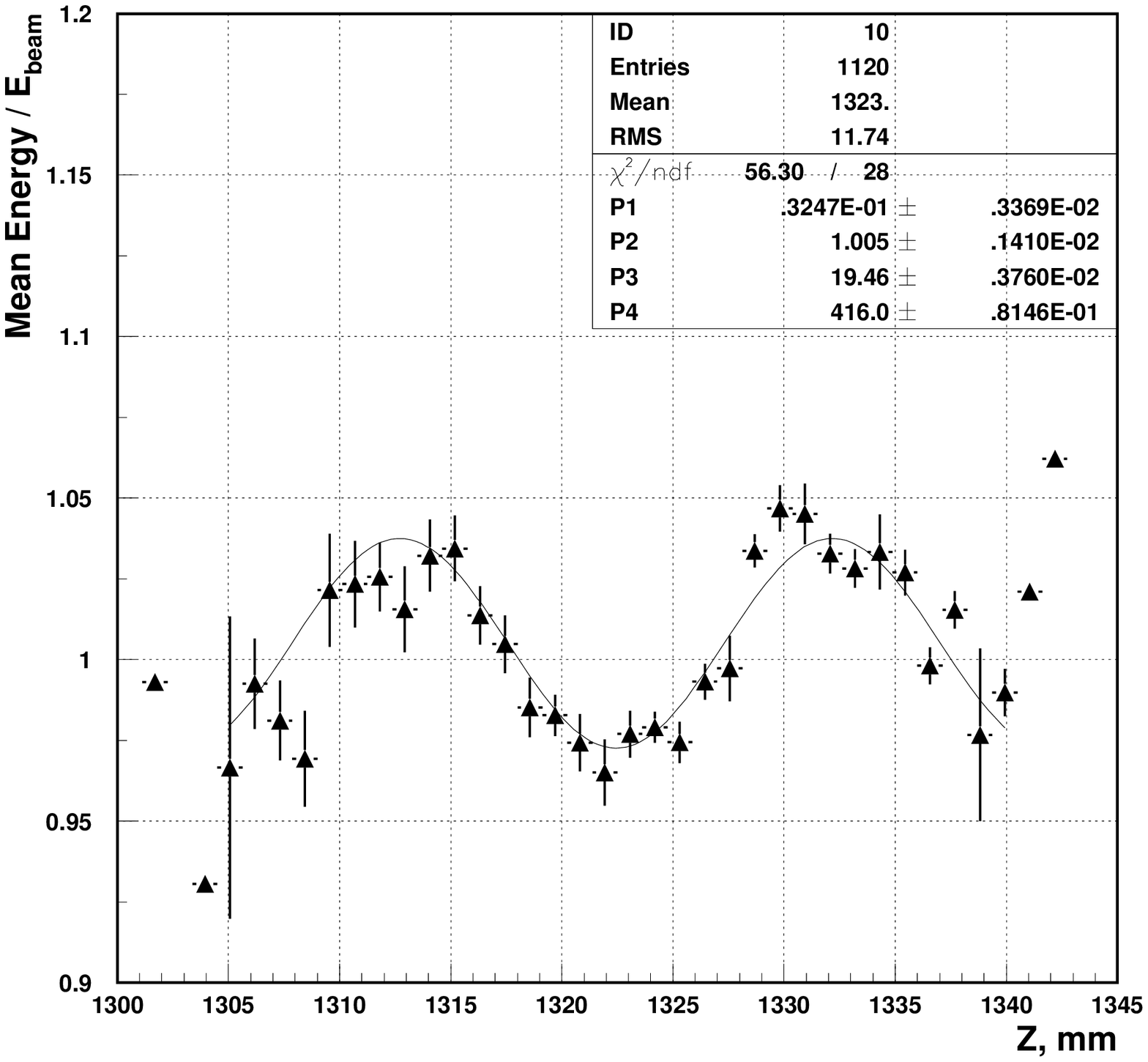,width=0.45\textwidth,height=0.25\textheight}}
        &
        \mbox{\epsfig{figure=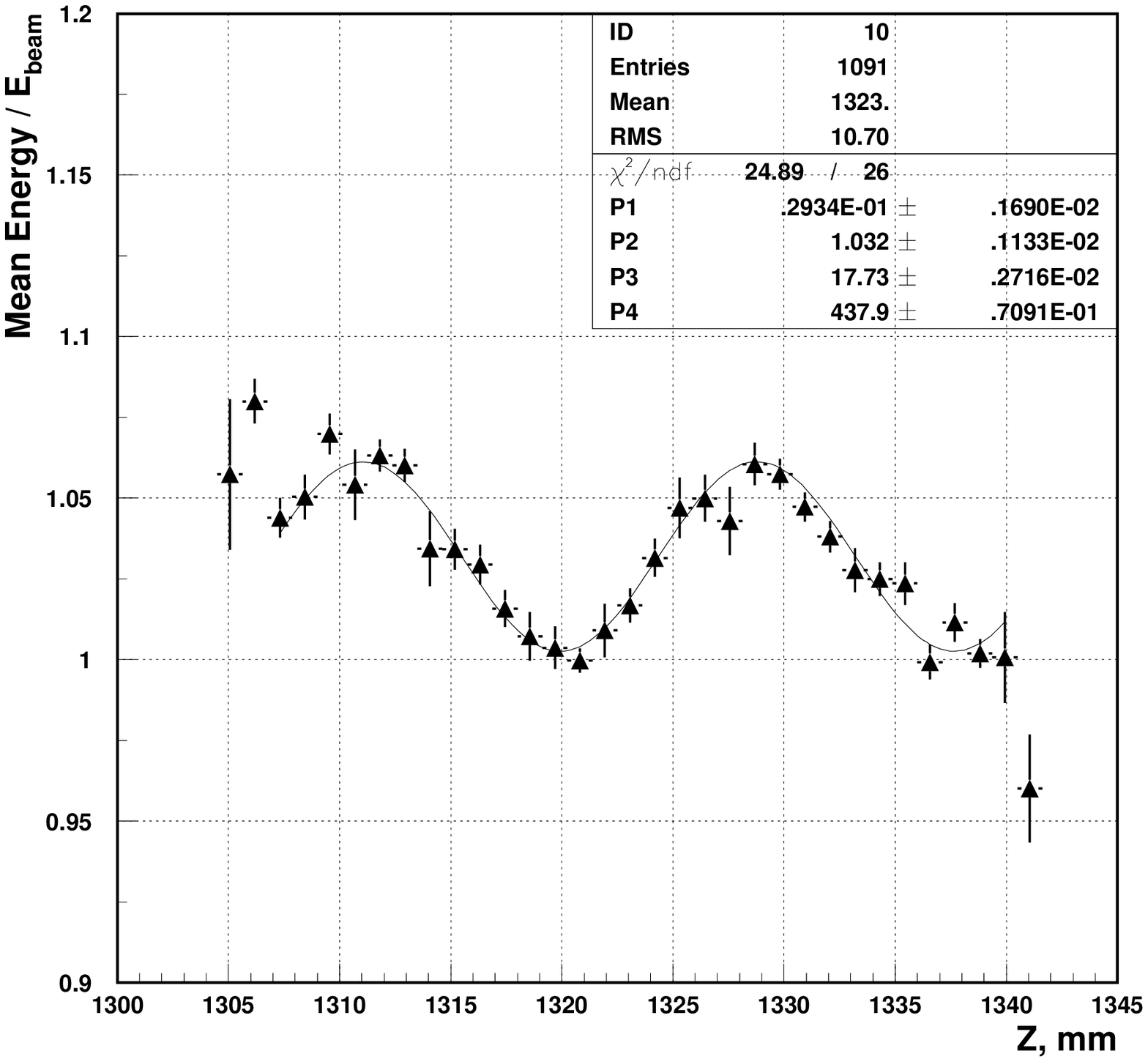,width=0.45\textwidth,height=0.25\textheight}}
        \\
        \end{tabular}
     \end{center}
       \caption{
	The normalized electron response ($E_e / E_{beam}$)
 	for E = 10, 60, 100 GeV (left column, up to down)
  	and
	E = 20, 80, 180 GeV (right column, up to down)
 	at $\eta = -0.55$
	as a function of impact point $Z$ coordinate.
       \label{fv37}}
\end{figure*}

Therefore an electron response ($R = E_e / E_{beam}$) is rather complicated
function of $E_{beam}$,  $\Theta$ and $Z$.
The energy response spectrum for given run (beam has the transversal spread
$\pm 10\ mm$) as a rule is non-Gaussian
(Fig.~\ref{fv34} and Fig.~\ref{fv35}), 
since it is a superposition of
different response spectra, 
but it becomes Gaussian
for given E,  $\Theta$,  Z values.
Fig.~\ref{fv36} and Fig.~\ref{fv37} show
the normalized electron response  for
E = 10, 20, 60, 80, 100, 180 GeV
 at $\eta = -0.25$ and $-0.55$
as a function of the impact point $Z$ coordinate.
One can see the clear periodical structure of the response
 with 18~mm period.
The mean values (parameter $P_2$)
 and the amplitudes(parameter $P_1$)
 of these spectra have been extracted
by fitting the sine function:
\begin{equation}
	f(Z) = P_2 + P_1 \sin{(2\pi Z/P_3+P_4)}
\label{ev4}
\end{equation}

\begin{figure*}[tbph]
     \begin{center}
        \begin{tabular}{c}
     \mbox{\epsfig{figure=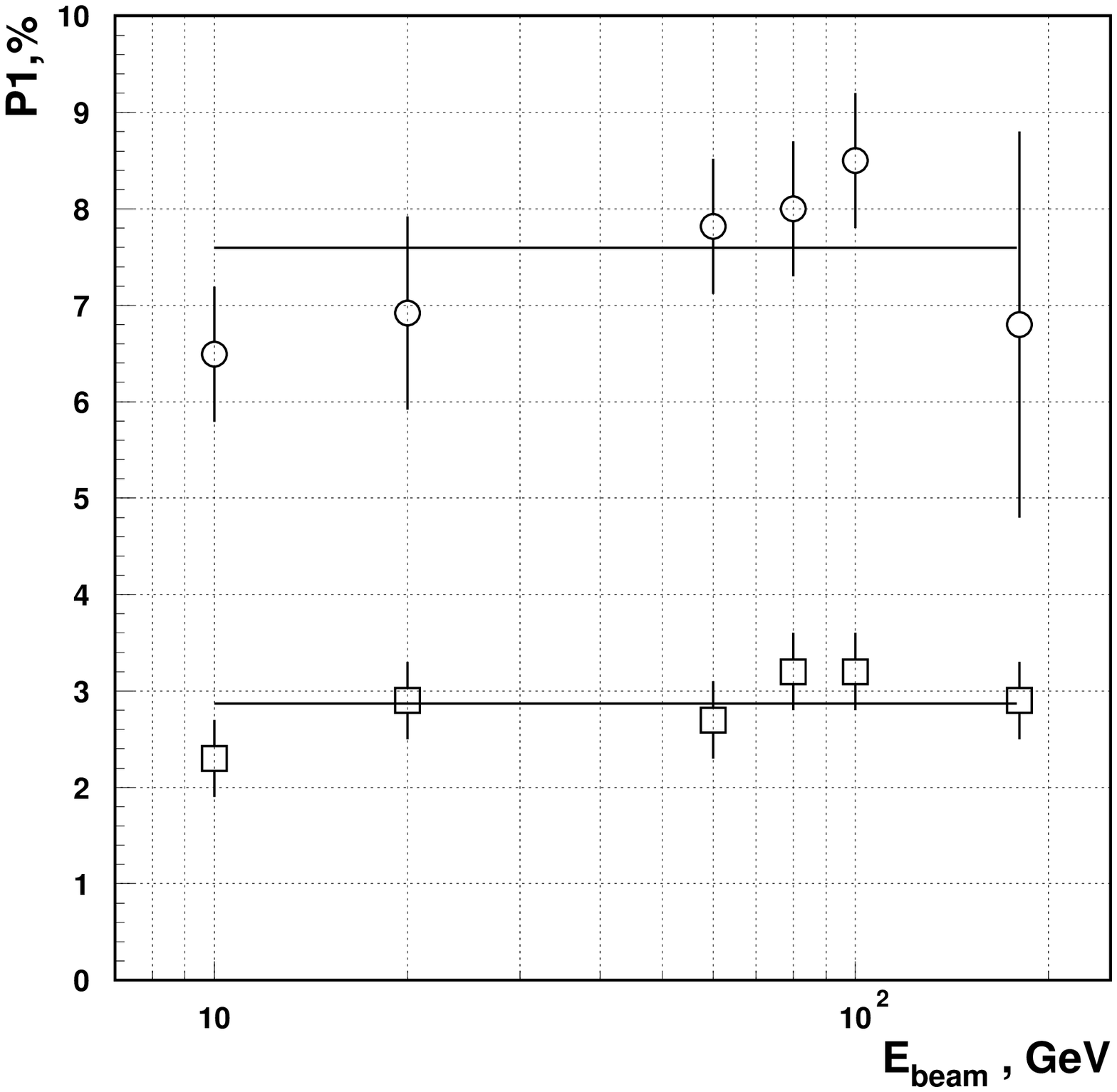,width=0.9\textwidth,height=0.4\textheight}}
\\
     \mbox{\epsfig{figure=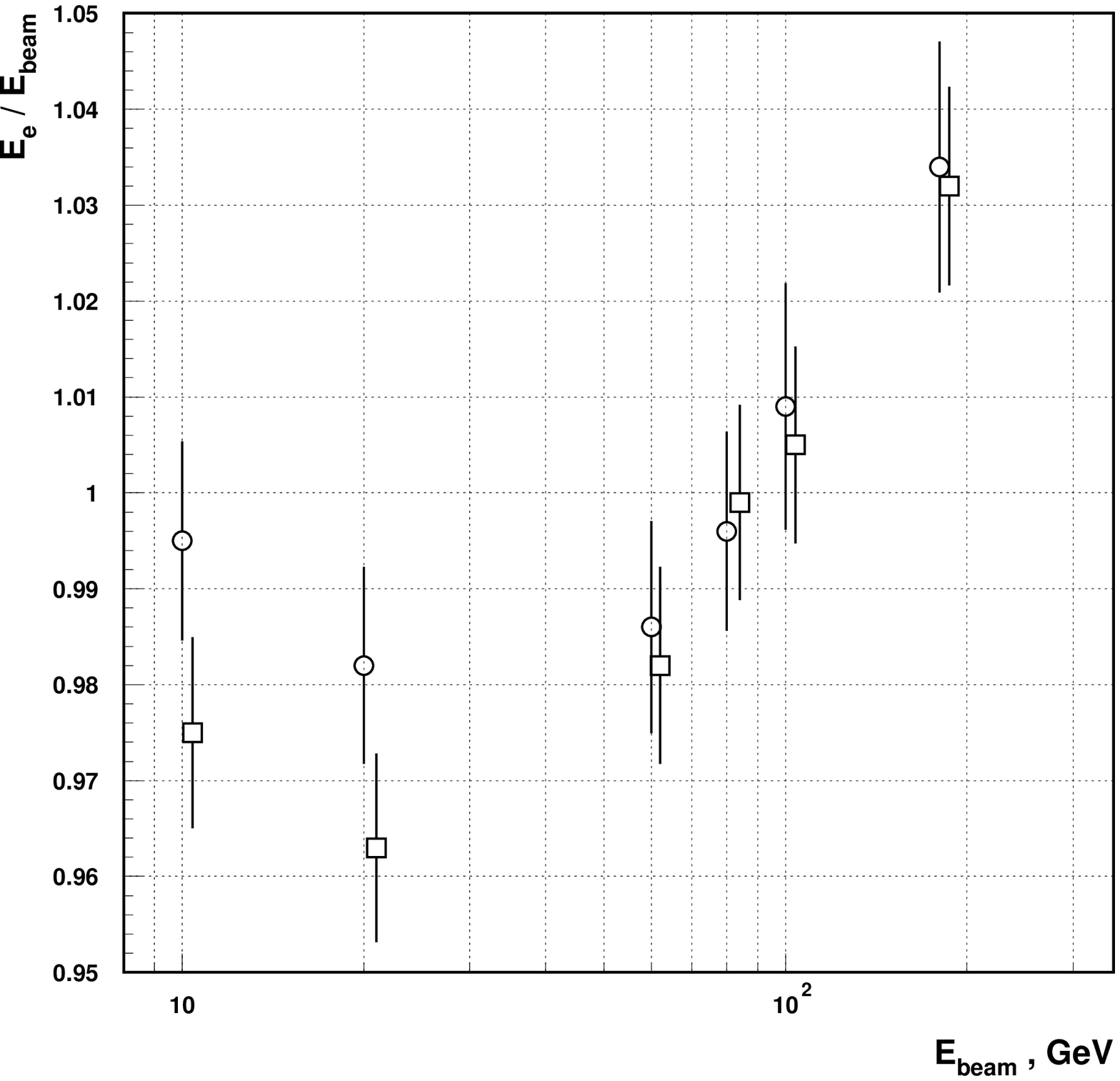,width=0.9\textwidth,height=0.4\textheight}}
\\
        \end{tabular}
     \end{center}
     \caption{
	Top: The amplitude (parameter $P_1$) of the electron response
	as a function of the beam energy.
	Bottom: 
	The mean normalized electron response as a function of the beam energy.
	$\bigcirc$ ($\square$) are the data for $\eta = -0.25$ ($-0.55$).
       \label{fv27}}
\end{figure*}

Fig.~\ref{fv27} (top) shows the parameter $P_1$ as a function of the 
beam energy.
As can be seen this parameter does not depend from the beam energy within
errors and decreases with increasing of $\eta$ from
$(7.6\pm0.3) \% $ at $\eta = -0.25$ to $(2.9\pm0.2) \% $ at 
$\eta = -0.55$.

Fig.~\ref{fv27} (bottom)
shows the mean normalized electron response as a function of
energy for two values of $\eta$.
As can be seen there is 
some increase of the mean normalized electron response with
increasing of energy.
%But these values can be in the $\pm 2\ \%$ range taking into account their
%errors.
There is no difference between ones for various values of $\eta$.
Note that there are the additional systematic errors in these values
(not given in this Figure) due to the uncertainties in the average beam 
energies. These  uncertainties are determined by the expression
$$\frac{\Delta E_{beam}}{E_{beam}} = \frac{25 \%}{E_{beam}} \oplus 0.5 \%$$
and range from 2.5 \% for $E_{beam} = 10$ GeV to 0.5 \% 
for $E_{beam} = 180$ GeV.

\begin{figure*}[tbph]
     \begin{center}
     \mbox{\epsfig{figure=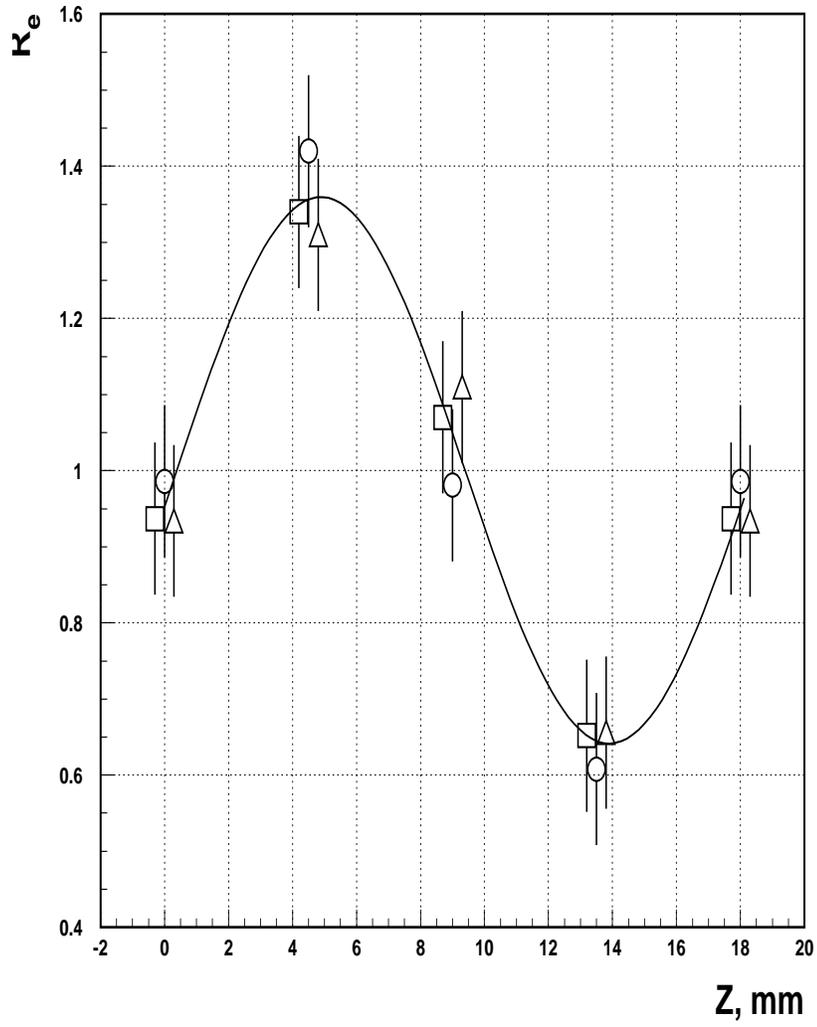,width=0.9\textwidth,height=0.8\textheight}}
     \end{center}
     \caption{
	The electron response as a function of $Z$ coordinate (calculations).
	$\bigcirc$ are for 10 GeV energy,
	$\square$ are for 100 GeV energy,
	$\triangle$ are for 100 GeV energy.
       \label{fv38}}
\end{figure*}

We attempted to explain the electron response as a function of $Z$ coordinate
calculating the total number of shower electrons (positrons)
crossing scintillator tiles
taking into account the arrangement of tiles and its sizes and
 using the shower curve
(the number of particles in the shower $N_{e}$ as a function of the
longitudinal shower development).
 which is given in \cite{abshire}.
These calculations were performed for
some energies and angles for the trajectories entering into four
different elements of calorimeter periodic structure ---
spacer,  master,  tile,  master. 
The results for E = 10, 100, 180 GeV at $\eta = -0.25$
are shown in Fig.~\ref{fv38}.
There is a maximum at the impact point corresponding to tile and a
minimum at the spacer plate.
Such simple calculations are in agreement with experimental
 data as to
non-dependence from energy
and
the periodicity in the electron response.
But
these calculations
do not reproduce the values of the amplitude.
The latter is connected  with non-taking into account
 the shower lateral spread.

%%%%%%%%%%%%%%%%%%%%%%%%%%%%%%%%%%%%%%%%%%%%%%%%%%%%%%%%%%%%%%%%%%%%%%%%%%
\section{Electron Energy Resolution}

\begin{figure*}[tbph]
     \begin{center}
       \mbox{\epsfig{figure=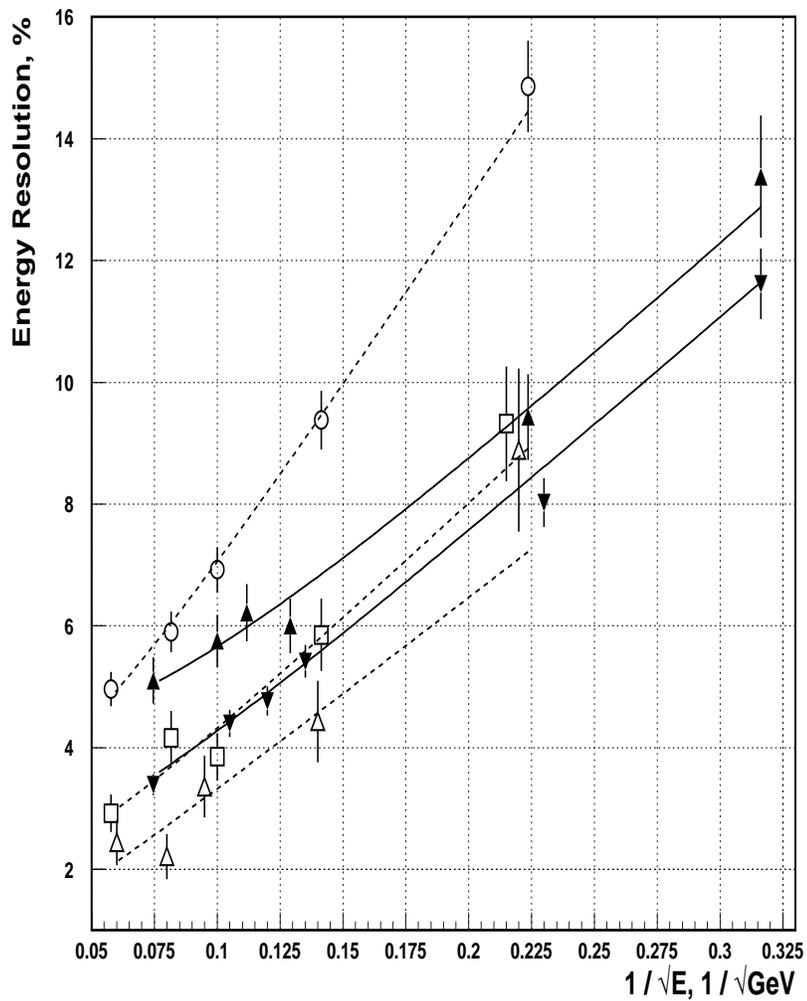,width=0.9\textwidth,height=0.8\textheight}}
    \end{center}
       \caption{
       The energy resolution for electrons as a function of energy.
       The black points are the Module-0 data
	($\blacktriangle$ are the $14^{o}$ data,
	$\blacktriangledown$ are the $30^{o}$ data),
       the open points are the 1m prototype modules data
       ($\bigcirc$ -- $10^{o}$, $\square$ --  $20^{o}$, 
	$\triangle$ --  $30^{o}$).
       The lines are fits of eq.\ (\ref{qs}).
       \label{fv11}}
\end{figure*}

The relative electron energy resolutions, 
extracted from the energy distributions
(Fig.~\ref{fv34} and  Fig.~\ref{fv35}), are shown in Fig.~\ref{fv11} 
together with the 
1m prototype data as a function of $1 / \sqrt{E}$.
Fit of these data by the expression (\ref{qs})
produced the parameters $a_{exp}$ and $b_{exp}$
given in Table~\ref{Tb11}
together with the data 
for various iron-scintillator calorimeters.
\begin{equation}
\label{qs}
\frac{\sigma}{E}=\frac{a}{\sqrt{E}}\oplus b, 
\end{equation}

We compared our results on the energy resolution with the parameterization
suggested in  \cite{delpe}:
\begin{equation}
\label{dpe}
\frac{\sigma}{E} = \frac{a}{\sqrt{E}} =
\frac{{\sigma}_{o}}{\sqrt{E}} \cdot
{\Biggl( \frac{t}{X_t} \Biggr) }^{\gamma}
\cdot
{\Biggl( \frac{s}{X_s} \Biggr)}^{- \delta}, 
\end{equation}
where ${\sigma}_{o} = 6.33\ \% \cdot \sqrt{GeV}$,  $\gamma$ = 0.62, 
$\delta$ = 0.21 are the parameters, 
$X_t$ and $X_s$ are the radiation lengths of iron and scintillator
respectively.
In our case the values of $t$ and $s$ are equal to:
$t = 14~mm / \sin \Theta $,  $s = 3~mm / \sin \Theta$.
This formula is purely empirical and the parameters
${\sigma}_{o},  \gamma ,  \delta$ were determined by 
fitting the Monte Carlo data.

The results of calculations are given in Table~\ref{Tb11}.
As can be seen from this Table
the energy resolutions
obtained for ``ideal'' calorimeter are more accurate
(about a factor 1.5)
than the experimental ones.

%%%%%%%%%%%%%%%%%%%%%%%%%%%%%%%%%%%%%%%%%%%%%%%%%%%%%%%%%%%%%%%%%%%%%%%%%%
\section{Pion Response}

\begin{figure*}[tbph]
     \begin{center}
        \begin{tabular}{cc}
        \mbox{\epsfig{figure=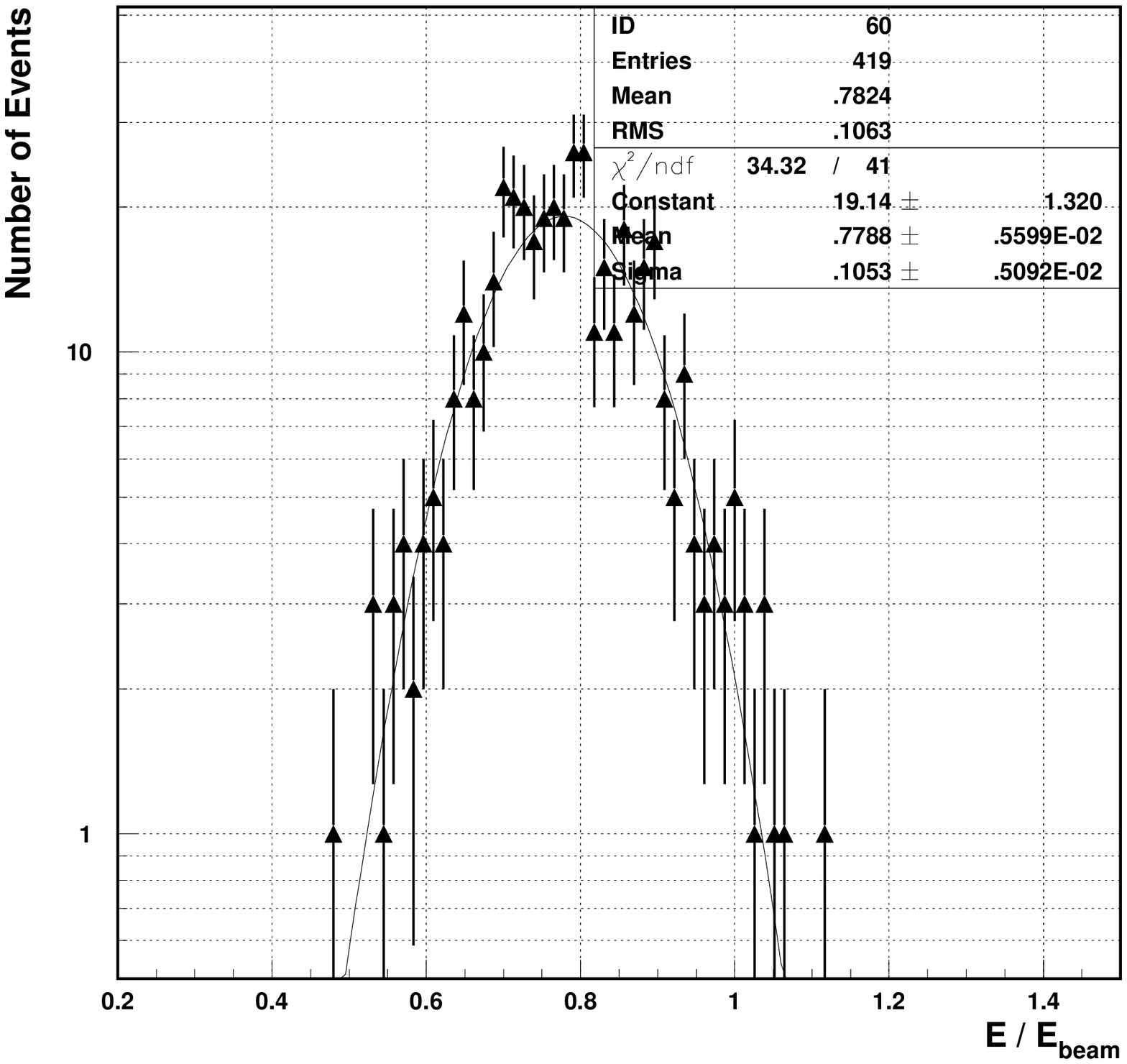,width=0.45\textwidth,height=0.25\textheight}}
        &
        \mbox{\epsfig{figure=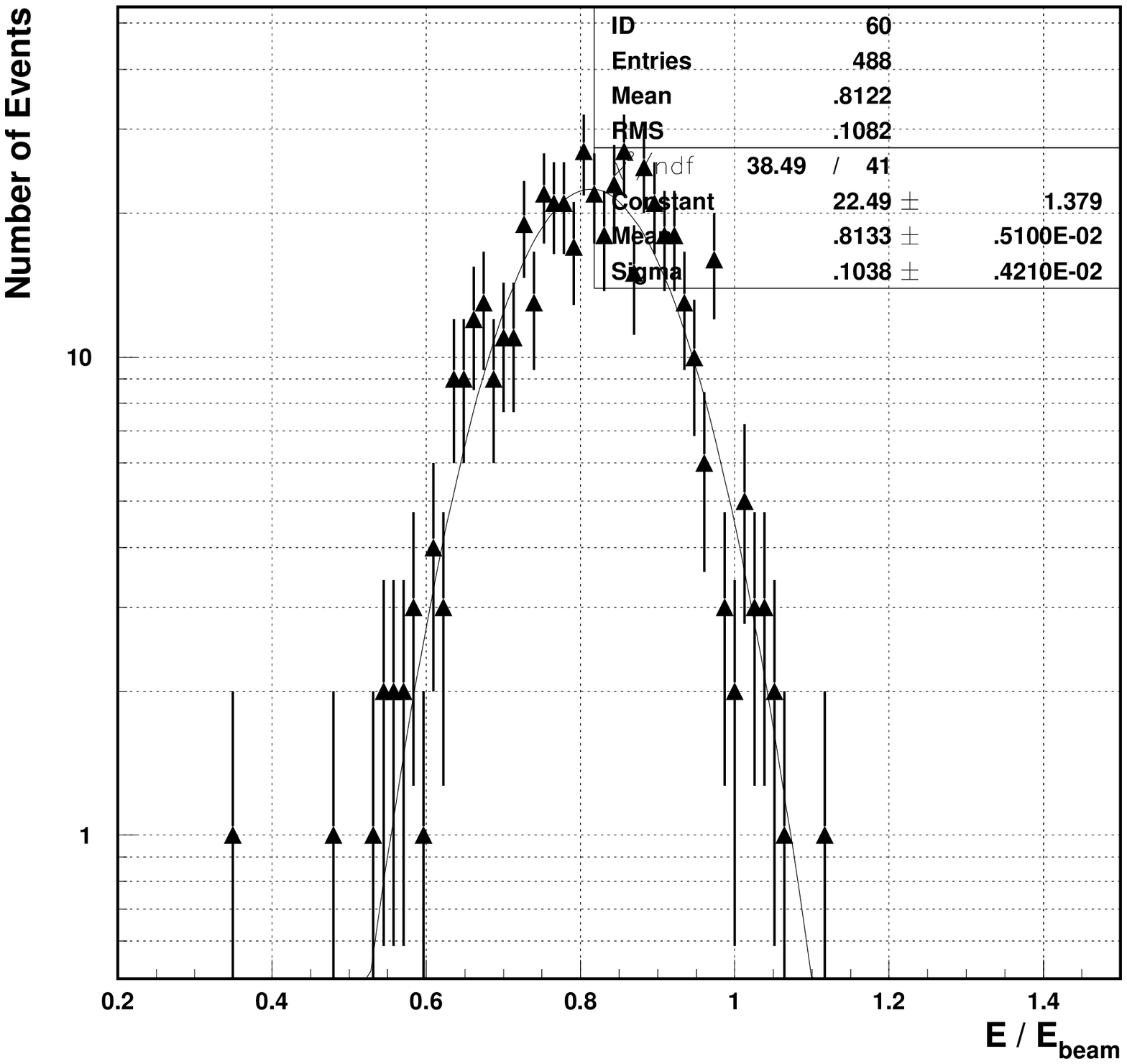,width=0.45\textwidth,height=0.25\textheight}}
        \\
        \mbox{\epsfig{figure=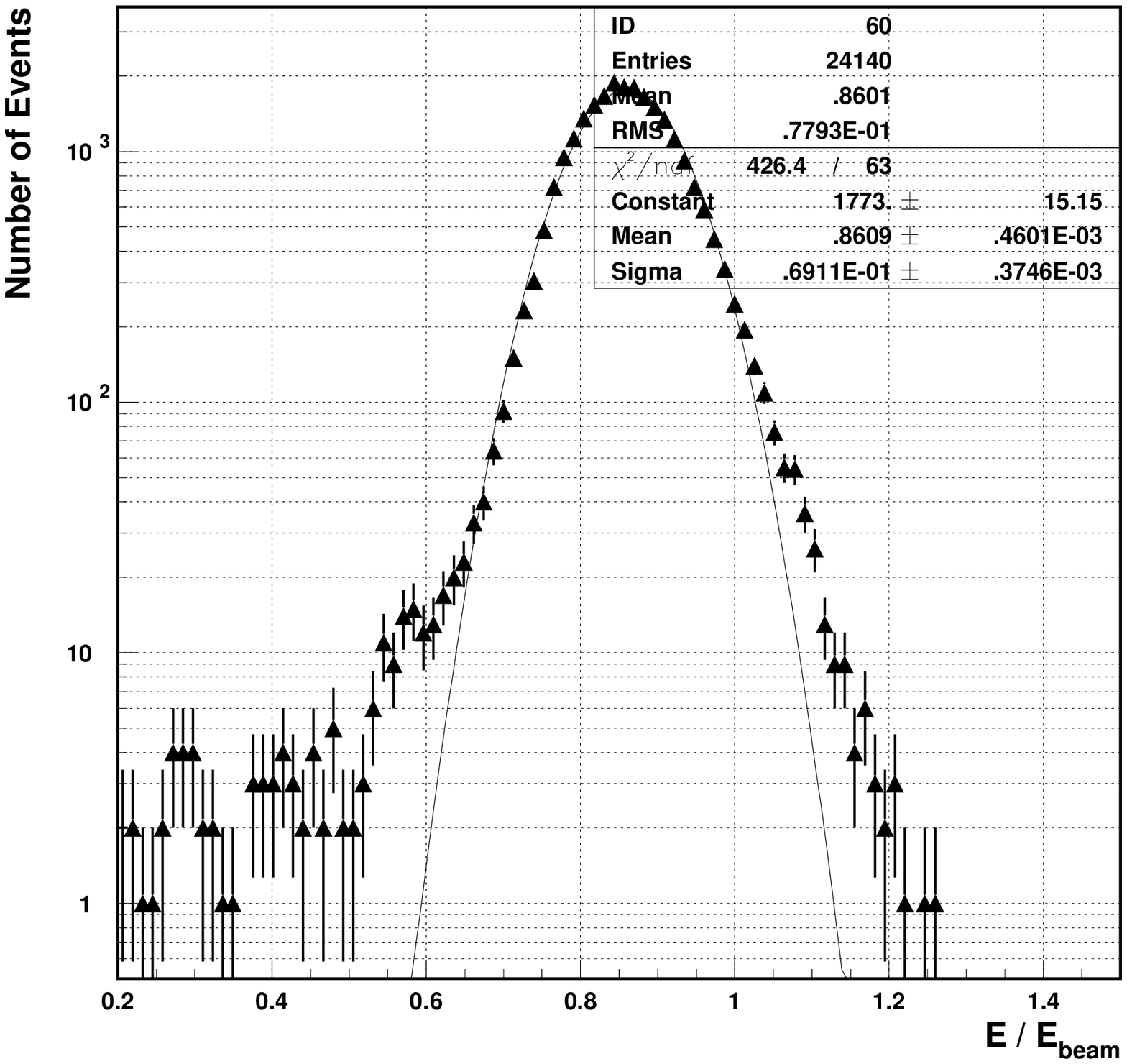,width=0.45\textwidth,height=0.25\textheight}}
        &
        \mbox{\epsfig{figure=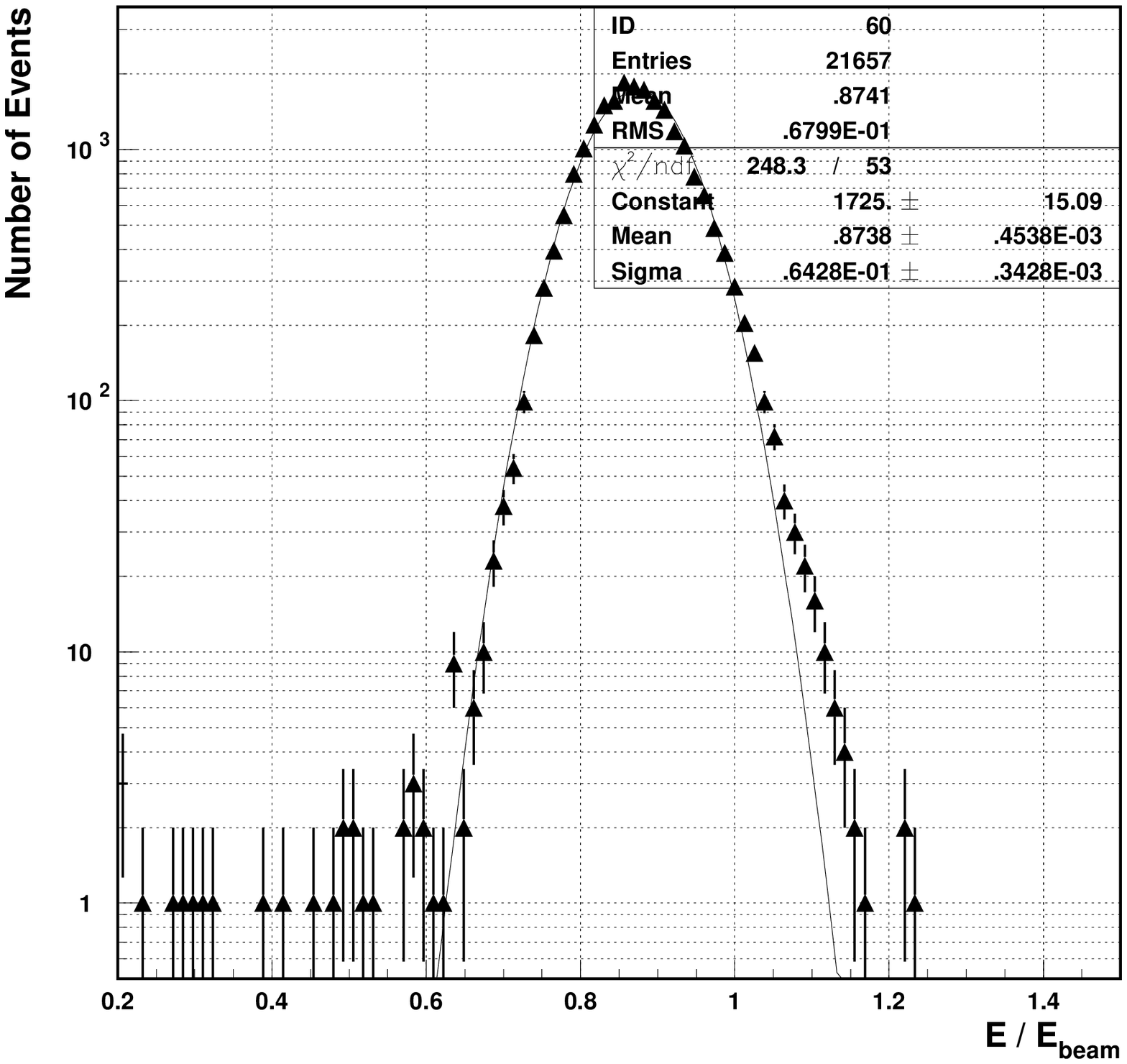,width=0.45\textwidth,height=0.25\textheight}}
        \\
        \mbox{\epsfig{figure=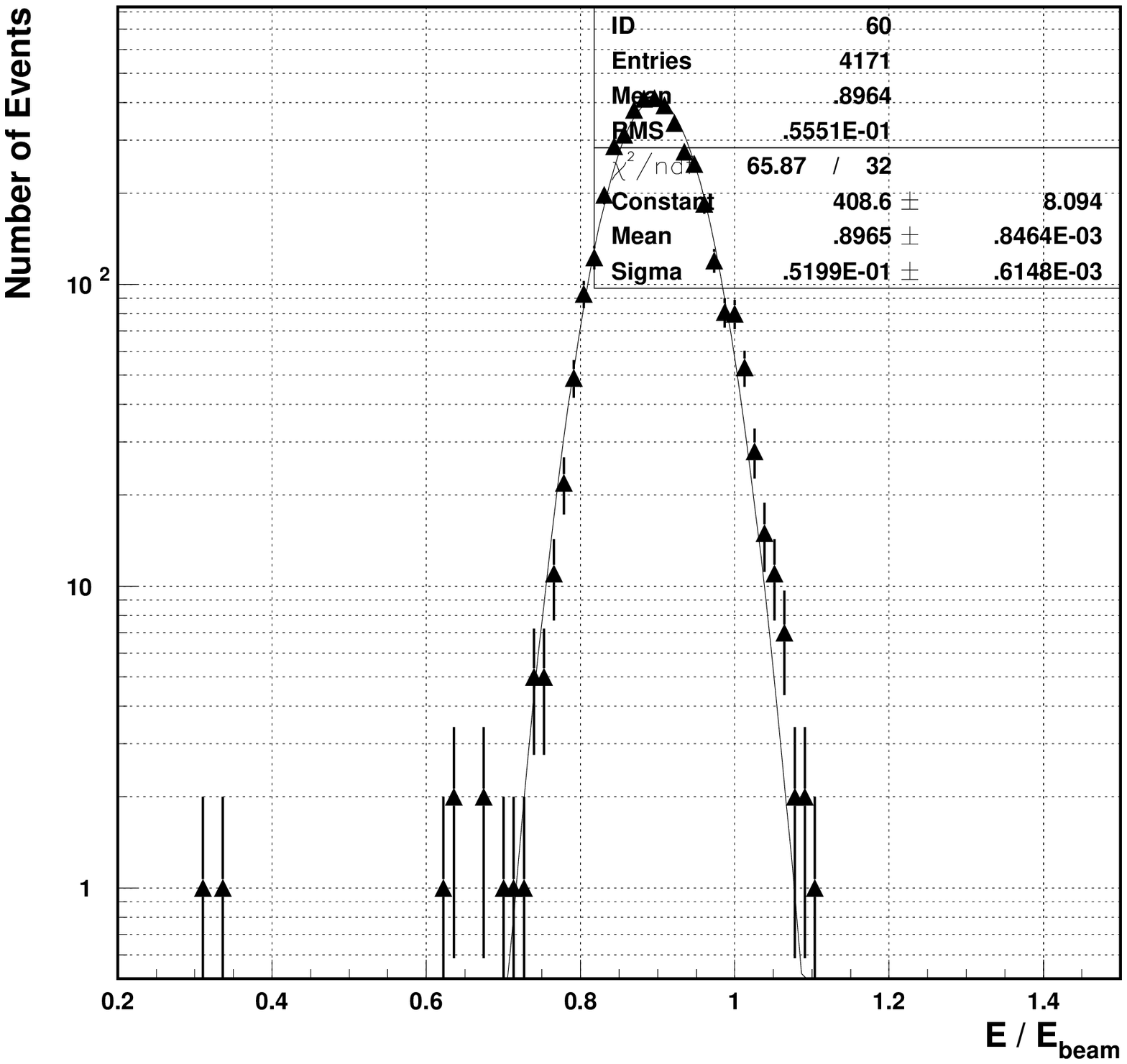,width=0.45\textwidth,height=0.25\textheight}}
        &
        \mbox{\epsfig{figure=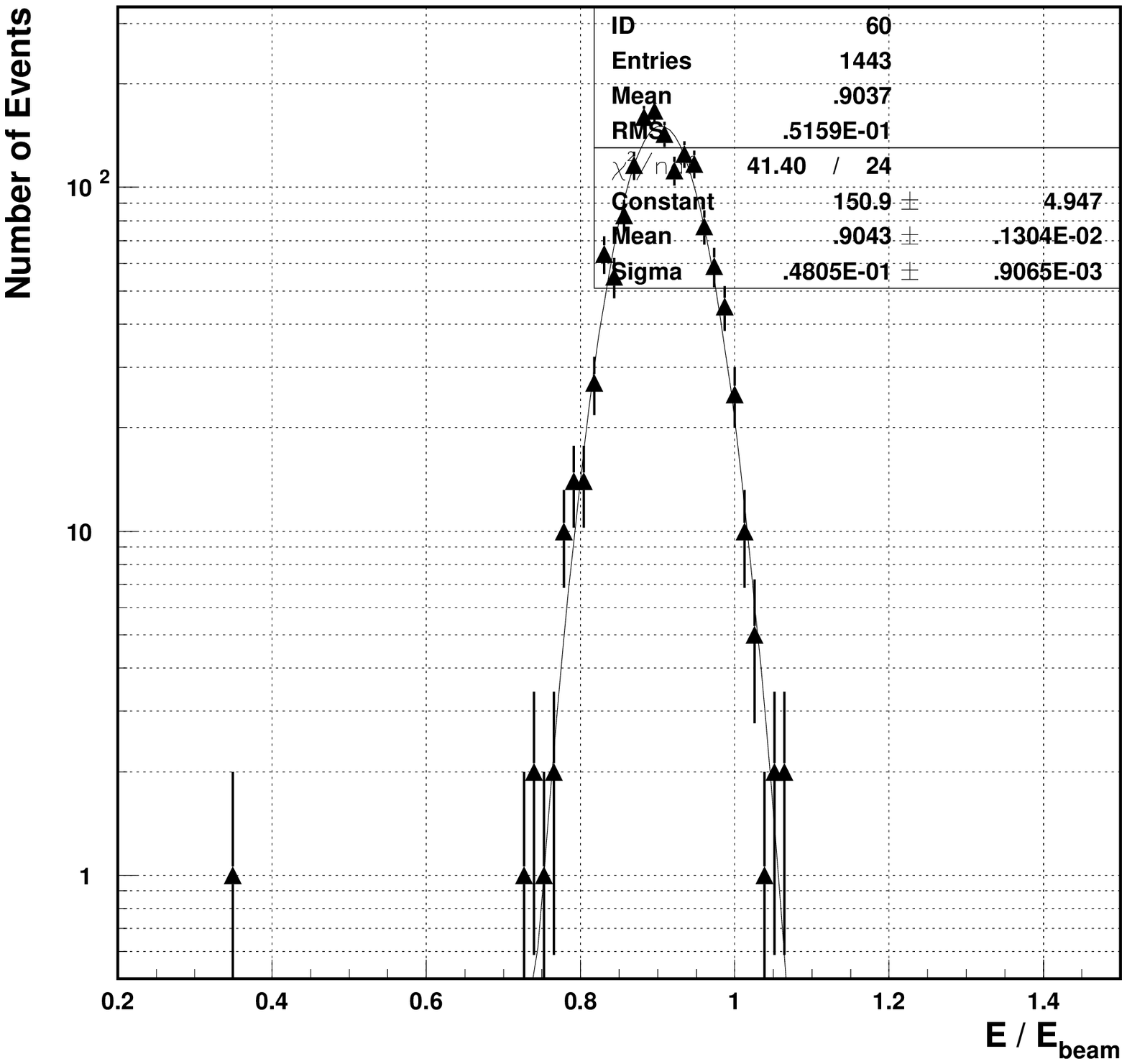,width=0.45\textwidth,height=0.25\textheight}}
        \\
        \end{tabular}
     \end{center}
       \caption{
	The distributions of
	the normalized pion response ($E_{\pi} / E_{beam}$)
 	for E = 20, 100, 180 GeV
 	at $\eta = -0.25$ (left column, up to down)   and
 	at $\eta = -0.55$    (right column, up to down).
       \label{fv40}}
\end{figure*}

\begin{figure*}[tbph]
     \begin{center}
        \begin{tabular}{cc}
        \mbox{\epsfig{figure=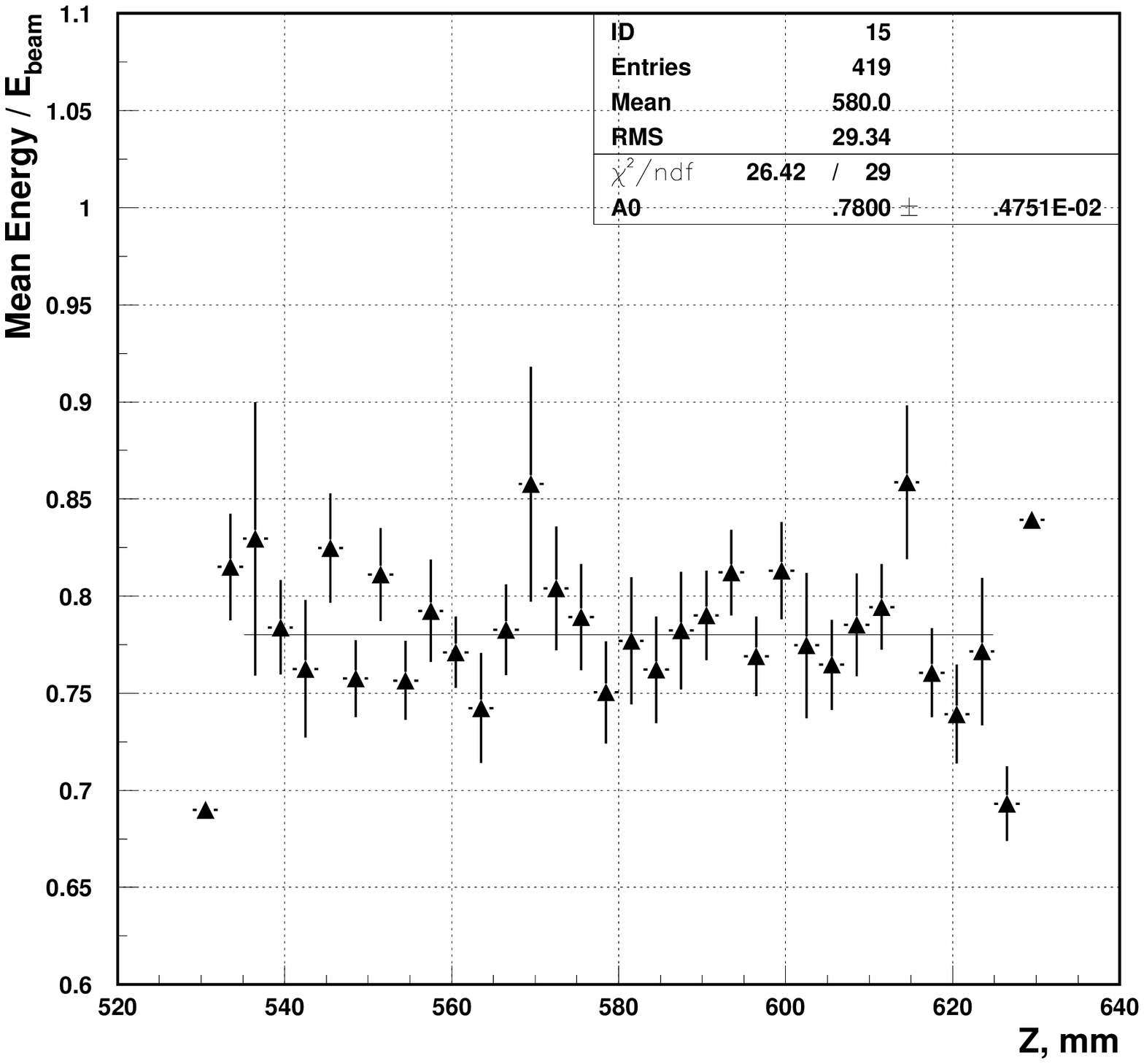,width=0.45\textwidth,height=0.25\textheight}}
        &
        \mbox{\epsfig{figure=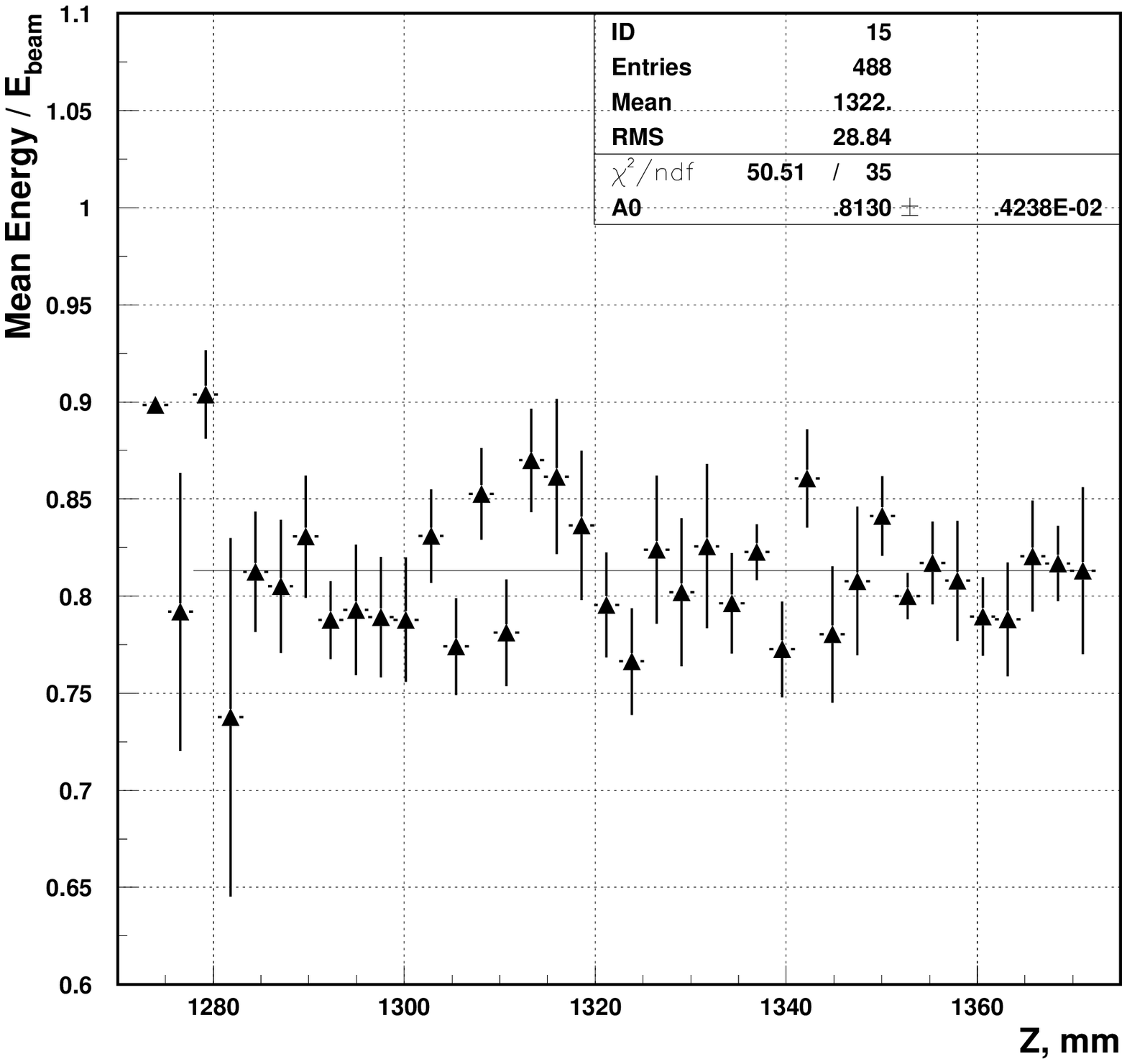,width=0.45\textwidth,height=0.25\textheight}}
        \\
        \mbox{\epsfig{figure=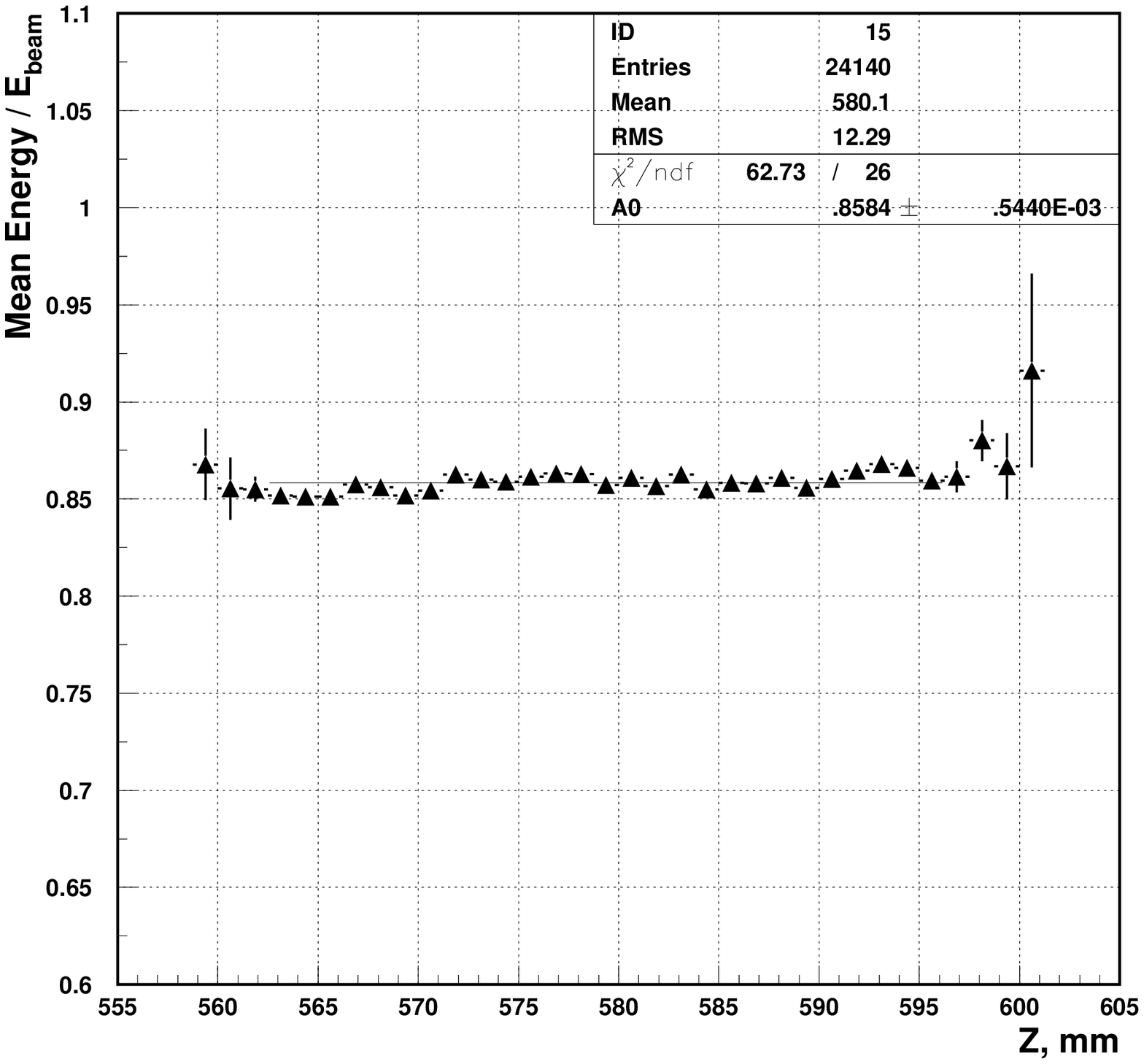,width=0.45\textwidth,height=0.25\textheight}}
        &
        \mbox{\epsfig{figure=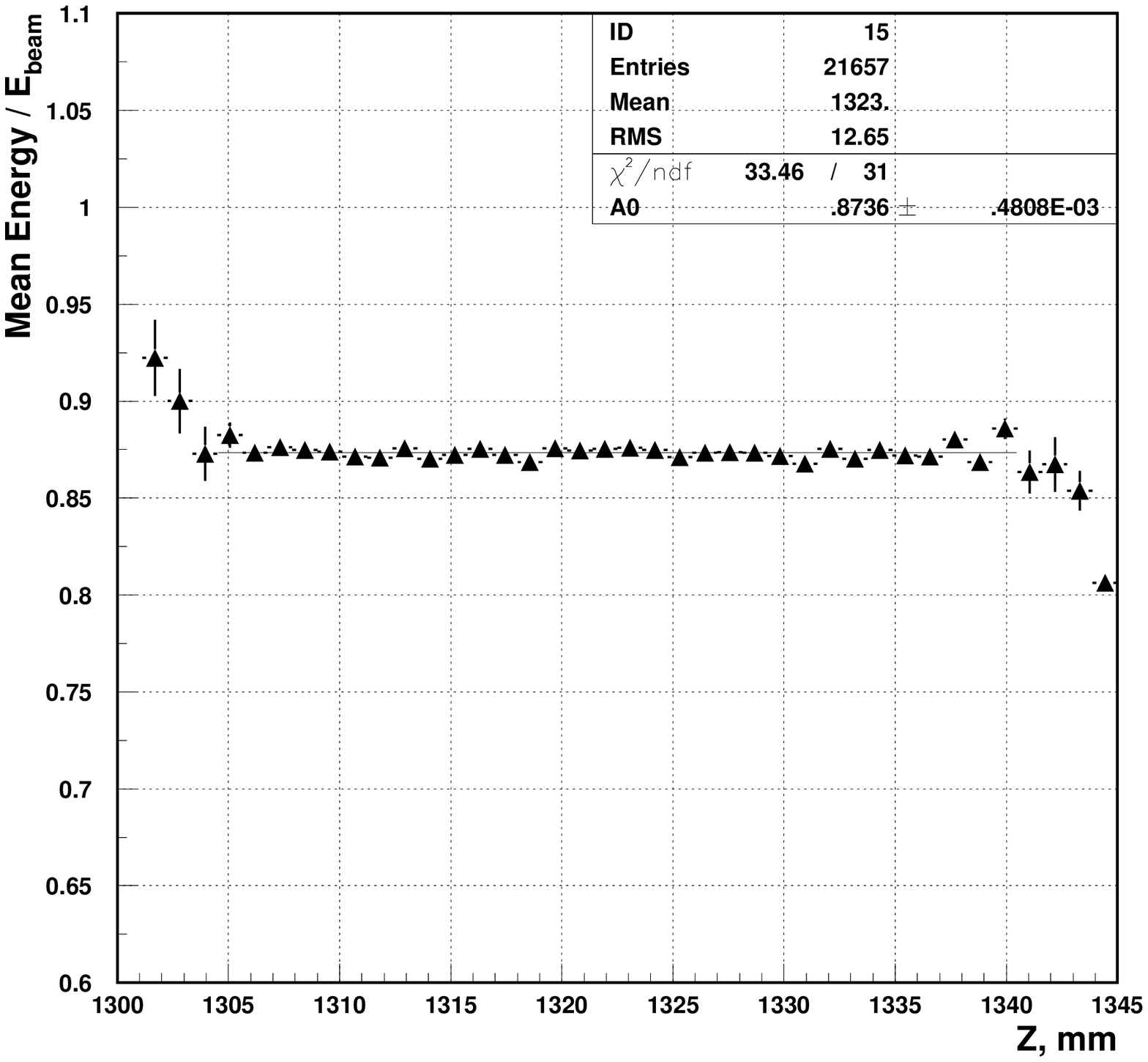,width=0.45\textwidth,height=0.25\textheight}}
        \\
        \mbox{\epsfig{figure=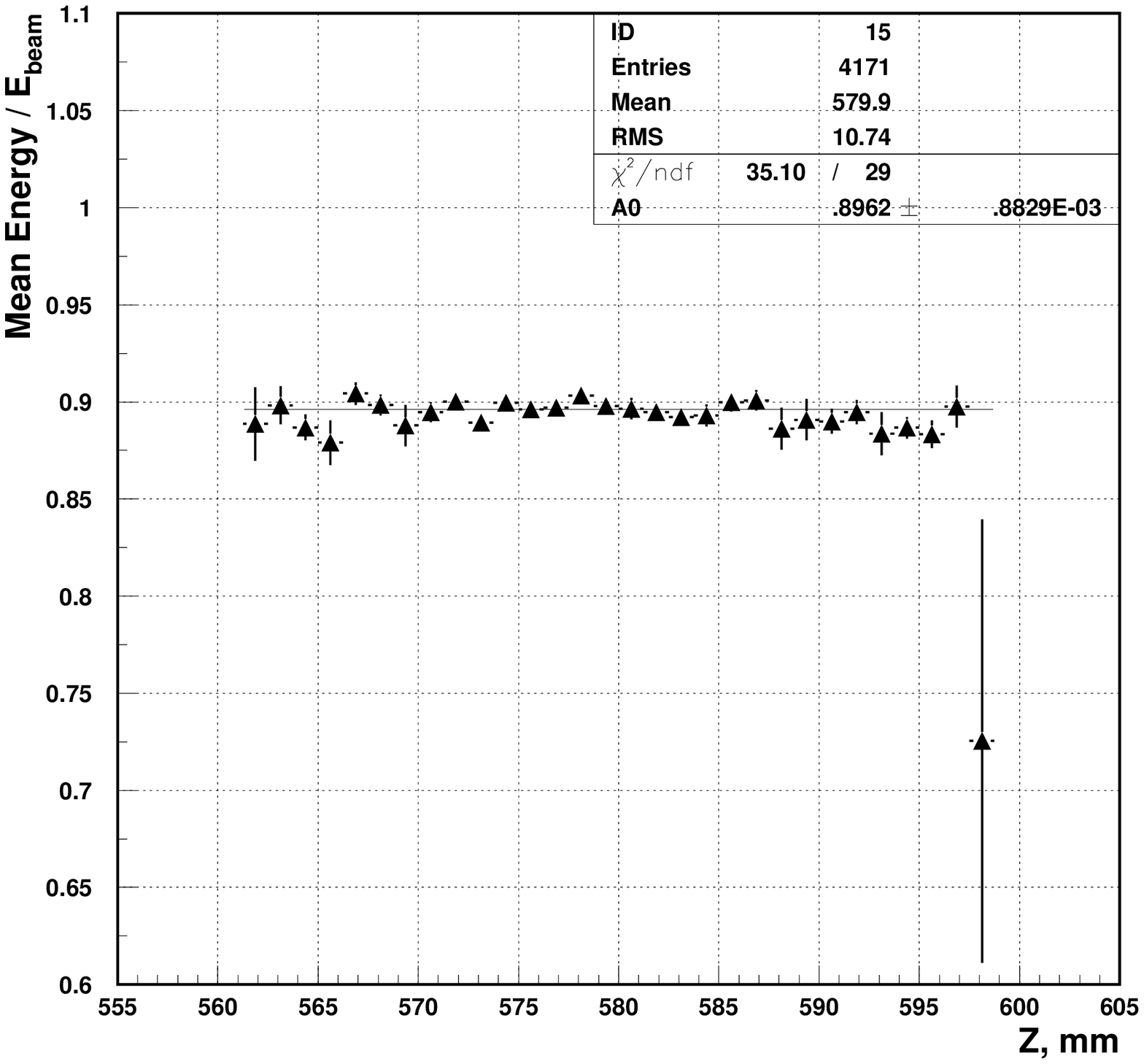,width=0.45\textwidth,height=0.25\textheight}}
        &
        \mbox{\epsfig{figure=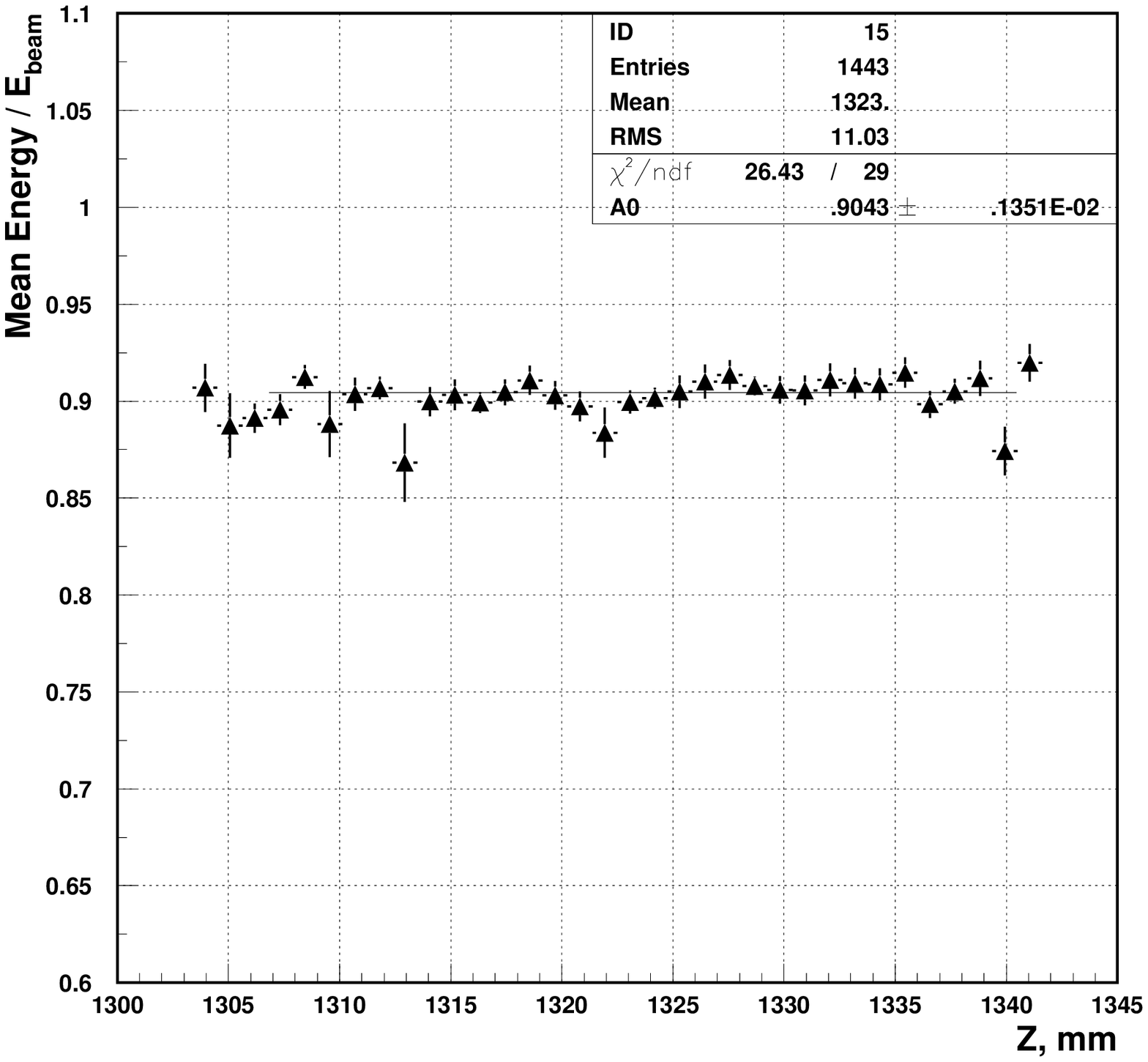,width=0.45\textwidth,height=0.25\textheight}}
        \\
        \end{tabular}
     \end{center}
       \caption{
	The normalized pion response ($E_{\pi} / E_{beam}$)
 	for $E_{beam}$ = 20, 100, 180 GeV
 	at $\eta = -0.25$
	(left column, up to down)
  	and
 	at $\eta = -0.55$    (right column, up to down)
 	as a function of $Z$ coordinate.
       \label{fv41}}
\end{figure*}

Fig.~\ref{fv40}  shows
the normalized pion response ($E_{\pi} \ E_{beam}$) 
for $E_{beam}$ = 20, 100, 180 GeV
 at $\eta = -0.25$ and $-0.55$.
Fig.~\ref{fv41}  shows
the normalized pion response  for $E_{beam}$ = 20, 100, 180 GeV
 at $\eta = -0.25$ and $-0.55$
as a function of impact point $Z$ coordinate.
Contrary to electrons these pion Z-dependences do not show any significant
periodical structure.

Fig.~\ref{fv42} shows the mean normalized pion response,
extracted from Fig.~\ref{fv41}, as a function of
energy for two values of $\eta$.
The meaning of lines is given below.
As can be expected,  since the  $e / \pi$ ratio is not equal to 1, 
the mean normalized pion response increases with the beam energy increasing.

As can be seen the pion response is different for various $\eta$. 
The values of the pion response for $\eta = -0.55$ are larger than ones for
$\eta = -0.25$.
We tried to explain if the reason of this difference is the lateral leakage
through gaps between the 1m prototype modules.
We estimated the lateral leakages to the gaps taking into account the 
longitudinal energy deposition and the spatial radial deposition.
It turned out that the leakage for $\eta = -0.25$ is larger than for 
$\eta = -0.55$ but it is unsufficient, less than 1 \%, in order to explain
the observed difference in the pion responses.

%%%%%%%%%%%%%%%%%%%%%%%%%%%%%%%%%%%%%%%%%%%%%%%%%%%%%%%%%%%%%%%%%%%%%%%%%%
\section{$e / h$ Ratio}

\begin{figure*}[tbph]
     \begin{center}
        \begin{tabular}{cc}
        \mbox{\epsfig{figure=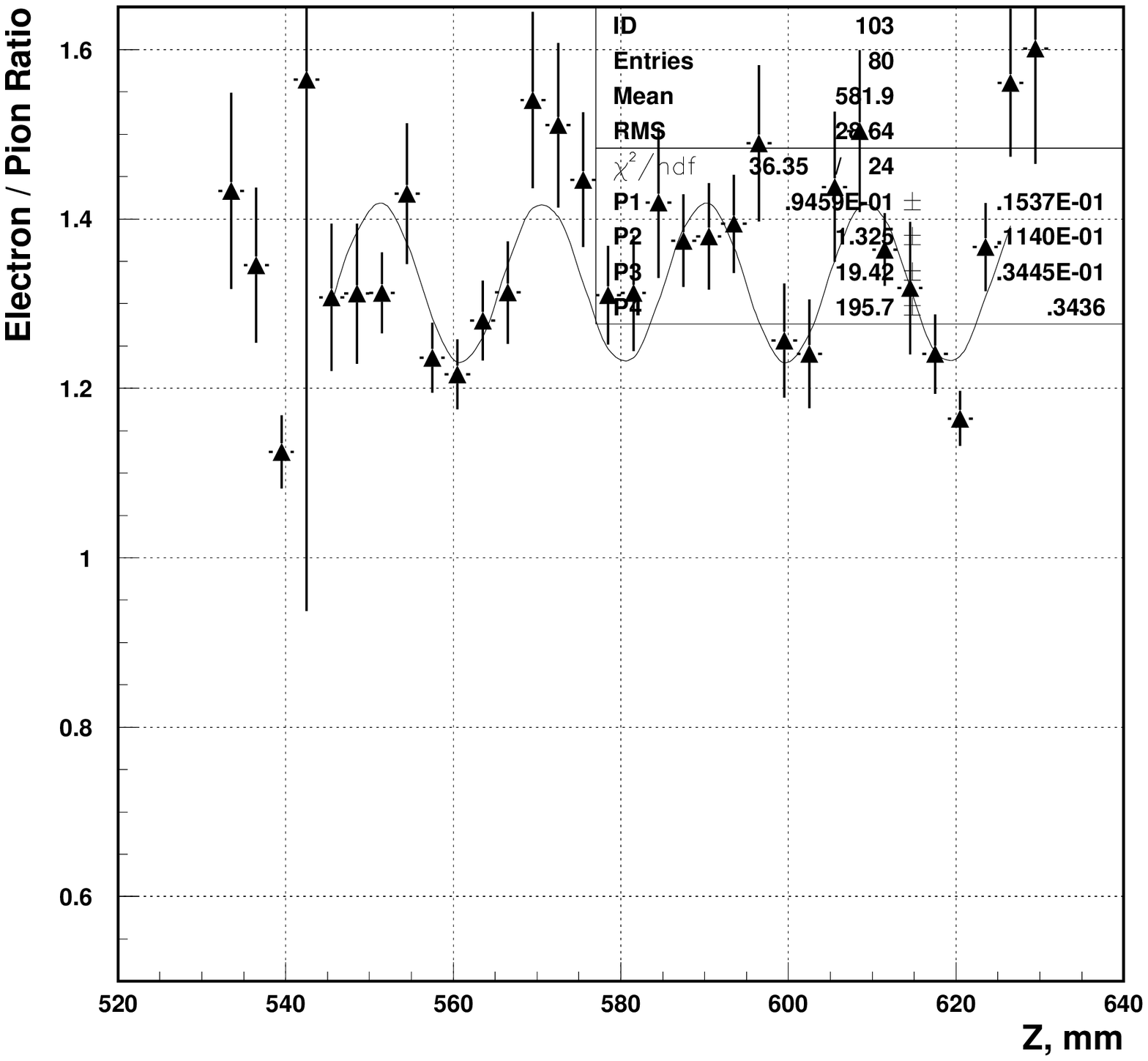,width=0.45\textwidth,height=0.25\textheight}}
        &
        \mbox{\epsfig{figure=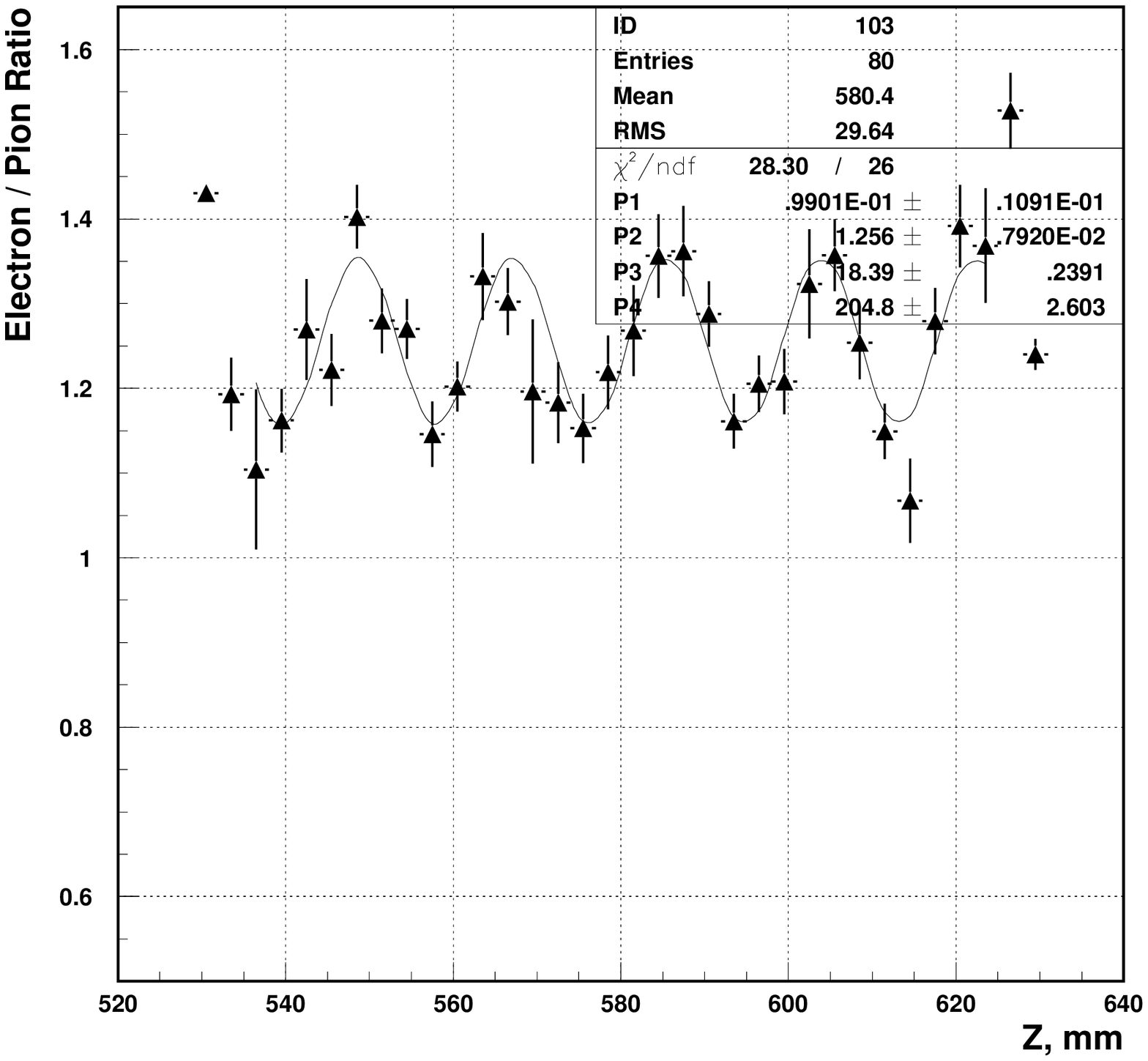,width=0.45\textwidth,height=0.25\textheight}}
        \\
        \mbox{\epsfig{figure=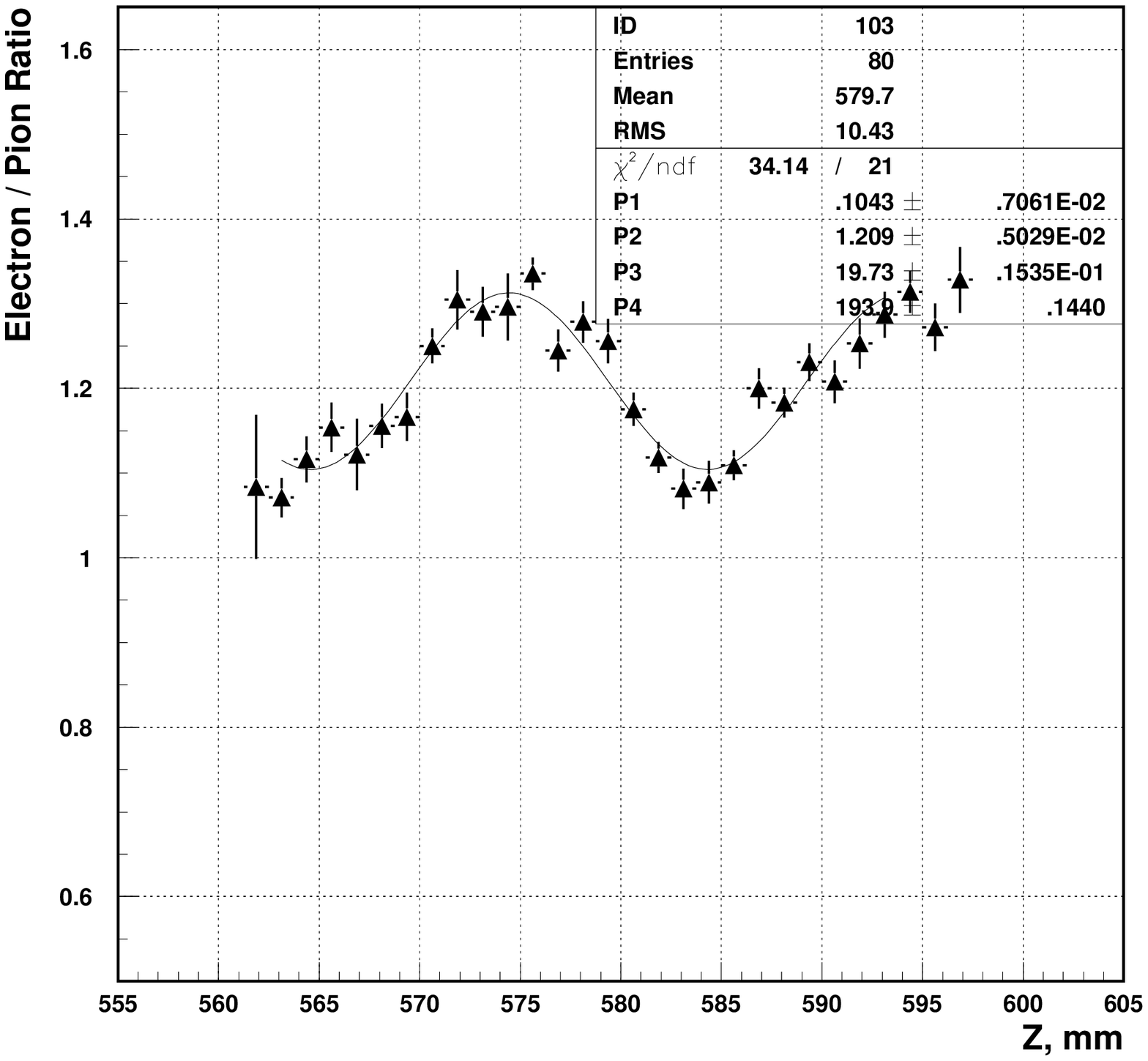,width=0.45\textwidth,height=0.25\textheight}}
        &
        \mbox{\epsfig{figure=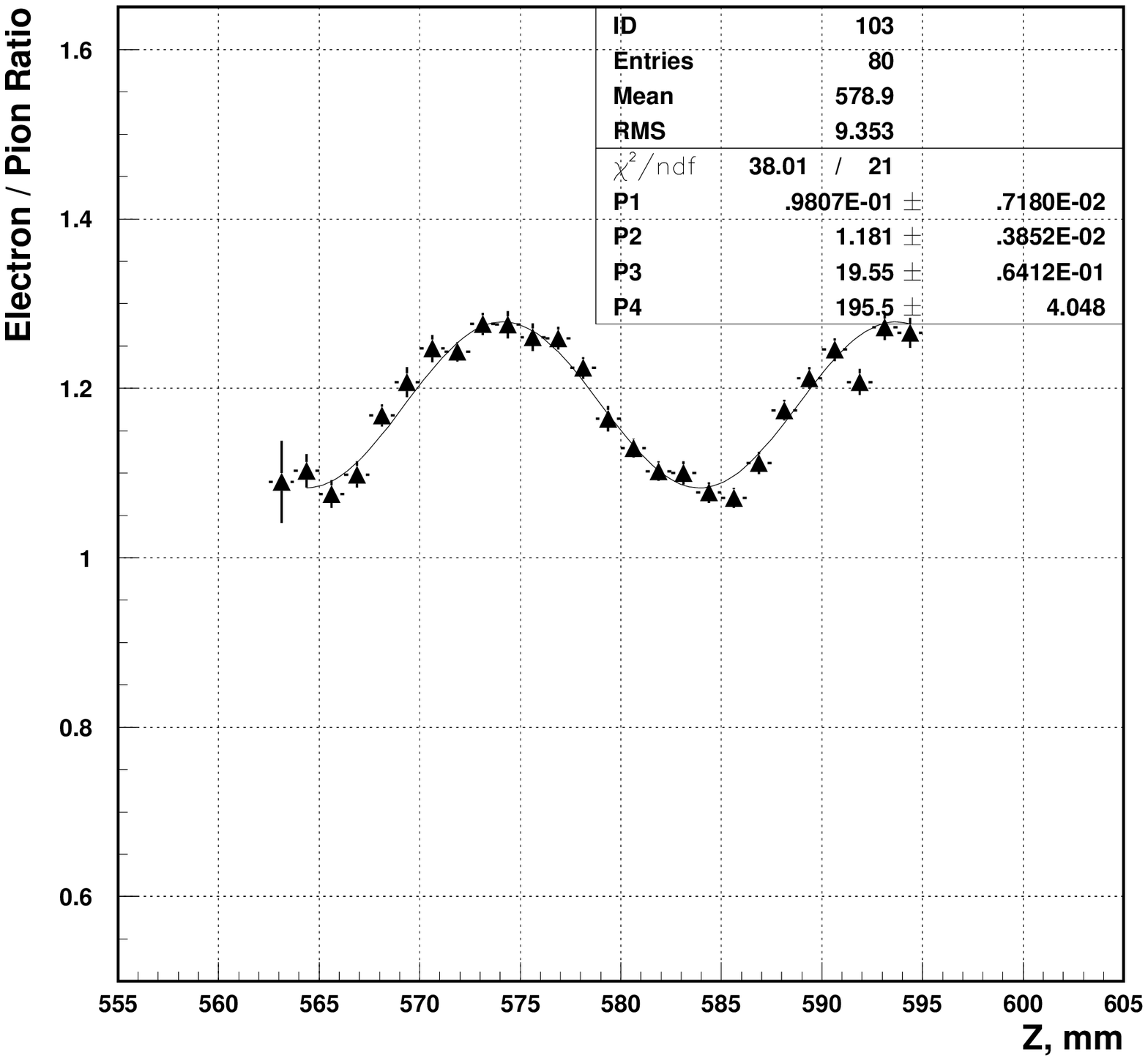,width=0.45\textwidth,height=0.25\textheight}}
        \\
        \mbox{\epsfig{figure=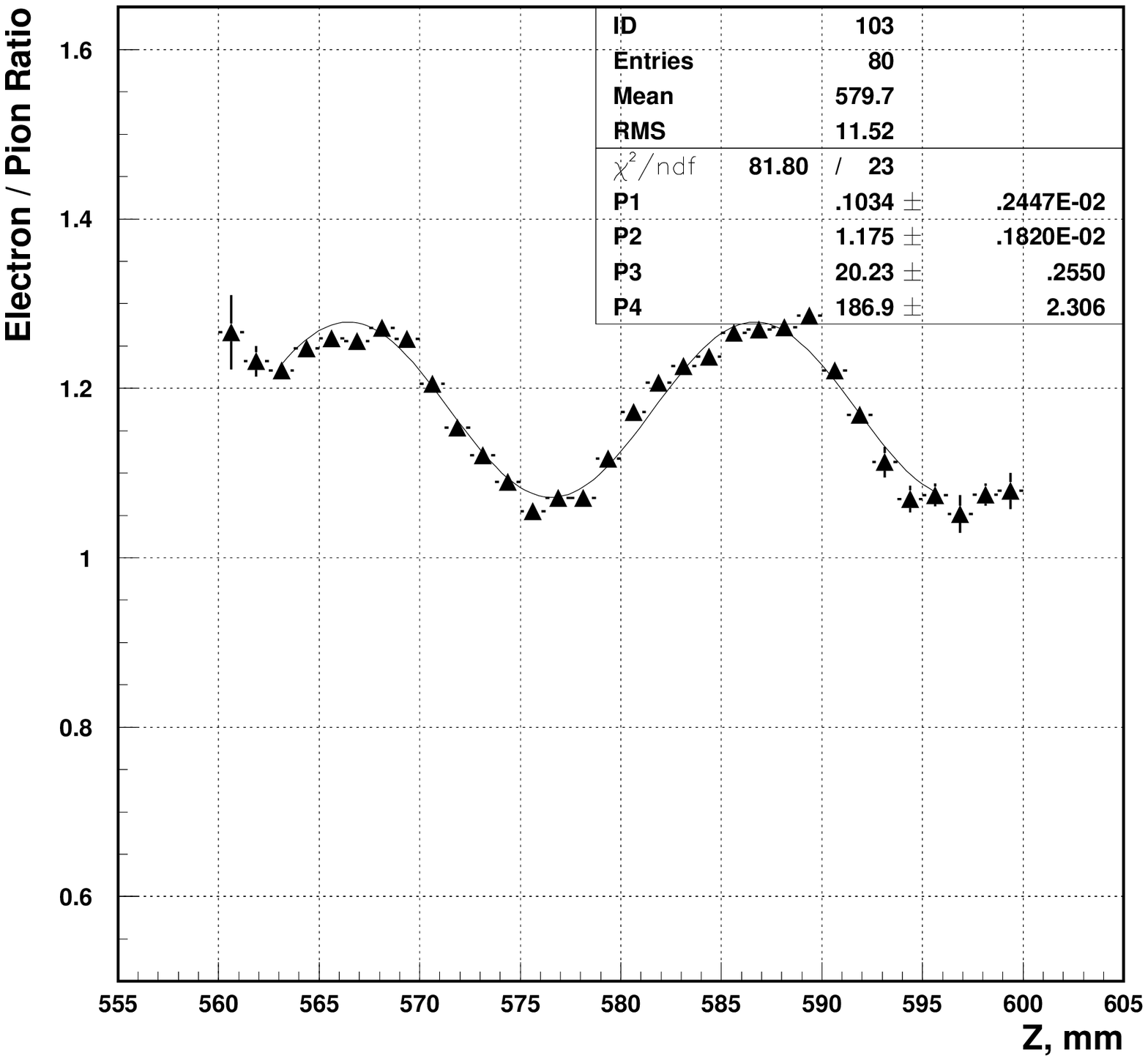,width=0.45\textwidth,height=0.25\textheight}}
        &
        \mbox{\epsfig{figure=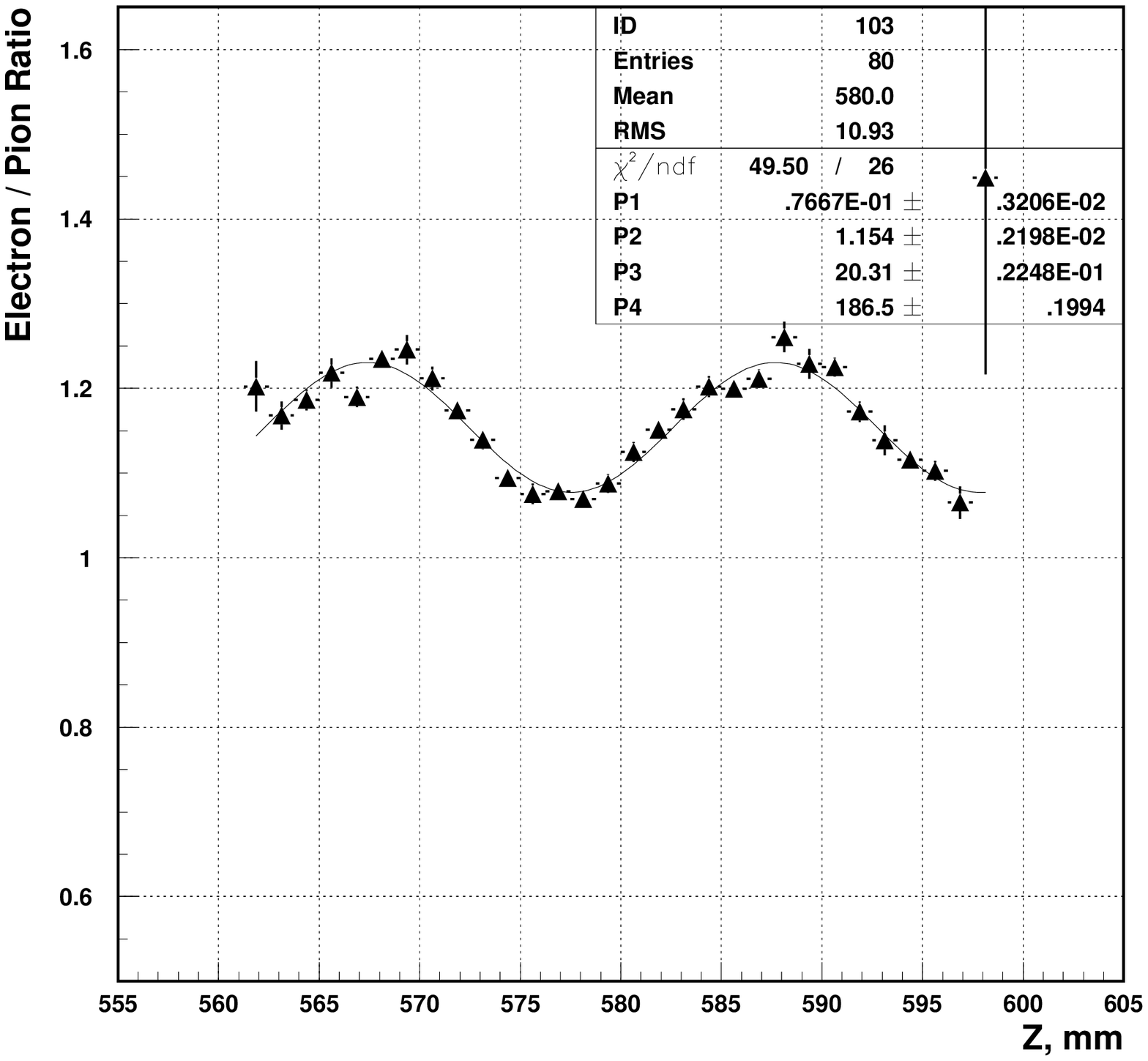,width=0.45\textwidth,height=0.25\textheight}}
        \\
        \end{tabular}
     \end{center}
       \caption{
	        The $e / \pi$ ratio for Module-0
                as a function of $Z$ coordinate
 		for E = 10, 60, 100 GeV (left column, up to down)
  		and
		E = 20, 80, 180 GeV (right column, up to down)
 		at $\eta = -0.25$
       \label{fv30}}
\end{figure*}

\begin{figure*}[tbph]
     \begin{center}
        \begin{tabular}{cc}
        \mbox{\epsfig{figure=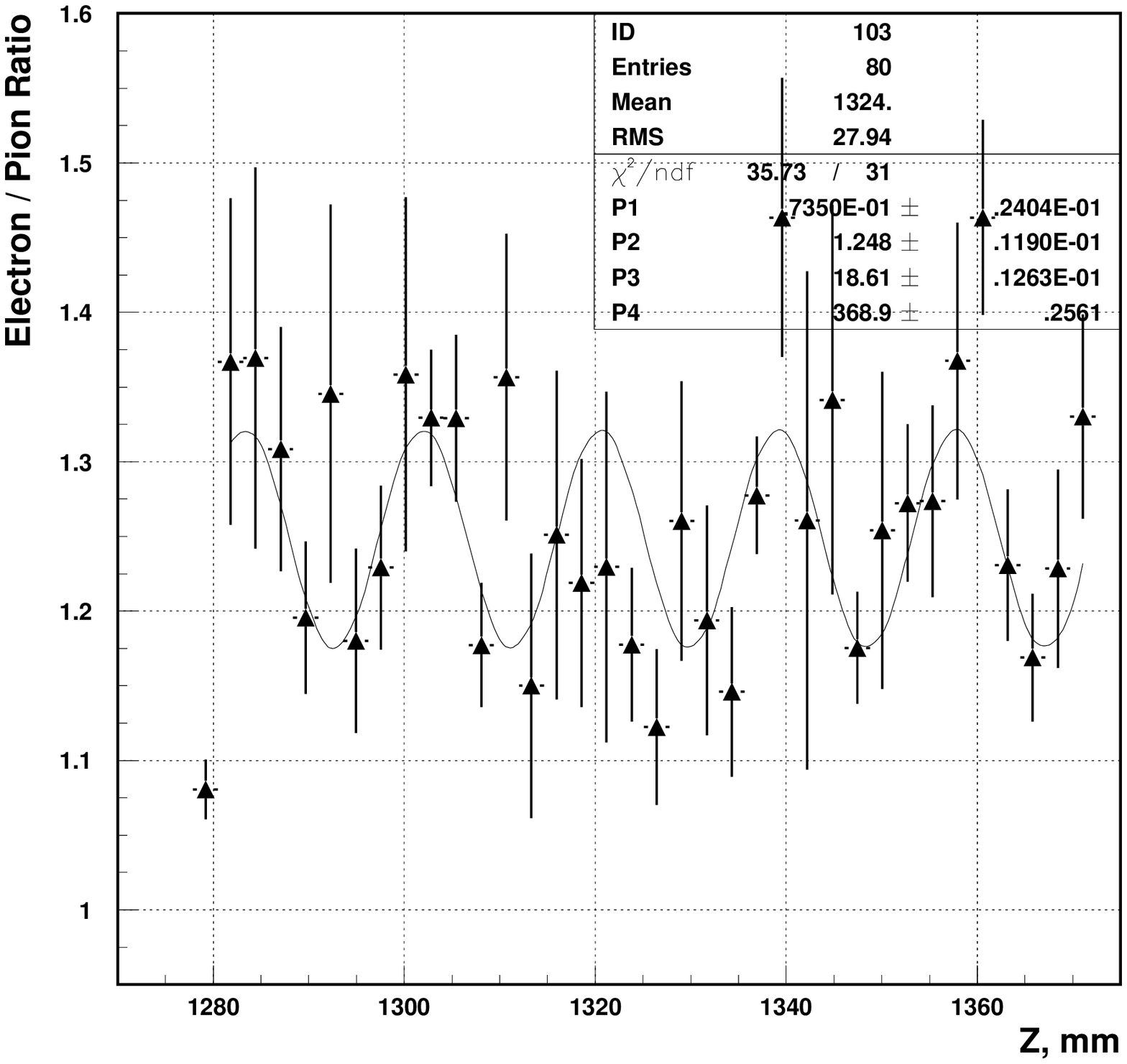,width=0.45\textwidth,height=0.25\textheight}}
        &
        \mbox{\epsfig{figure=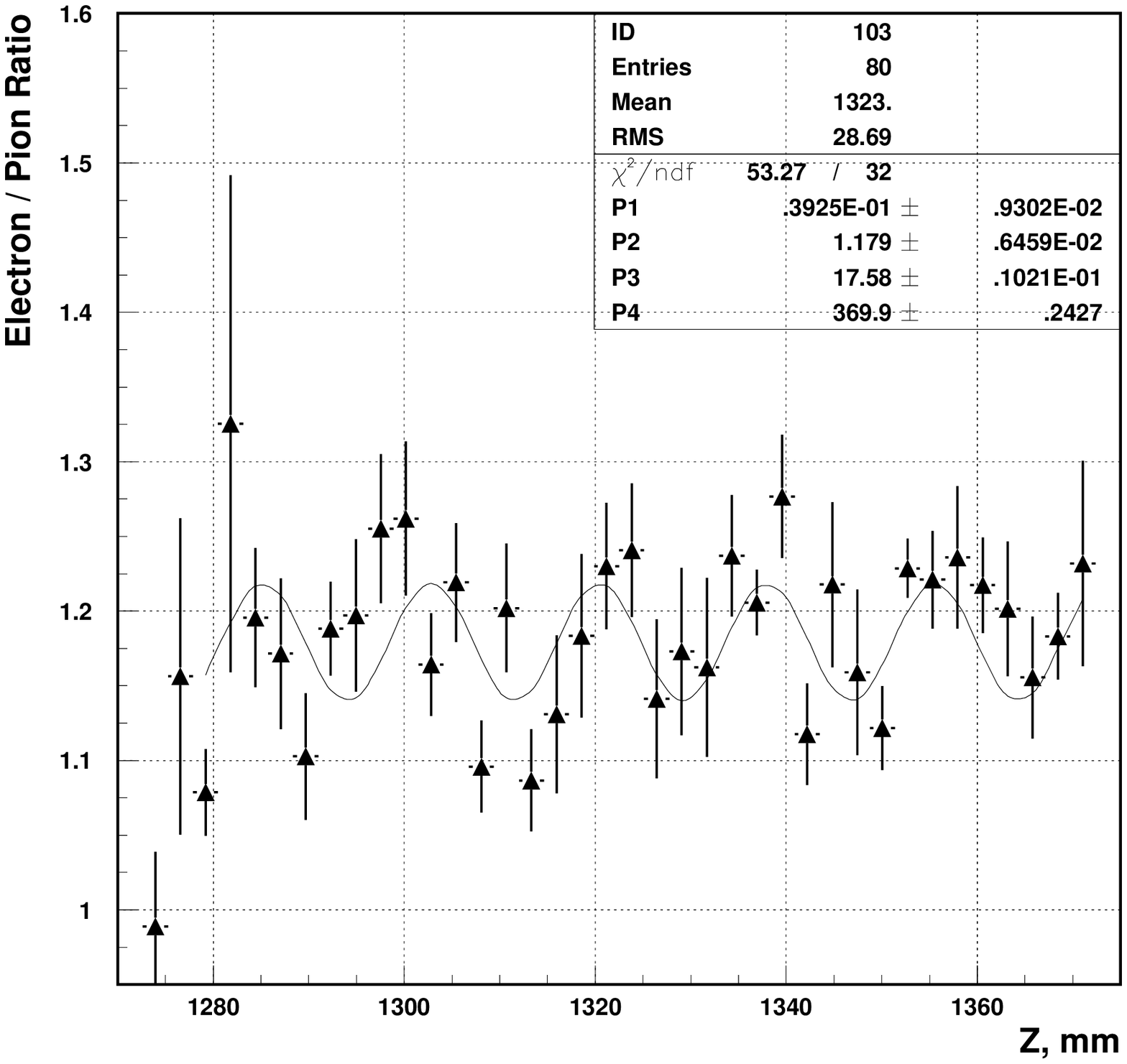,width=0.45\textwidth,height=0.25\textheight}}
        \\
        \mbox{\epsfig{figure=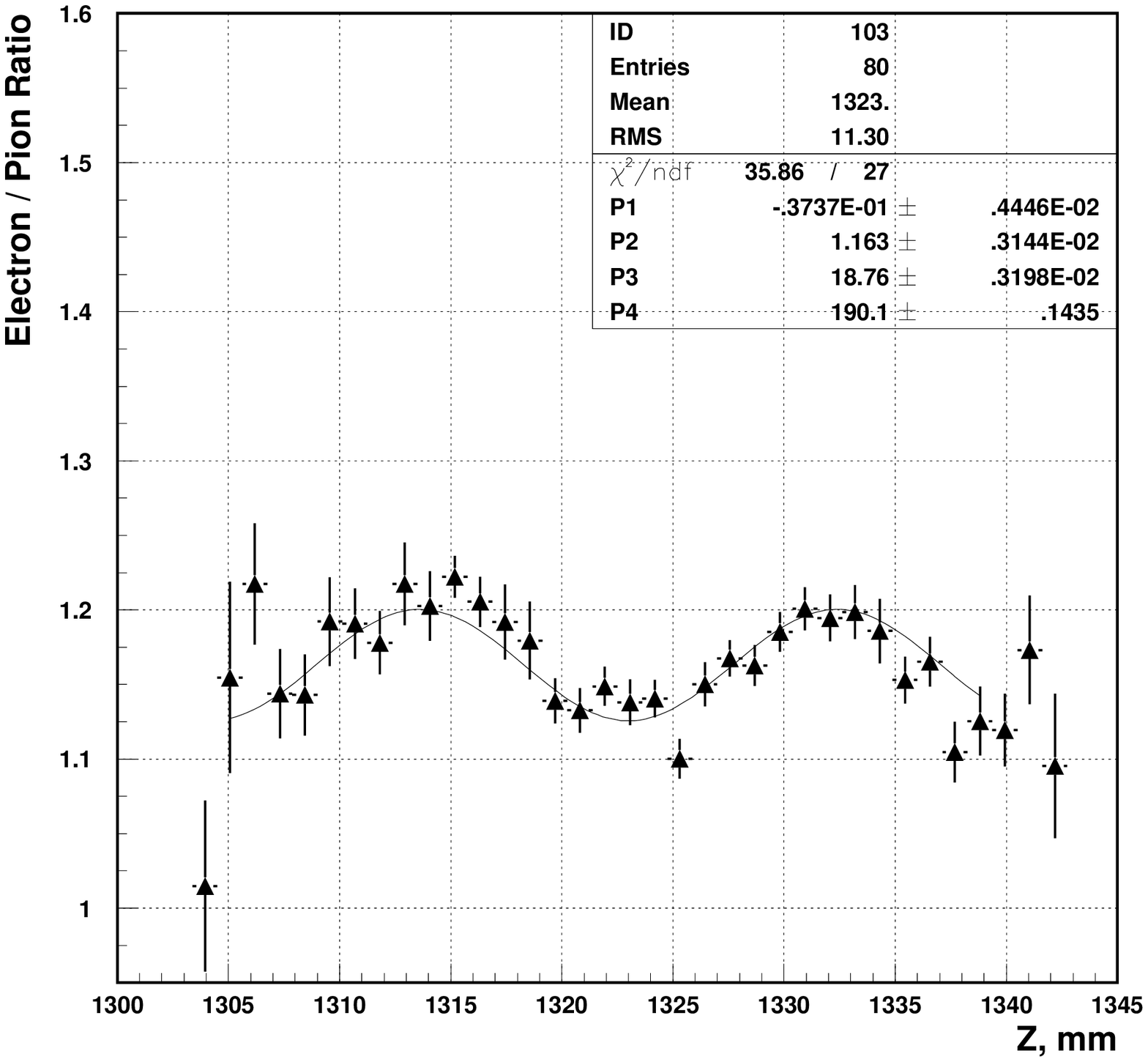,width=0.45\textwidth,height=0.25\textheight}}
        &
        \mbox{\epsfig{figure=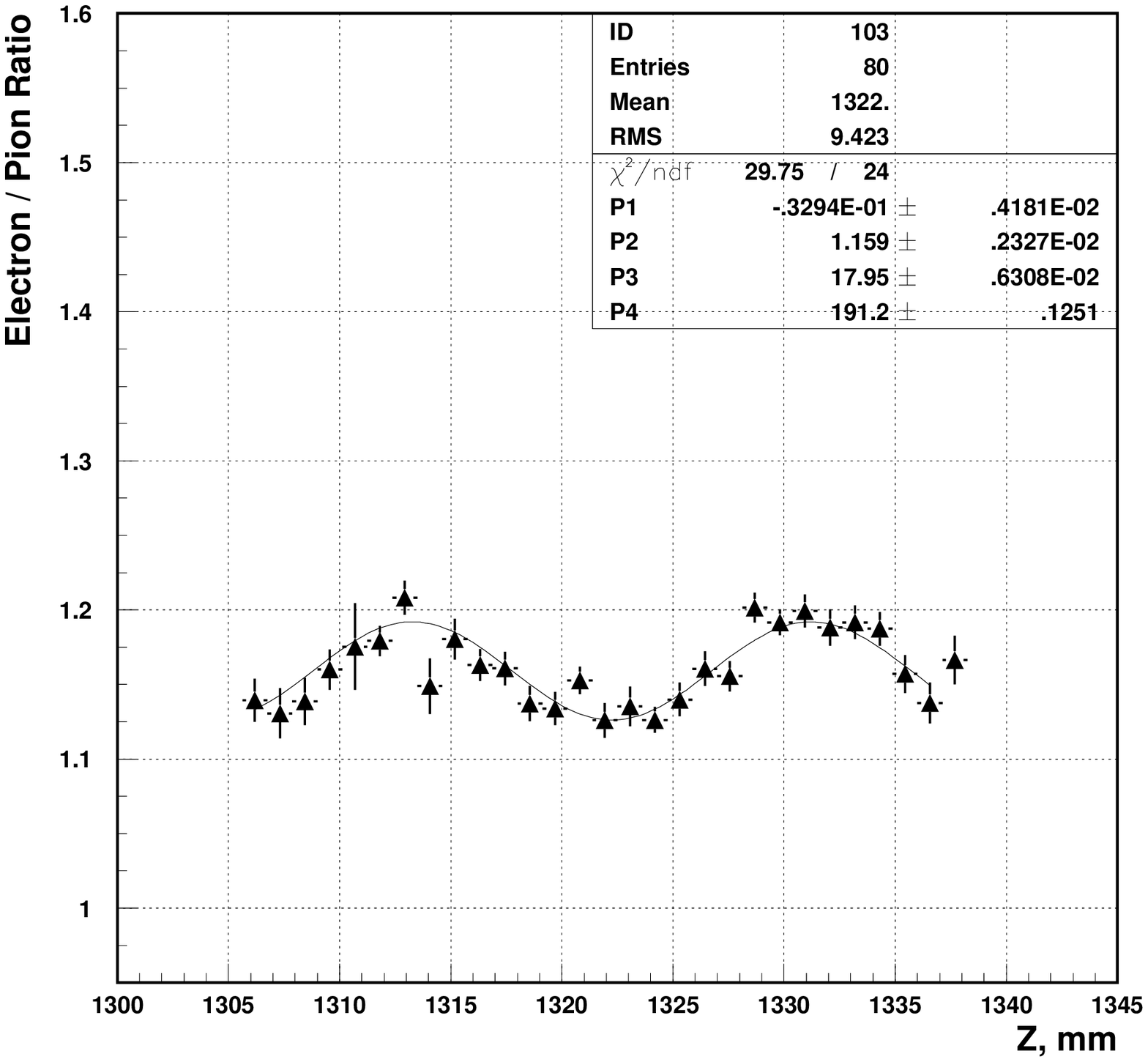,width=0.45\textwidth,height=0.25\textheight}}
        \\
        \mbox{\epsfig{figure=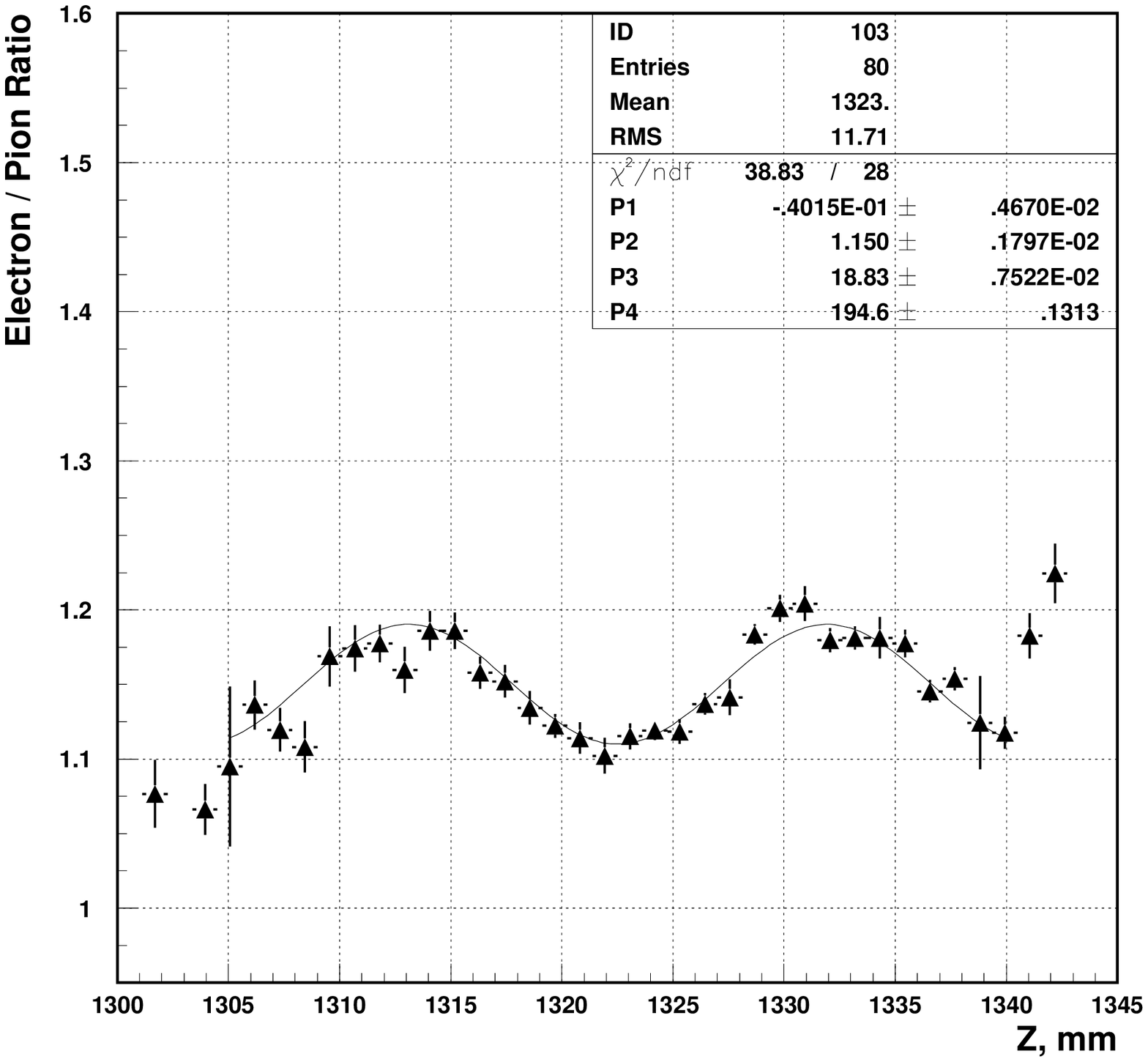,width=0.45\textwidth,height=0.25\textheight}}
        &
        \mbox{\epsfig{figure=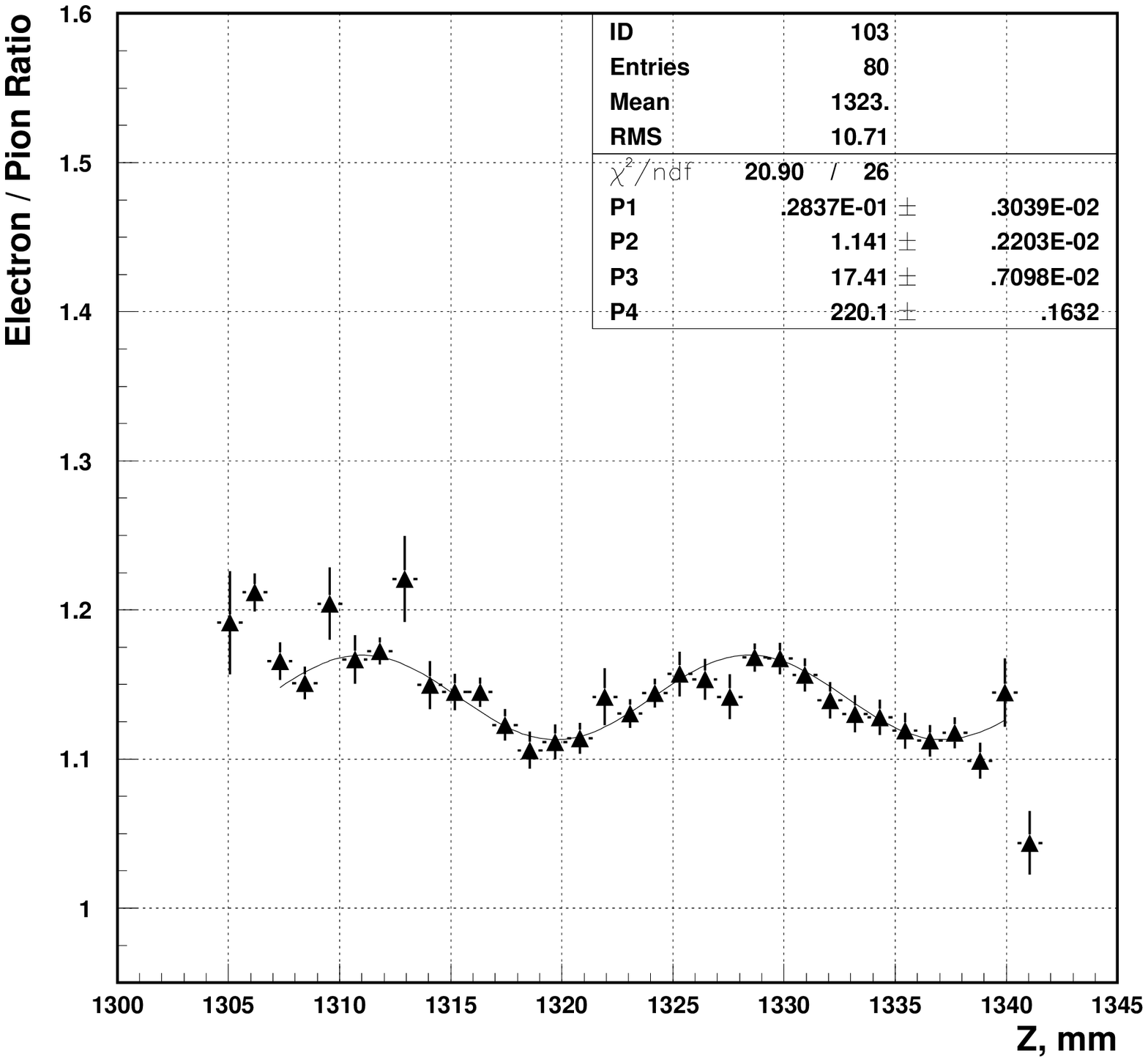,width=0.45\textwidth,height=0.25\textheight}}
        \\
        \end{tabular}
     \end{center}
       \caption{
		The $e / \pi$ ratio for the Module-0
                as a function of $Z$ coordinate
 		for E = 10, 60, 100 GeV (left column, up to down) and
		E = 20, 80, 180 GeV (right column, up to down)
 		at $\eta = -0.55$.
       \label{fv31}}
\end{figure*}

The responses obtained for $e$ and $\pi$ give the possibility to determine
the  $e / h$ ratio, 
an intrinsic non-compensation of a calorimeter.
In our case the electron -- pion ratios reveal complicated structures
$e / \pi  = f(E,  \Theta,  Z)$.
Fig.~\ref{fv30} and Fig.~\ref{fv31} show the $e/ \pi$ ratios
 for Module-0 for
E = 10,  20,  60,  80,  100 and 180  GeV at
$\eta = -0.25$ and $-0.55$
as a function of $Z$ coordinate.
If for the 1m prototype modules
the local compensation has been observed
(for some $Z$ points  at 20 GeV
 and  $\Theta = 10^{o}$,  see Fig.\ 4 given in \cite{budagov95})
 as to the Module-0 this is not this case.

\begin{figure*}[tbph]
    \begin{center}
       \mbox{\epsfig{figure=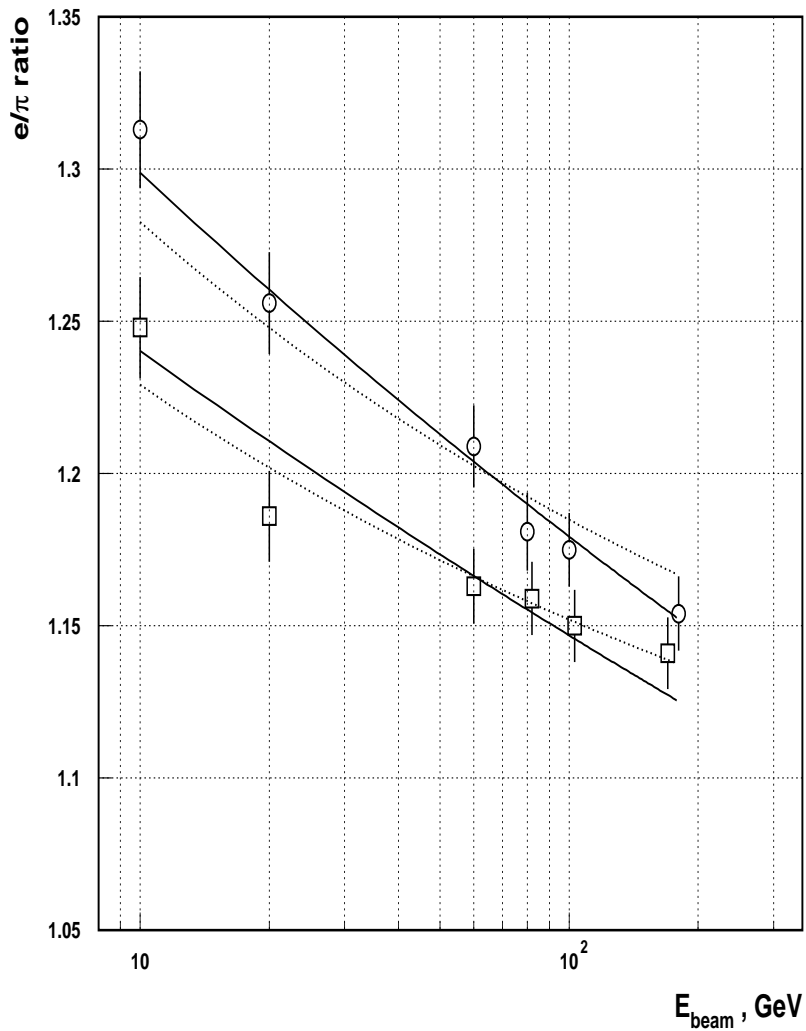,width=0.9\textwidth,height=0.8\textheight}}
     \end{center}
       \caption{
        The $e/ \pi$ ratio as a function of the beam energy for the Module-0:
        the $\circ$ points are the $\eta = -0.25$ data,
        the $\square$ points are the $\eta = -0.55$ data.
	The solid  (dashed) lines are the fits of equation (\ref{ev1}) with
	the Wigmans (Groom) parameterization of $f_{\pi^{o}}(E)$.
       \label{fv4}}
\end{figure*}

\begin{figure*}[tbph]
    \begin{center}
       \mbox{\epsfig{figure=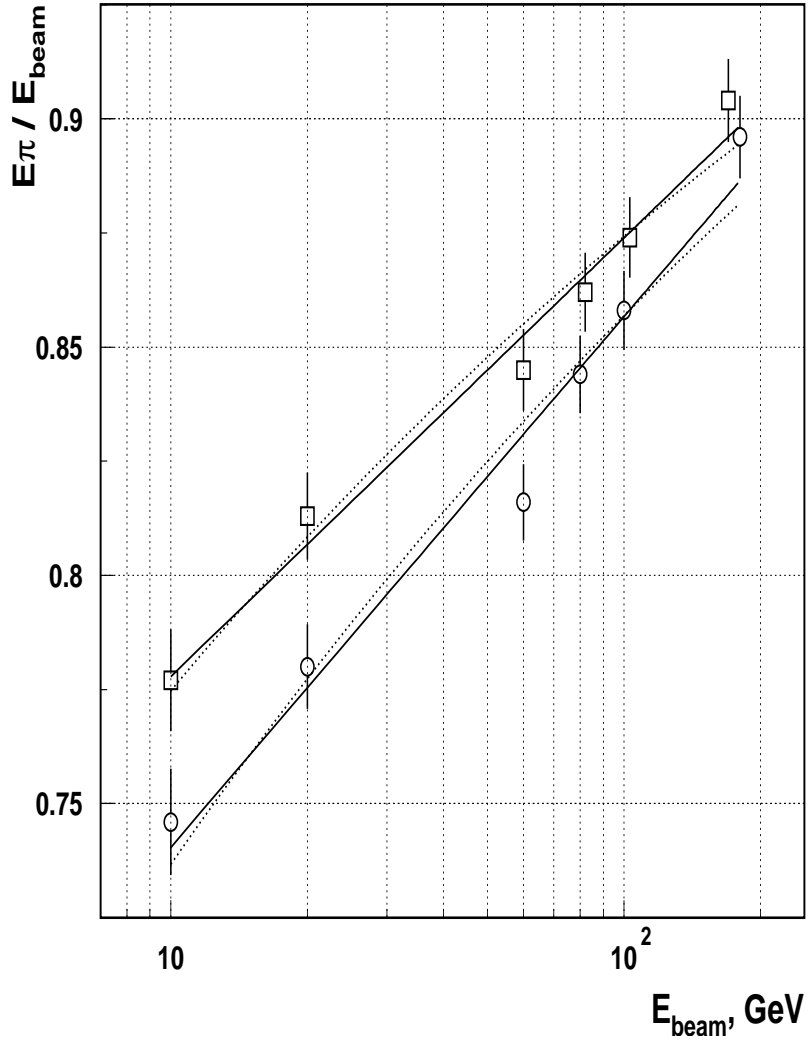,width=0.9\textwidth,height=0.8\textheight}}
     \end{center}
       \caption{
        The mean normalized
	pion response ($E_{\pi} / E_{beam}$)
	as a function of the beam energy for the Module-0:
        the $\circ$ points are the $\eta = -0.25$ data,
        the $\square$ points are the $\eta = -0.55 $ data.
	The solid  (dashed) lines are the fits of equation (\ref{ev1}) with
	the Wigmans (Groom) parameterization of $f_{\pi^{o}}(E)$.
       \label{fv42}}
\end{figure*}

The $e/ \pi$ ratios,  averaged over two 18 mm period,  are
shown in Fig.~\ref{fv4} as a function of the
beam energy.
The errors include statistical errors and a systematic error of 1 \%, 
added in quadrature.

For extracting the $e / h$ ratio we have used two methods:
the standard $e / \pi$ method and the pion response method.

In the first method, the relation between 
the $e / h$ ratio and  the $e / \pi$ ratio
is:
\begin{equation}
\label{ev1}
e/ \pi  = \frac{< E_e >}{< E_{\pi} >}
	= \frac{ e/h }{ 1 + (e/h - 1) \cdot f_{\pi^0} }, 
\end{equation}
where $f_{\pi^0}$ is the average fraction of the energy 
of the incident hadron
going into $\pi^0$ production \cite{wigmans88}. 

In the second method, the relation between the $e / h$ ratio and
the pion response, $< E_{\pi} >$,  is:
\begin{equation}
\label{ev2}
\frac{ < E_{\pi} > }{ E_{beam} } =
\frac{ e }{ e/h } (1 + (e/h - 1) \cdot f_{\pi^0}) , 
\end{equation}
where $e$ is the efficiency for the electron detecting.
Note that usually this is two parameters fit \cite{juste95} 
with parameters $e$ and
$e / h$. 
In principle,  the $e$ value can be determined from the ratio
$e = <E_e> / E_{beam}$.

There are two analytic forms for the intrinsic $\pi^{o}$
fraction suggested by Groom
\cite{groom90}
\begin{equation}
        f_{\pi^{o}} =
                1 - { \biggl( \frac{E}{E_o^{\prime}} \biggr) }^{m-1}
\label{e506-2}
\end{equation}
and Wigmans \cite{wigmans88}
\begin{equation}
        f_{\pi^{o}} =
                k \cdot ln \biggl( {\frac{E}{E_o^{\prime}}} \biggr), 
\label{ev3}
\end{equation}
where $E_o^{\prime} = 1$ GeV,  $m = 0.85$,  $k = 0.11$.

We used both parameterizations.
Fig.~\ref{fv4} shows the $e/ \pi$ ratio as a function of the beam energy
for Module-0 and its fitting of equation (\ref{ev1}) with
the Wigmans (Groom) parameterization of $f_{\pi^{o}}(E)$.

Fig.~\ref{fv42} shows the pion response as a function of the beam energy
for the Module-0 and its fitting of equation (\ref{ev2}) with
the Wigmans (solid line) and
Groom (dashed line) parameterizations of $f_{\pi^{o}}(E)$.

The confidence levels of the fits for these parameterizations
are good,  i.e., $\chi^2$ is less then the numbers of
degrees of freedom.
So, we could obtain four values for the $e / h$ ratio.
The results are presented in Table~\ref{tv8}.

As can be seen, the $e / h$ ratios obtained by the pion response method have
the errors about 10 times larger than obtained by the $e/ \pi$ method.
In addition,  there is some systematic difference:
the $e / h$ ratios, obtained by the pion response method, are 
of 20 -- 40 \% 
larger than ones, obtained by the
$e/ \pi$ method.
This can be explained by some increase in the electron response 
in the 60 -- 180 GeV energy range.
This systematic is cancelled in the $e/ \pi$ method.
It is remarkable that in \cite{juste95}, in which the $e / h$ ratio for the 
1m prototype modules have been determined,  obtained the contrary result
concerning  advantages in using these methods.
Advantage have been observed for the pion response method.
In their case the standard $e/ \pi$ method led to
a larger error (about a factor 2)
than the pion response method called  in this work the non-linearity method.
This can be explained by different scale of errors 
in the corresponding input data.
In their work the $e/ \pi$ ratios had 3 \% errors and the 
pion response values
had 0.3 \% errors.
In our case,  errors in the $e/ \pi$ ratios and the pion response values
have errors at the same 1 \% level.

\begin{figure*}[tbph]
     \begin{center}
       \mbox{\epsfig{figure=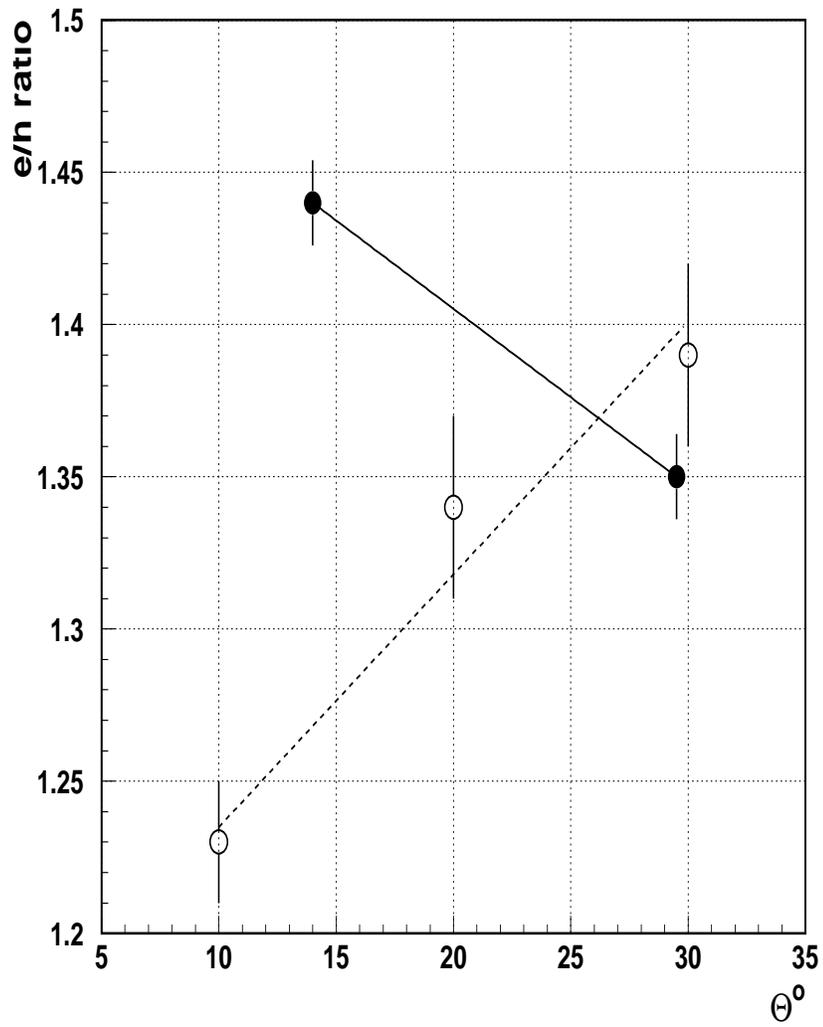,width=0.9\textwidth,height=0.8\textheight}}
     \end{center}
       \caption{
        The $e / h$ ratios for the 
	Module-0 (black points) and the 1m propotype modules
         (open points) as a function of $\Theta$ angle.
       \label{fv20}}
\end{figure*}

\begin{figure*}[tbph]
     \begin{center}
        \begin{tabular}{c}
       \mbox{\epsfig{figure=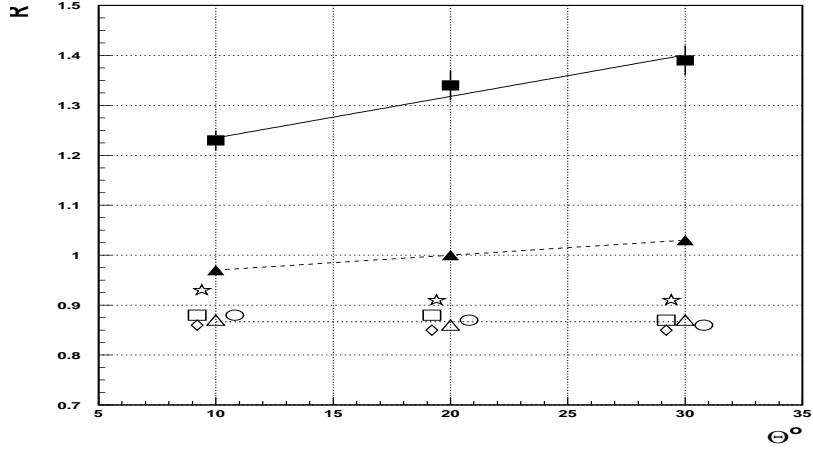,width=0.9\textwidth,height=0.35\textheight}}
\\
       \mbox{\epsfig{figure=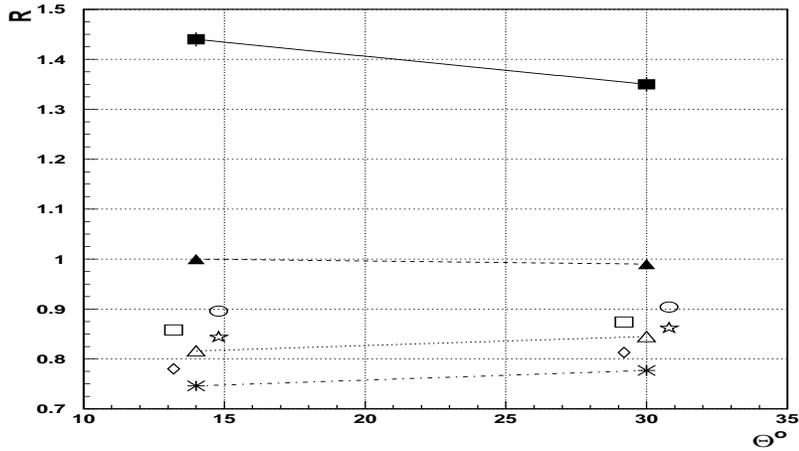,width=0.9\textwidth,height=0.35\textheight}}
\\
        \end{tabular}
     \end{center}
       \caption{
        Top: The $e / h$ ratios, the electron and pion responses for the
	1m propotype modules as a function of $\Theta$ angle.
	$\blacksquare$ are the $e / h$ ratios, $\blacktriangle$ are the 
	electron  response, the rest is the pion response for
	20 ($\lozenge$), 50 ($\triangle$), 100 ($\square$), 
	150 ($\bigcirc$), 
	300 ($\star$) GeV. 
        Bottom: The $e / h$ ratios, the electron and pion responses 
	for the Module-0
        as a function of $\Theta$ angle. $\blacksquare$  
	are the $e / h$ ratios,
	$\blacktriangle$ are the electron  response,
	the rest is the pion response for 10 ($\ast$), 20 ($\lozenge$), 
	60 ($\triangle$), 80 ($\star$), 100 ($\square$), 
	180 ($\bigcirc$) GeV.
	The lines are the results of linear fits.
       \label{fv21}}
\end{figure*}

We made preference to the $e/ \pi$ method and our final results are:
$e / h = 1.45 \pm 0.014$ for $\eta = -0.25$ and
$e / h = 1.36 \pm 0.014$ for $\eta = -0.55$.
Fig.~\ref{fv20} 
shows these values together with ones for the 1m prototype modules
as a function of $\Theta$ angle.
The difference in $\Theta$ behavior is observed.
This can be explained by different behaviour for the electron and pion
responses as a function of $\Theta$
for these two calorimeters as shown in Fig.~\ref{fv21}.
For the Module-0 it is observed slight decrease of the electron response 
and some increase of the pion response.
As a result of the $e/h$ ratio has 6 \% decrease.
 
The simple calculations of the responses by counting 
of the energy depositions
in crossing tiles along the shower axes taking into account the arrangement
and sizes of tiles and the longitudinal shower profiles confirmed these
observations.

\begin{figure*}[tbph]
     \begin{center}
        \begin{tabular}{c}
     \mbox{\epsfig{figure=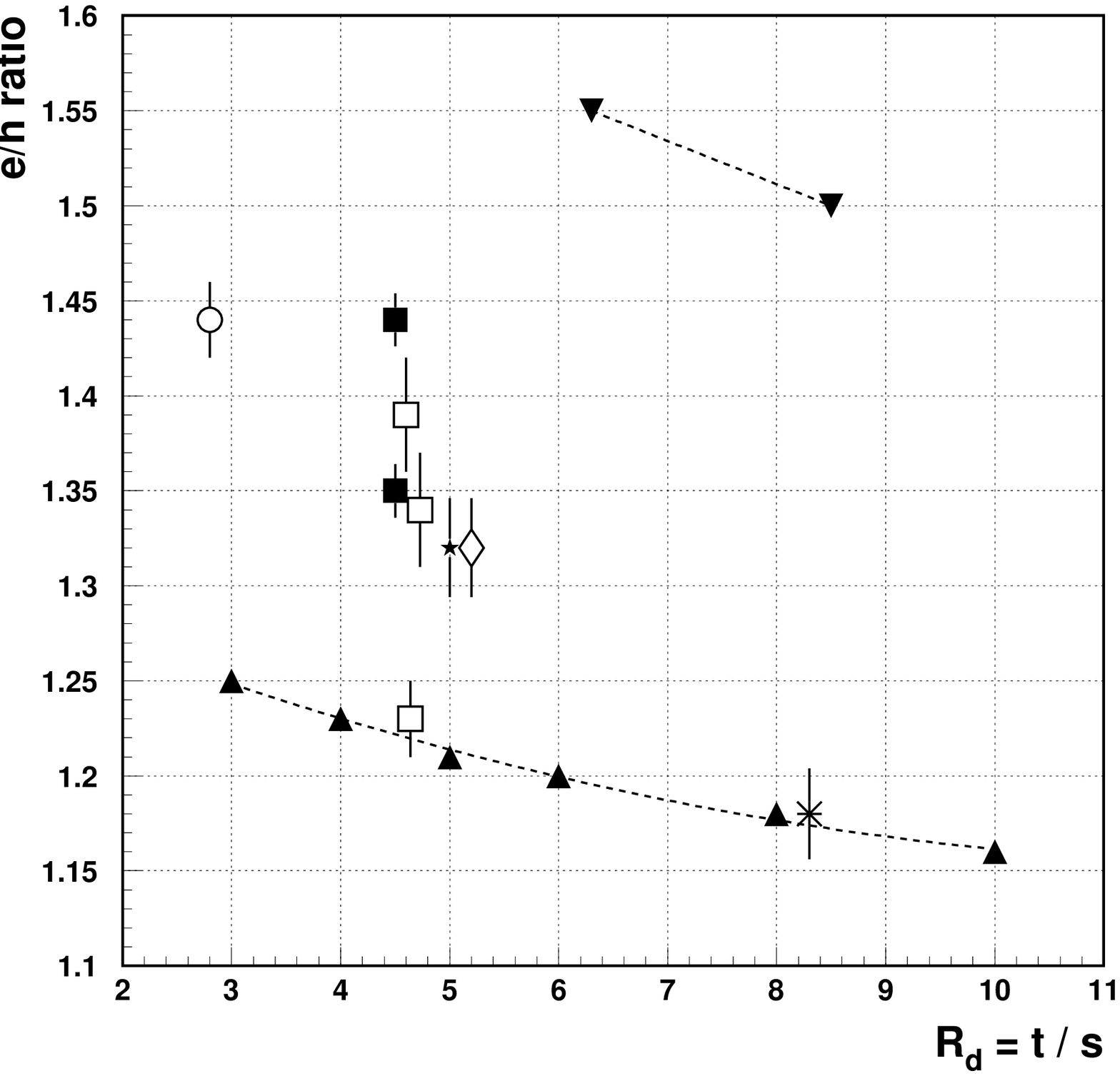,width=0.9\textwidth,height=0.4\textheight}}
\\
     \mbox{\epsfig{figure=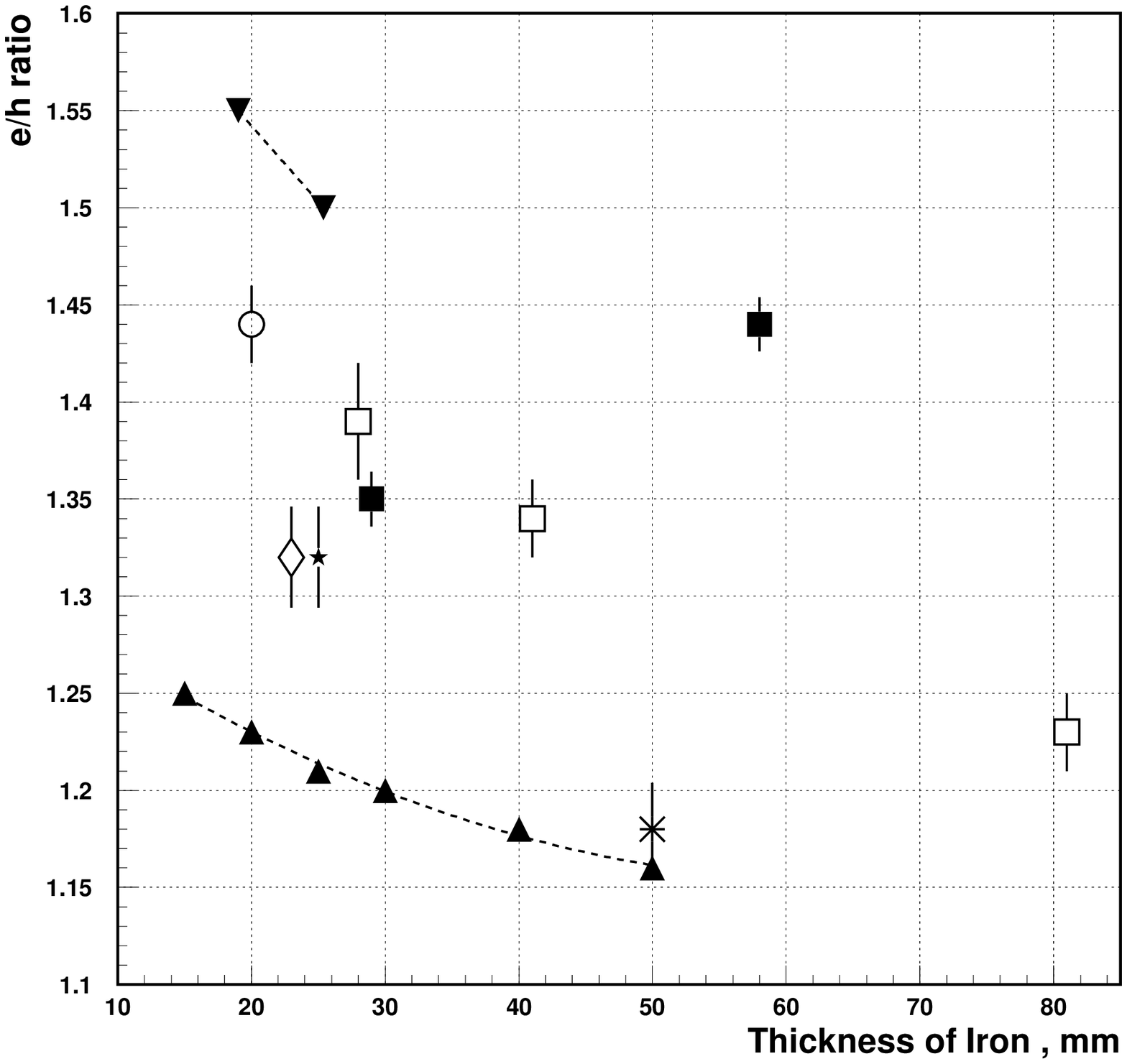,width=0.9\linewidth,height=0.4\textheight}}
\\
        \end{tabular}
     \end{center}
     \caption{
       Top: The $e/ h$-ratios as a function of $R_d$.
       Bottom: The $e/ h$-ratios as a function of iron thickness.
       \label{fv5}}
\end{figure*}

The obtained  $e / h$ values are given in
Table~\ref{tv2}   with the other existing
experimental data and the Monte Carlo calculations
for various iron-scintillator
calorimeters.
The corresponding values of the thickness of the iron absorber
($t$), the thickness of the readout scintillator layers ($s$),
the ratio $R_{d} = t / s$ and the used symbols are also given.
These  $e / h$ values are also shown in Fig.~\ref{fv5} 
as a function of $R_{d}$  ratio and the iron thickness.
As can be seen the $e / h$ ratio has very complicated behaviour being
the function of the thickness of the passive (iron) layers, 
the sampling fraction and,  in our case, 
from the $\Theta$
angle and the sizes and replacement of the scintillator tiles.

Besides,  the considerable disagreement between different
Monte Carlo calculations \cite{wigmans},  \cite{gabriel} 
and experimental data
is observed.

%%%%%%%%%%%%%%%%%%%%%%%%%%%%%%%%%%%%%%%%%%%%%%%%%%%%%%%%%%%%%%%%%%%%%%%%%%
\section{Conclusions}

The detailed experimental information about the electron and pion responses,
the electron energy resolution
and the $e / h$ ratios as a function of the incident energy $E$,  
the impact point $Z$
and the incidence
angle $\Theta$ of the Module-0 of the ATLAS iron-scintillator barrel hadron
calorimeter with
the longitudinal tile configuration is obtained.
The results are compared with the existing experimental data,
obtained for the 1m prototype modules and    
the various iron-scintillator calorimeters, and with the
Monte Carlo calculations.
It is shown that the $e / h$ ratio has very complicated behaviour being
the function of the thickness of the passive (iron) layers, 
the sampling fraction and,  in our case, 
from the $\Theta$ angle and the sizes and 
replacement of the scintillator tiles.

%%%%%%%%%%%%%%%%%%%%%%%%%%%%%%%%%%%%%%%%%%%%%%%%%%%%%%%%%%%%%%%%%%%%%%%%%%
\section{Acknowledgments}

This work is the result of the efforts of many people from the ATLAS
Collaboration.
The authors are greatly indebted to all Collaboration
for their test beam setup and data taking.
The authors are thankful M.~Nessi and J.~Budagov for their attention
and support of this work.

%%%%%%%%%%%%%%%%%%%%%%%%%%%%%%%%%%%%%%%%%%%%%%%%%%%%%%%%%%%%%%%%%%%%%%%%%%

%%%%%%%%%%%%%%%%%%%%%%%%%%%%%%%%%%%%%%%%%%%%%%%%%%%%%%%%%%%%%%%%%%%%%%%%%%
% TABLE %
%%%%%%%%%%%%%%%%%%%%%%%%%%%%%%%%%%%%%%%%%%%%%%%%%%%%%%%%%%%%%%%%%%%%%%%%%%
%1
\begin{table}[tbph]
\caption{
        The values of parameter 
	$a_{exp}$ and $b_{exp}$ of the electron energy resolution
 	for various
        iron-scintillator calorimeters.
$a_{th}$ is the prediction of the parameterization of Del Peso et al.
\label{Tb11}}
\begin{center}
\begin{tabular}{|l|l|c|c|c|c|c|}
\hline
Author     & Ref.             &$t$&$s$&$a_{exp}$ &$b_{exp}$&$a_{th}$ \\
\hline
\hline
Stone      & \cite{stone}     &4.8&6.3& 10. &         & 7.0            \\
\hline
Antipov    & \cite{antipov}   &20.&5.0& 27. &         & 17.               \\
\hline
Abramovicz & \cite{abram}     &25.&5.0& 23.  & &20.            \\
\hline
Mod.0, 30$^o$ &      &28.&6.0&$33.\pm 2.$&$1.\pm 0.3 $& 20.     \\
\hline
1m pr., 30$^o$ &\cite{budagov95} &28.&6.0& $33.\pm 9.$&$0.1\pm 0.8 $& 20.  \\
\hline
1m pr., 20$^o$ &\cite{budagov95} &41.&9.& $36.\pm 5.$&$0.8\pm 5.0 $& 20.    \\
\hline
Mod.0, 14$^o$ &      &58.&12.0&$ 32.\pm 4.$&$2.5\pm 0.5 $& 27.     \\
\hline
1m pr., 10$^o$ & \cite{budagov95} &81.&17.& $58.\pm 4.$&$1.4 \pm 0.4 $& 32.\\
\hline  
\end{tabular}
\end{center}
\end{table}

%2
\begin{table}[tbph]
\caption{
        The values of the $e / h$ ratio for different methods and 
	$f_{\pi^0} (E)$
        parameterizations 
	(W -- the Wigmans parameterization,  
	 G -- the Groom parameterization).
\label{tv8}}
\begin{center}
\begin{tabular}{|c|c|c|c|}
\hline
 Method & $f_{\pi^{o}}(E)$ & \multicolumn{2}{|c|}{$e / h$}  \\
\hline
        &                  & $\eta=-0.25$  & $\eta=-0.55$   \\
\hline
$e/\pi$ &   W              & 1.45$\pm$0.014&1.35$\pm$0.013  \\
\cline{2-4}
        &   G              & 1.45$\pm$0.015&1.36$\pm$0.013  \\
\hline
$\pi$   &   W              & 1.72$\pm$0.11 &1.56$\pm$0.07   \\
\cline{2-4}
        &   G              & 2.00$\pm$0.19 &1.76$\pm$0.11   \\
\hline
\end{tabular}
\end{center}
\end{table}

%3
\begin{table}[tbph]
\caption{
        The $e/h$ ratios for our and various iron-scintillator calorimeters.
	$t$  is the thickness of the iron absorber, 
	$s$ is the thickness of the readout scintillator layers and
	the ratio $R_{d} = t / s$.
\label{tv2}}
\begin{center}
\begin{tabular}{|l|l|c|c|c|c|c|}
\hline
Author     & Ref.\                &$R_d$&$t$,  mm&$s$,  mm&$e / h$ & Symb.\ \\
\hline
\hline
Bohmer  & \cite{bohmer}~$^{**}$   &2.8  &20.    &7.0    &1.44$\pm$0.03&
{ $\circ$} \\
\hline
Wigmans     & \cite{wigmans}~$^*$ &3.0  &15.    &5.0    &1.25   &
$\blacktriangle$ \\
\hline
Wigmans     & \cite{wigmans}~$^*$ &4.0  &20.    &5.0    &1.23   &
$\blacktriangle$ \\
\hline
Module-0,  30$^o$&               &4.7  &28.    &6.0    &$1.36 \pm 0.014$ &
$\blacksquare$ \\
\hline
1m prot.,  30$^o$&\cite{budagov95}&4.7  &28.    &6.0    &$1.39 \pm 0.03$ &
$\square$ \\
\hline
1m prot.,  20$^o$&\cite{budagov95}&4.7  &41.    &9.0    &$1.34 \pm 0.03$ &
$\square$ \\
\hline
Module-0,  14$^o$&               &4.7  &58.    &12.    &$1.45 \pm 0.014$ &
$\blacksquare$ \\
\hline
1m prot.,  10$^o$&\cite{budagov95}&4.7  &81.    &17.    &$1.23 \pm 0.02$ &
$\square$ \\
\hline
Wigmans     & \cite{wigmans}~$^*$ &5.0  &25.    &5.0    &1.21   &
$\blacktriangle$      \\
\hline
Abramovicz  & \cite{abram}~$^{**}$        &5.0  &25.&5.0&1.32$\pm$0.03   &
$\lozenge$  \\
\hline
Vincenzi    & \cite{vince}~$^{**}$        &5.0  &25.&5.0&1.32$\pm$0.03   &
$\bigstar$ \\
\hline
Wigmans     & \cite{wigmans}~$^*$ &6.0  &30.    &5.0    &1.20   &
$\blacktriangle$      \\
\hline
Gabriel     & \cite{gabriel}~$^*$ &6.3  &19.    &3.0    &1.55   &
$\blacktriangledown$       \\
\hline
Wigmans     & \cite{wigmans}~$^*$ &8.0  &40.    &5.0    &1.18   &
$\blacktriangle$      \\
\hline
Holder      & \cite{holder}~$^{**}$       &8.3  &50.&6.0&1.18$\pm$0.02   &
{\Large $\ast$}  \\
\hline
Gabriel     & \cite{gabriel}~$^*$ &8.5  &25.4   &3.0    &1.50   &
$\blacktriangledown$      \\
\hline
Wigmans     & \cite{wigmans}~$^*$ &10.  &50.    &5.0    &1.16   &
$\blacktriangle$      \\
\hline
\multicolumn{7}{l}{}\\[-2mm]
\cline{1-3}
\multicolumn{7}{l}{}\\[-4mm]
\multicolumn{7}{l}{$^*$~Monte Carlo calculations}\\
\multicolumn{7}{l}{$^{**}$~The our estimate of 2 \% error is given}\\
\end{tabular}
\end{center}
\end{table}

%%%%%%%%%%%%%%%%%%%%%%%%%%%%%%%%%%%%%%%%%%%%%%%%%%%%%%%%%%%%%%%%%%%%%%%%%%%%
% FIGURES
%%%%%%%%%%%%%%%%%%%%%%%%%%%%%%%%%%%%%%%%%%%%%%%%%%%%%%%%%%%%%%%%%%%%%%%%%%%%

%1
\clearpage
\newpage

%2
\clearpage
\newpage

%3
\clearpage
\newpage

%4
\clearpage
\newpage

%5
\clearpage
\newpage

%6
\clearpage
\newpage

%7
\clearpage
\newpage

%8
\clearpage
\newpage

%9
\clearpage
\newpage

%10
\clearpage
\newpage

%11
\clearpage
\newpage

%12
\clearpage
\newpage

%13
\clearpage
\newpage

%14
\clearpage
\newpage

%15
\clearpage
\newpage

%16
\clearpage
\newpage

%17
\clearpage
\newpage

%18
\clearpage
\newpage

%19
\clearpage
\newpage

%20
\clearpage
\newpage

%%%%%%%%%%%%%%%%%%%%%%%%%%%%%%%%%%%%%%%%%%%%%%%%%%%%%%%%%%%%%%%%%%%%%%%%%%%%

\begin{thebibliography}{99}
%1
\bibitem{atcol}
ATLAS Collaboration,  
ATLAS Technical Proposal for a General-Pur\-po\-se
pp Experiment at the Large Hadron Collider,  
CERN/ LHCC/ 94-93, CERN,   Geneva,   Switzerland,  1994.
%2
\bibitem{lhcnews}
LHC News,  $N^{\b{o}}$ 7 September 1995,  CERN,  Geneva,  Switzerland.
%3
\bibitem{gild-91}
O.~Gildemeister,  F.~Nessi-Tedaldi and M.~Nessi, 
Proc.~2nd Int. Conf. on Cal.~in HEP,  Capri,  1991.
%4
\bibitem{tdr96}
ATLAS Collaboration, 
ATLAS TILE Calorimeter Technical Design Report, 
CERN/ LHCC/ 96-42,  ATLAS TDR 3,  CERN,   Geneva,   Switzerland,  1996.
%5
\bibitem{budagov95}
J.A.\ Budagov,   Y.A.\ Kulchitsky,   V.B.\ Vinogradov $et\ al.$, 
JINR,  E1-95-513,  Dubna,  Russia,  1995;
ATLAS Internal note,  TILECAL-No-72,  CERN,  Geneva,  Switzerland,  1995.
%6
\bibitem{berger}
E.~Berger et.\ al., 
CERN/LHCC 95-44, CERN,  Geneva,  Switzerland.
%7
\bibitem{ariz-94}
F.~Ariztizabal et.\ al.,  NIM A349 (1994) 384.
%9
\bibitem{juste95}
A.\ Juste, 
ATLAS Internal note,  TILECAL-No-69,  1995,  CERN,  Geneva,  Switzerland.
%10
\bibitem{abshire}
G.\ Abshire et.\ al., NIM 164 (1979) 67.
%10-1
\bibitem{delpe}
J.~Del~Peso,  E.~Ros,  NIM A276 (1989) 456.
%11
\bibitem{groom90}
D.\ Groom, Proceedings of the Workshop on Calorimetry for the Supercollides, 
Tuscaloosa, Alabama, USA, 1990.
%12
\bibitem{wigmans88}
R.~Wigmans,  NIM A265 (1988) 273.
%13
\bibitem{wigmans}
R.~Wigmans,  NIM A259 (1987) 389.
%14
\bibitem{gabriel}
T.~A.~Gabriel et.\ al.,  NIM A295 (1994) 336.
%15
\bibitem{stone}
S.~L.~Stone et.\ al.,  NIM 151 (1978) 387.
%16
\bibitem{antipov}
Y.~A.~Antipov et.\ al.,  NIM 180 (1990) 81.
%17
\bibitem{abram}
H.~Abramowicz et.\ al.,  NIM 180 (1981) 429.
%18
\bibitem{bohmer}
V.~Bohmer et.\ al.,  NIM 122 (1974) 313.
%19
\bibitem{vince}
M.~De~Vincenze et.\ al.,  NIM A243 (1986) 348.
%20
\bibitem{holder}
M.~Holder et.\ al.,  NIM 151 (1978) 69.
\end{thebibliography}
\end{document}